%% file: Characterisation_181114.tex
\documentclass[a4paper,english]{elsarticle}
\usepackage[T1]{fontenc}
\usepackage[latin9]{inputenc}
\usepackage{varioref}
\usepackage{units}
\usepackage{amsmath}
\usepackage{amssymb}
\usepackage{graphicx}
\usepackage{esint}
\usepackage{caption}
\usepackage{subcaption}

\makeatletter

\pdfpageheight\paperheight
\pdfpagewidth\paperwidth

\journal{Elsevier}
\usepackage{upgreek}
\usepackage[bookmarks,bookmarksopen,bookmarksdepth=6]{hyperref}
\usepackage{cases}
\usepackage{url}
\usepackage{lineno}

\usepackage{a4wide}

\makeatother

\usepackage{babel}
\begin{document}

\title{Characterisation of SiPMs}

\author[]{R.~Klanner \corref{cor1}}

  \cortext[cor1]{Email: Robert.Klanner@desy.de, Tel. +49 40 8998 2558.}
   \address{Institute for Experimental Physics, University of Hamburg,
   \\Luruper Chaussee 149, D\,22761, Hamburg, Germany.}

\begin{abstract}

 Silicon photomultipliers, thanks to their excellent performance, robustness and relatively simple use, are the photon-detectors of choice for many present and future applications.
 This paper gives an overview of  methods to characterise SiPMs.
 The different SiPM parameters are introduced and generic setups for their determination presented.
 Finally, ways to extract the parameters from the measurements are discussed and the results  shown.
 If a parameter can be obtained from different measurements, the results are compared and recommendations given, which is considered to be the most reliable.
 The characterisation of SiPMs, in particular for high light intensities and in high radiation fields, is presently a field of intensive research with many open questions and problems which will be discussed.

\end{abstract}


\maketitle
 \tableofcontents
 \pagenumbering{arabic}


\section{Introduction}
 \label{sect:Introduction}

 In this contribution an overview of different methods of characterising SiPMs is given.
 After a short discussion of the most relevant parameters and their relation to the electrical parameters of SiPMs, generic measurement setups are presented.
 Finally, methods how the SiPM parameters can be determined with the different setups are presented, and their advantages, disadvantages and limitations discussed.

 Several groups have developed  methods of characterising SiPMs and most of them are well documented in publications.
 As it is not possible to do justice to all this work, only generic setups and analysis methods are presented.
 One complication is that the different groups use different symbols for the technical terms.
 In addition, these are not always clearly defined.
 The next section is an attempt to give clear definitions and to summarise the symbols used in this paper in a table.
 Clearly a common nomenclature is more than welcome, and efforts towards this goal are important for the advancement of the field\,\cite{ICASiPM}.

 The emphasis of the paper is on the characterisation of Analog SiPMs, on which most of the work has been done so far.
 This in no way means that the development of Digital SiPMs is not appreciated by the author.
 In fact the opposite is true, and given the impressive developments of microelectronics and 3-D integration, Digital SiPMs may well surpass in the future Analog SiPMs in many applications.

 The paper does not cover the excellent timing performance of SiPMs and its measurements, which however is discussed in other articles of this Special Issue (\cite{Piemonte:2018, Acerbi:2018, Marzocca:2018, Simon:2018}).

 As this is a review paper, most of the results are  based on discussions with colleagues or on published papers, for which the sources are quoted.
 If no reference is given, the results are from measurements by members of the Hamburg Detector Laboratory with the analysis performed by the author.
 Most of these studies used SiPMs produced by KETEK, as for these devices we have access to the technological information to perform simulations.

 \section{SiPM parameters}
  \label{sect:Parameters}

  Silicon Photomultipliers, also referred to as SiPM (Silicon photomultiplier or Silicon Photo Multiplier), MPPC (Multi Pixel Photon Counter) or G-APD (Geiger Mode Avalanche Photo Diode, which however is mainly used for single pixel devices) are two dimensional arrays of 100 to several 10 000 single photon avalanche diodes (SPAD), called pixels, with typical dimensions between $10\,\upmu \mathrm{m} \times 10\,\upmu \mathrm{m}$ and $100\,\upmu \mathrm{m} \times 100\,\upmu \mathrm{m}$.
  The pixels are operated in limited Geiger mode and every pixel gives approximately the same signal, independent of the number of photons which have produced simultaneously electron-hole pairs in the amplification region of the pixel.
  The sum of the pixel signals is proportional to the number of pixels with Geiger discharges, from which the number of incident photons is deduced.
  As the output charge for a single Geiger discharge is typically larger than $10^5$ elementary charges, 0, 1, 2, and more Geiger discharges can be easily distinguished, enabling the detection of single optical photons with high efficiency and sub-nanosecond timing.

  Two types of SiPMs have been developed: Analog and Digital.
  In Analog SiPMs the individual SPADs are connected via quenching resistors to a common readout and the SiPM delivers the summed analog signal.
  In Digital SiPMs each pixel has its own quenching circuit and a  digital switch to a multi-channel readout system.
  The output is the digitised pulse height and precise time information for the pixels with Geiger discharges.
  Digital SiPMs also allow disabling pixels with high dark-count rates.

\begin{figure}[!ht]
   \centering
   \begin{subfigure}[a]{0.45\textwidth}
    \includegraphics[width=\textwidth]{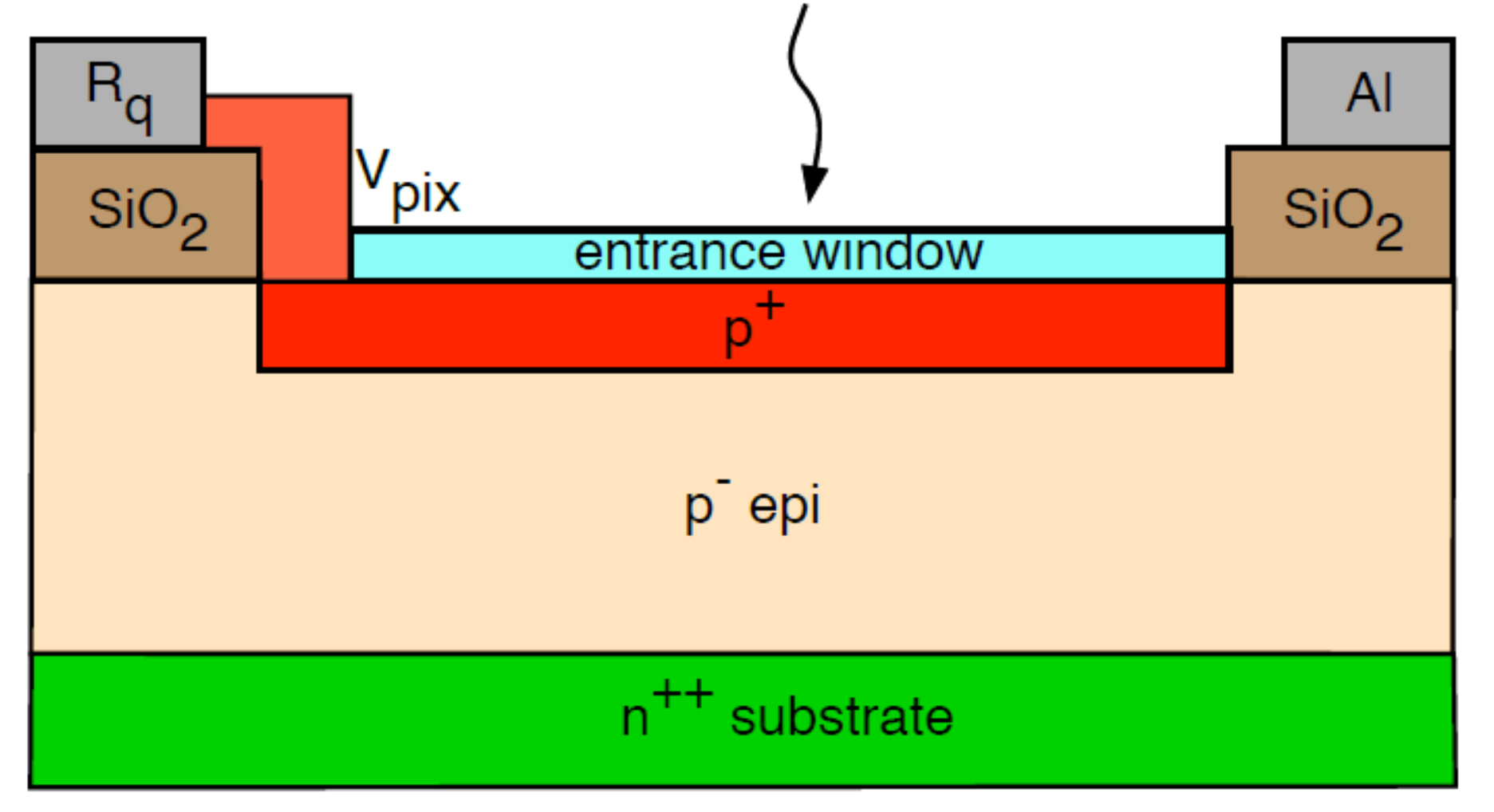}
    \caption{ }
    \label{fig:SchemaSiPM}
   \end{subfigure}%
    ~
   \begin{subfigure}[a]{0.4\textwidth}
    \includegraphics[width=\textwidth]{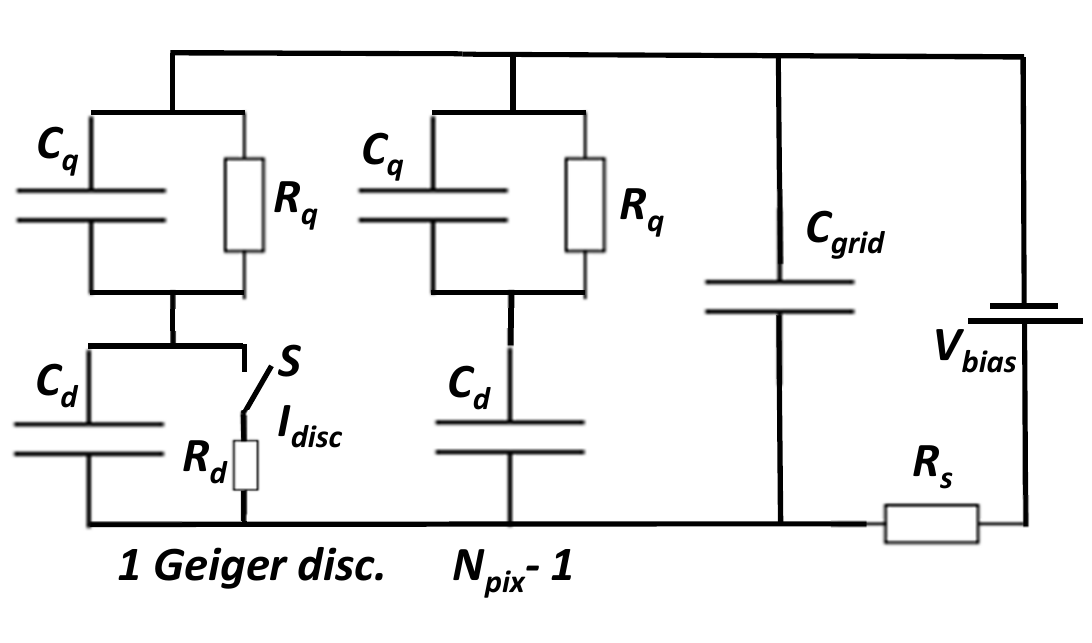}
    \caption{ }
    \label{fig:Emodel}
   \end{subfigure}%
   \caption{(a) Example of a possible cross section of a single pixel of a SiPM.
    $R_q$ represents the quenching resistance, and Al the biasing grid.
    (b) Electrical model for voltages above the breakdown voltage, $V_{bd}$, of a SiPM with $N_{pix}$ pixels and a single Geiger discharge.
   $I_{disc} (t)$ represents the discharge current of the pixel by the Geiger discharge, described by a switch and the resistor $R_d$.
   $I_{disc} (t)$ can also be simulated by a current source with a time-dependent current.
   $C_{grid}$ is the capacitance between the Al-grid which connects the individual pixels to the bias voltage and the substrate.
    The shunt resistor $R_s$ converts the current signal into a voltage, which is sensed by the readout.
    Most of the simulations found in the literature use a model with a voltage source $(V = V_{bd})$ in series with switch $S$ and the resistor $R_d$.
    For an explanation of the other symbols, see text.}

  \label{fig:SiPM}
 \end{figure}

 The basic functioning of a SiPM, as well as the  terms required for its description are explained with the help of Fig.\,\ref{fig:SiPM}, which shows an example of a possible cross section of a single pixel and the electrical diagram used by the author to simulate the pulses for a SiPM with $N_{pix}$ pixels.
 More realistic pixel layouts are discussed in Ref.\,\cite{Acerbi:2018}.
 Different to Fig.\,\ref{fig:Emodel}, in most of the literature (e.\,g. Refs.\,\cite{Piemonte:2018, Acerbi:2018, Marzocca:2018}) a voltage source with a value of the break-down voltage is implemented in series with the switch $S$ and the resistance $R_d$, or more general, the switch $S$ and the resistor $R_d$ are replaced by a time dependent current source.
 Which of the  models is the more appropriate one, is at present an open question.

\begin{figure}[!ht]
   \centering
   \begin{subfigure}[a]{0.5\textwidth}
    \includegraphics[width=\textwidth]{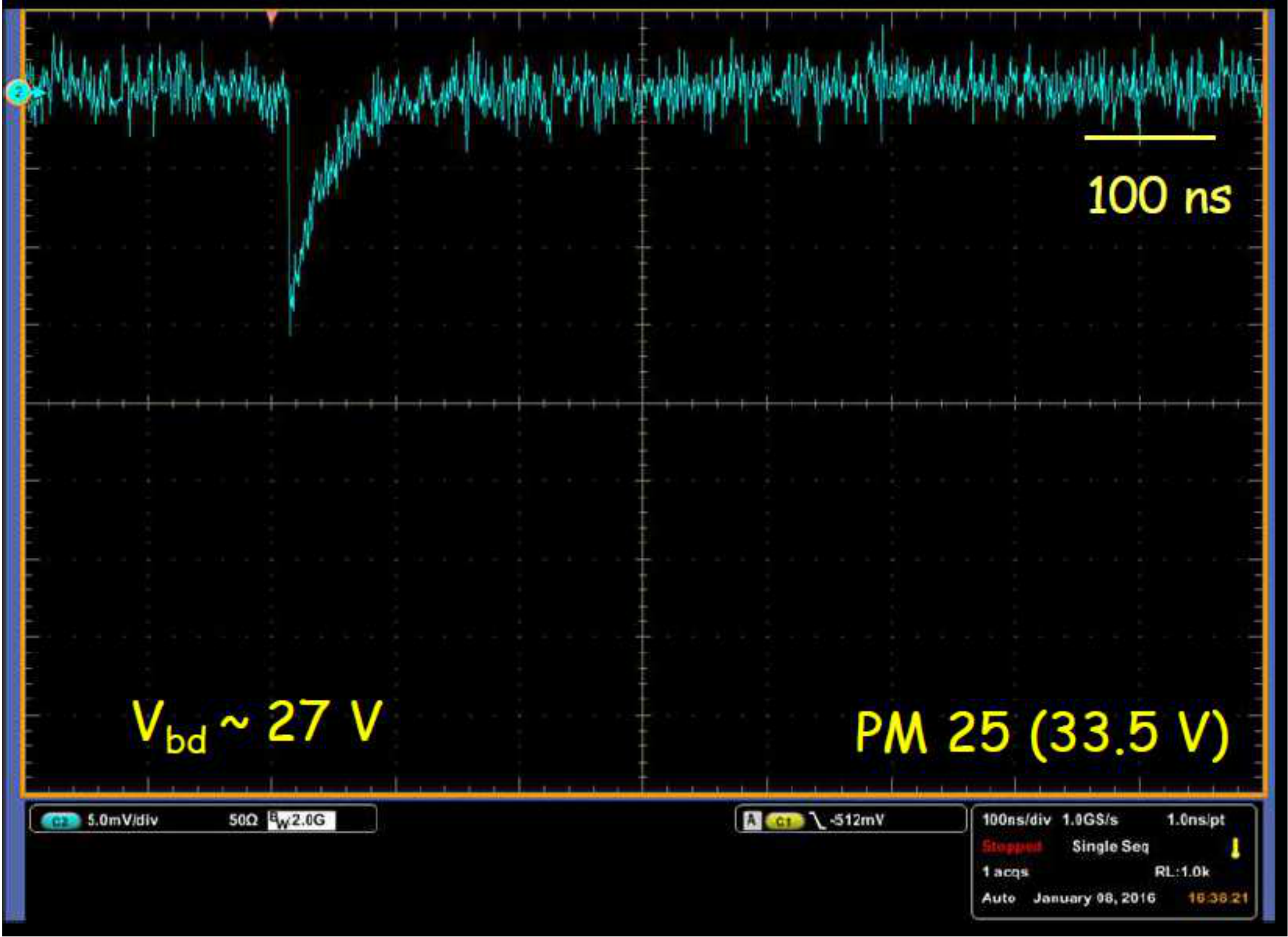}
    \caption{ }
    \label{fig:PulsePM25}
   \end{subfigure}%
    ~
   \begin{subfigure}[a]{0.5\textwidth}
    \includegraphics[width=\textwidth]{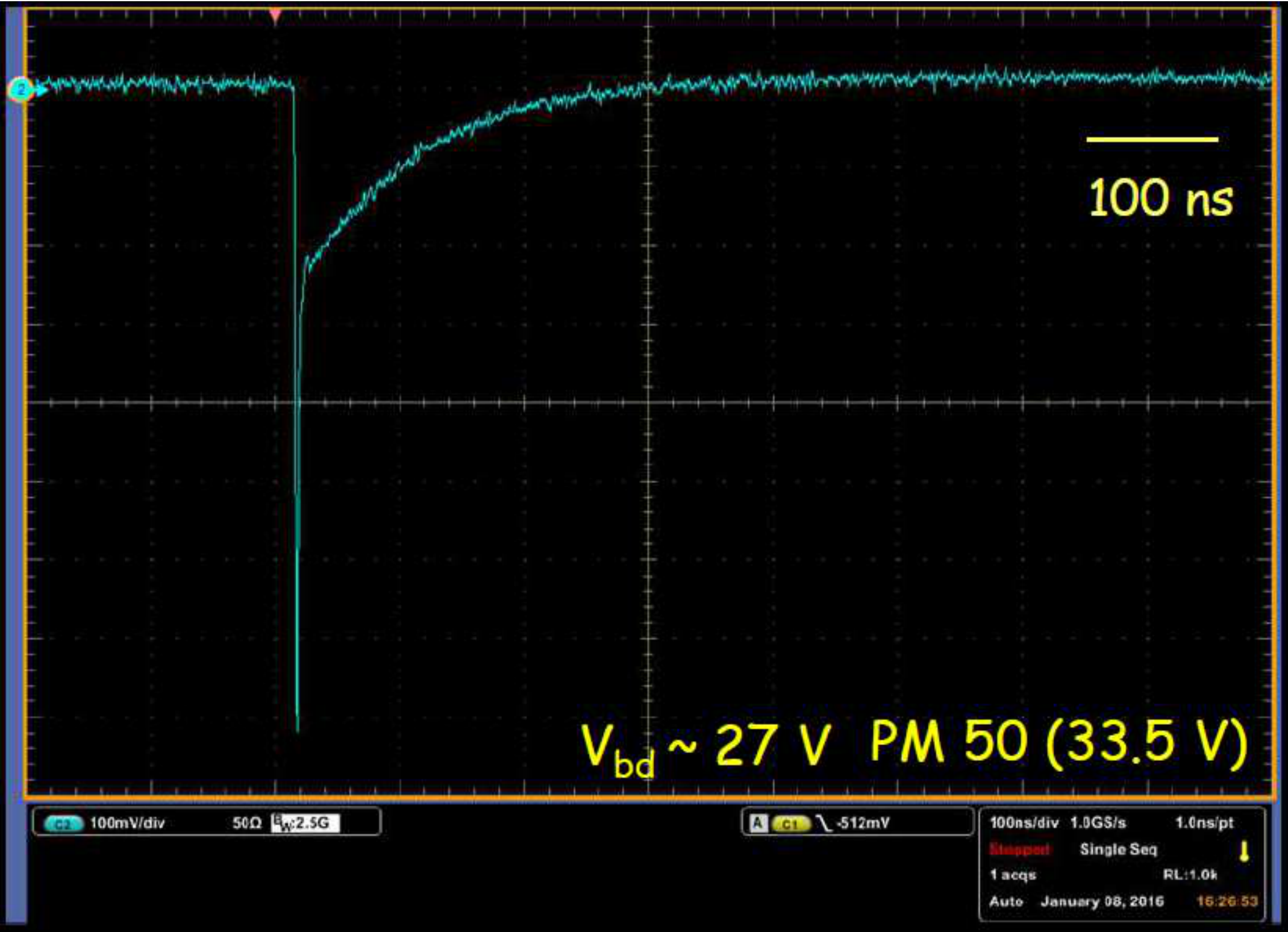}
    \caption{ }
    \label{fig:PulsePM50}
   \end{subfigure}%
   \caption{Pulse shape of a single Geiger discharge for (a) KETEK SiPM with 25\,$\mu$m, and (b) 50\,$\mu$m pixel size, measured at $V_{bias} = 33.5$\,V (colour online).}
  \label{fig:Pulse}
 \end{figure}

 The  terms used and the corresponding symbols are summarised in Table\,\ref{table:parameters}.
 Fig.\,\ref{fig:Pulse} shows examples of pulses from single Geiger discharges for a SiPM from KETEK with (a) a  pixel size of 25\,$\upmu \mathrm{m}$, and (b) of 50\,$\upmu \mathrm{m}$.

 Each pixel is connected to the power supply ($V_{bias}$) by the quenching resistance $R_q$.
 Parallel to $R_q$ there is a capacitance $C_q$.
 It is the parasitic capacitance of the quenching resistor to the Si-bulk of the pixel and can be intentionally increased to produce a narrow initial pulse allowing a better signal extraction.
 The capacitance of the diode corresponding to a single pixel is denoted by $C_d$.
 $R_s$ is the shunt resistor which converts the current signal into a voltage, which is sensed by the readout.
 The photon enters the SiPM through a window, which is typically covered by an anti-reflective coating (ARC).
 The ratio of the area of the entrance window to the pixel area is usually called fill factor, $FF$.

 The SiPM is biased by a voltage $V_{OV}$ above the breakdown voltage $V_{bd}$\,: $V_{OV} = V_{bias} - V_{bd} $.
 In the quiescent state no current flows through $R_q$ and the voltage over the pixel is $V_d = V_{bias}$.
 An $eh$ (electron hole) pair, produced either thermally, by a photon or by ionising radiation, initiates with the trigger probability $P_T$ a Geiger discharge by avalanche multiplication.
 $P_T$ is a function of the position where the $eh$\,pair is generated.
 A quantitative model for this dependence is given in Ref.\,\cite{McIntyre:1973}.
 The discharge takes place through a narrow ($\approx 10\,\upmu$m diameter) micro-plasma tube until the turn-off voltage $V_{off}$ is reached, when the multiplication is too low to maintain the micro-plasma.
 In the electrical model shown in Fig.\,\ref{fig:Emodel} the switch $S$ is closed at the start of the Geiger discharge, $C_d$ is discharged through $R_d$, and the switch opens when $V_d = V_{off}$.
 The observed differences $V_{bd} - V_{off}$ are small ($ < 1$\,V) and frequently compatible with zero, and in most of the literature just $V_{bd}$ is used.
 The only paper which reports a significant difference is Ref.\,\cite{Chmill1:2017}, and Ref.\,\cite{Marinov:2007} presents  a model calculation of a Geiger discharge and derives a formula for $V_{bd} - V_{off}$.
 The time constant of the pixel discharge
 is short compared to 1\,ns.
 It is responsible for the fast rise time of the output pulse seen in Fig.\,\ref{fig:PulsePM50}.
 As discussed in Refs.\,\cite{Corsi:2006, Piemonte:2007} the decay of the measured pulse has two time constants: a fast one $\tau _{in} \approx R_s \cdot C_{eq}$ and a slower one $\tau _r = R_q \cdot (C_d + C_q)$.
 $\tau _r$\,describes the recharging of the pixel after the switch $S$ in Fig.\,\ref{fig:Emodel} has opened at the end of the Geiger discharge.
 $\tau _{in}$ is associated with the shunt resistance $R_s$  and the total capacitance $C_{eq} \approx N_{pix} \big( 1/C_d + 1/C_q \big)^{-1}$ seen by the amplifier.
 The total charge of the SiPM pulse is  $Q = (V_{bias} - V_{off}) \cdot (C_d + C_q)$, and the voltage at the peak  $V_{max} \approx (Q/C_{eq}) \cdot \big(C_q/(C_q + C_d) \big)$.
 The latter formula assumes that the bandwidth of the readout is sufficiently high not to degrade the signal, which is a quite challenging requirement.
 From Fig.\,\ref{fig:Pulse} one sees that, contrary to the SiPM with 50\,$\upmu $m pixels, there is no evidence for a fast component for the 25\,$\upmu $m SiPM.
 It is concluded that in the latter case $C_q \ll C_d$.

 A parameter which is of particular relevance for the characterisation of SiPMs is the gain
  \begin{equation}\label{equ:Gain}
   G = \frac {(C_d + C_q ) \cdot (V_{bias} - V_{off}) } {q_0} \hspace{5mm} \mathrm{and} \hspace{5mm} G^\ast = G \cdot f_Q,
 \end{equation}
 The elementary charge is denoted $q_0$. If the entire signal is integrated, the gain is $G$.
 An integration window which is shorter than the pulse or pulse shaping by the readout electronics, result in a gain, $G^\ast $,  which is reduced by a factor $f_Q \leq 1$.

\begin{figure}[!ht]
   \centering
   \begin{subfigure}[a]{0.5\textwidth}
    \includegraphics[width=\textwidth]{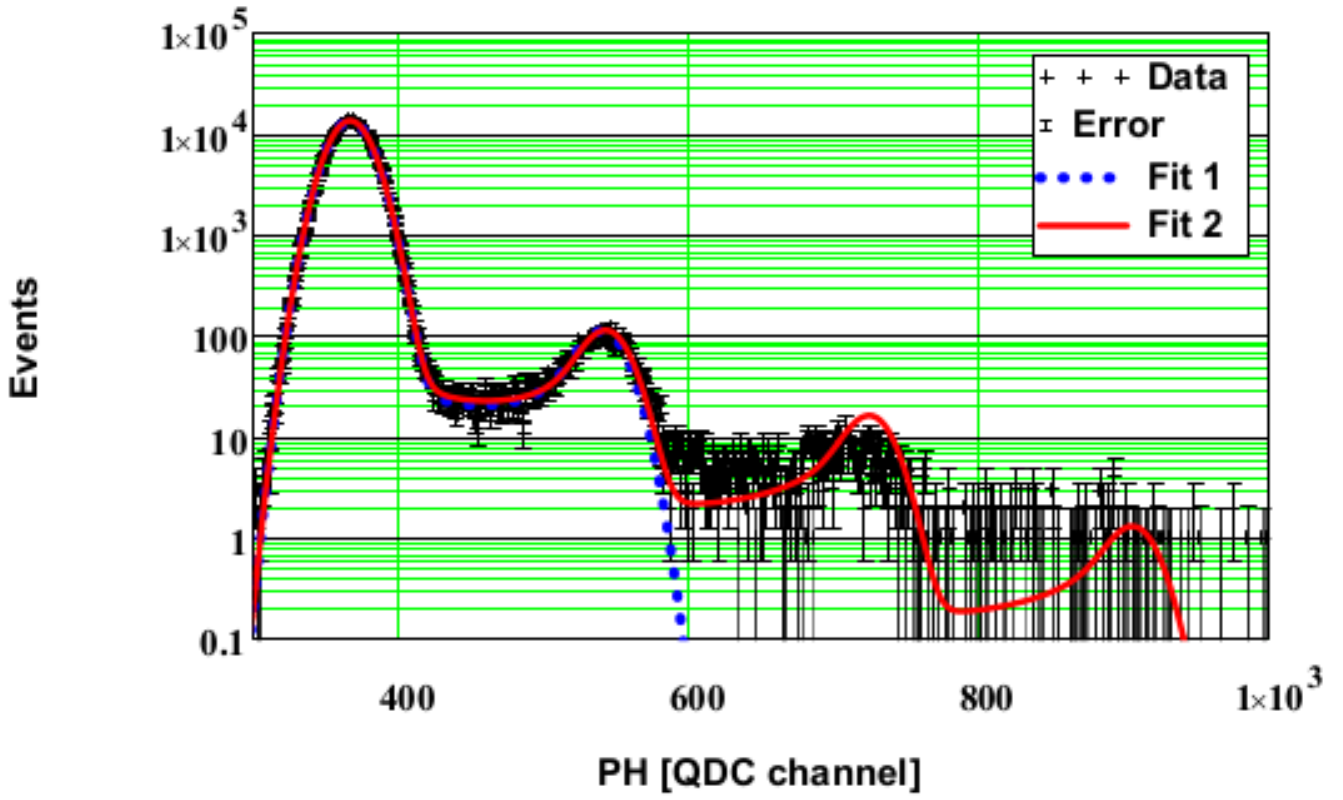}
    \caption{ }
    \label{fig:Dark33}
   \end{subfigure}%
    ~
   \begin{subfigure}[a]{0.5\textwidth}
    \includegraphics[width=\textwidth]{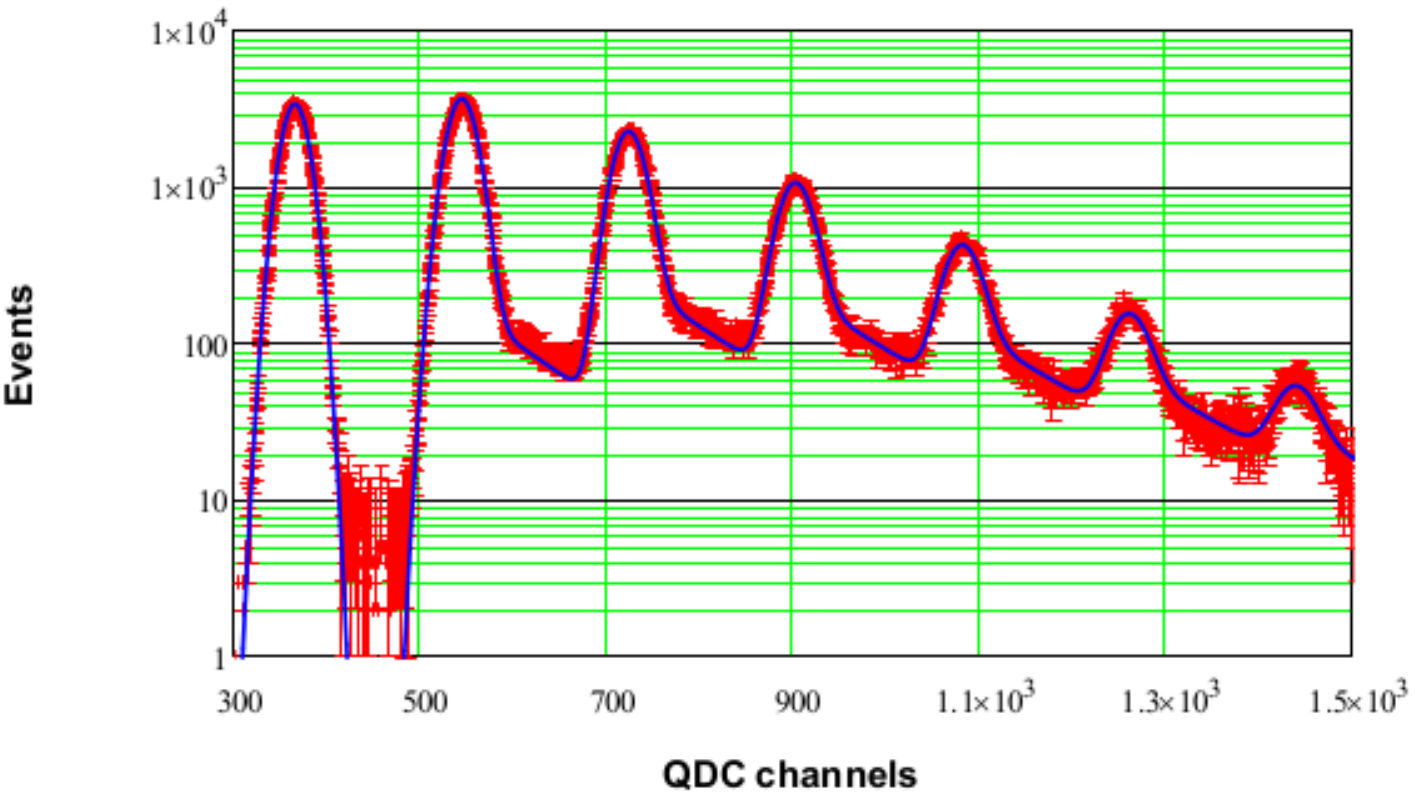}
    \caption{ }
    \label{fig:Light33}
   \end{subfigure}%
   \caption{(a) Charge spectra measured using a KETEK SiPM with a pixel size of 15\,$\upmu $m at $V_{bias} = 33$\,V (a) in the dark, and (b) with sub-nanosecond laser light. The figures are taken from Ref.\,\cite{Chmill:2017}.
   In (a), Fit\,1 considers only single dark counts randomly distributed in time without correlated noise, whereas for Fit\,2 correlated noise and multiple dark counts are included in the model fitted to the data.
   In (b) the fit function takes into account electronics noise, prompt and delayed correlated noise and gain fluctuations, but not dark counts. }
  \label{fig:PHspectra}
 \end{figure}

 As shown in Fig.\,\ref{fig:PHspectra}, SiPM charge spectra measured in the dark and  with low-intensity pulsed light, show peaks which correspond to $N_G$, the number of pixels with Geiger discharges.
 The lowest peak corresponds to $N_G = 0$, and the following to $N_G = 1$, 2, etc., with the distance between the peaks $q_0 \cdot G^\ast$.
 Following the convention from vacuum photomultipliers, it is customary to show charge spectra in units of photo-electrons (pe), by scaling the Q\,axis by $1/(q_0 \cdot G^\ast )$ and shifting the scale so that the $N_G = 0$ peak is at zero.
 As a result the $N_G = 1$, 2, ... $i$ peaks are at pe = 1, 2, ... $i$, independent of the value of $f_Q$.
 This is also valid if, instead of the charge, the amplitude of the SiPM signal is analysed.

 Another important parameter of a photon detector is the photon-detection efficiency, $PDE$.
 For a SiPM it is defined as the ratio of primary Geiger discharges due to the photons, $N_{pG,\, photo}$, to the number of photons hitting the SiPM, $N_ \gamma$:
 \begin{equation}\label{equ:PDE}
   PDE = \frac{\langle N_{pG, \, photo} \rangle } {\langle N_\gamma \rangle} = FF \cdot QE(\lambda) \cdot P_T(V_{bias}, \lambda ),
 \end{equation}
 where $FF$ is the fill factor (ratio of sensitive area to total area), $QE$ the efficiency of a photon with wavelength $\lambda $ entering the sensitive SiPM volume and producing an $eh$\,pair there, and $P_T$ the probability that the $eh$\,pair triggers a Geiger discharge.
 A primary Geiger discharge can produce correlated secondary Geiger discharges, which can be described by the excess charge factor, $ECF$, and the excess noise factor, $ENF$, which are defined below.
 From the  measured mean charge $\langle Q \rangle$, the number of primary Geiger discharges, $N_{pG}$, can be obtained using:
  \begin{equation}\label{equ:NpG}
   \langle N_{pG} \rangle = \frac{\langle Q \rangle} {q_0 \cdot G^\ast \cdot ECF}.
 \end{equation}
 For an absolute determination of $PDE$ the absolute value of $\langle N_\gamma \rangle $, $G^\ast$ and $ECF$ have to be known.
 The relative dependence of $PDE(V_{bias} )$ can be obtained more easily from the measured charge spectrum recorded at different $V_{bias}$\,values, as discussed in Sect.\,\ref{sect:Gain}.

\begin{figure}[!ht]
   \centering
   \begin{subfigure}[a]{0.5\textwidth}
    \includegraphics[width=\textwidth]{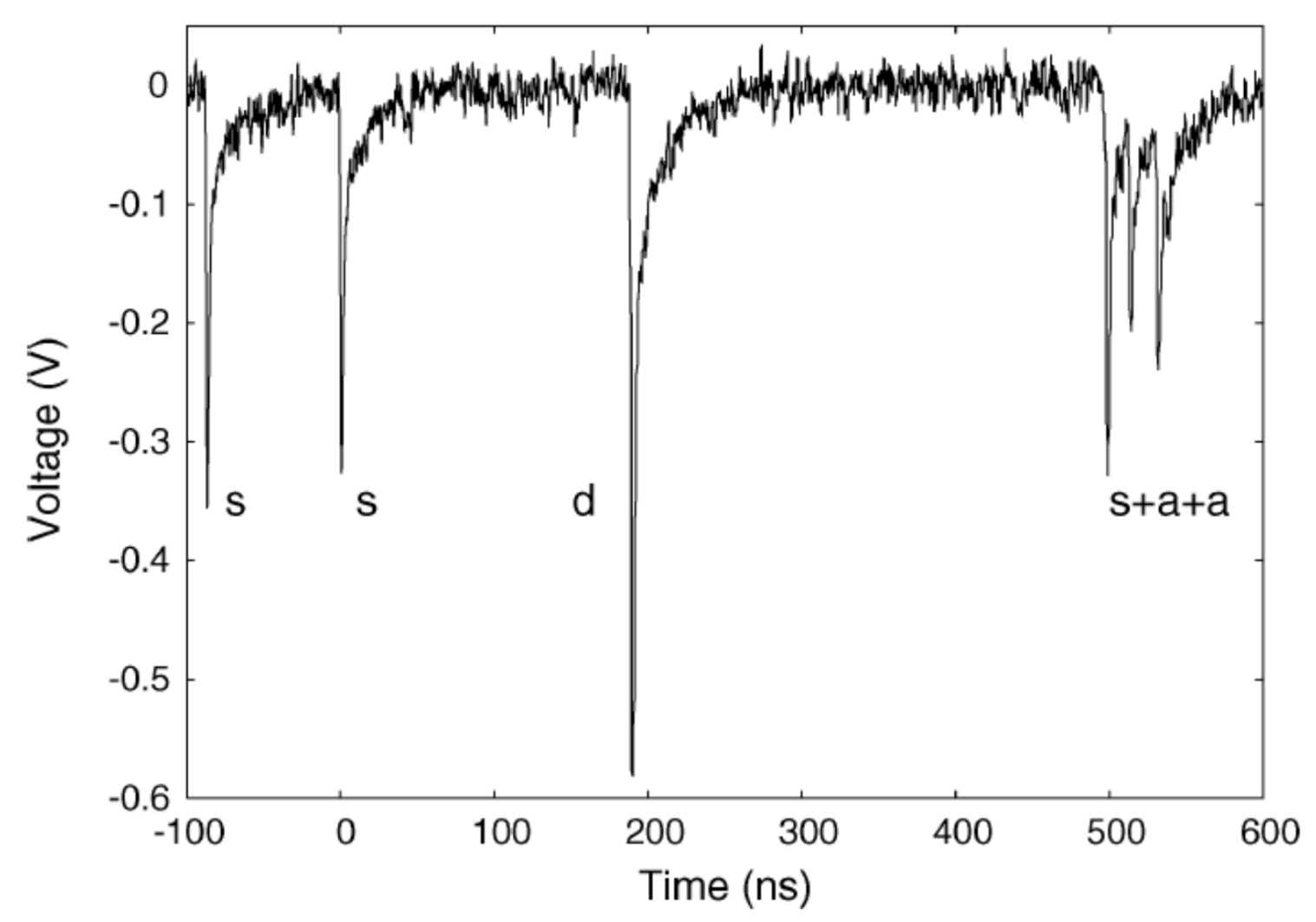}
    \caption{ }
    \label{fig:PulseMultiple}
   \end{subfigure}%
    ~
   \begin{subfigure}[a]{0.5\textwidth}
    \includegraphics[width=\textwidth]{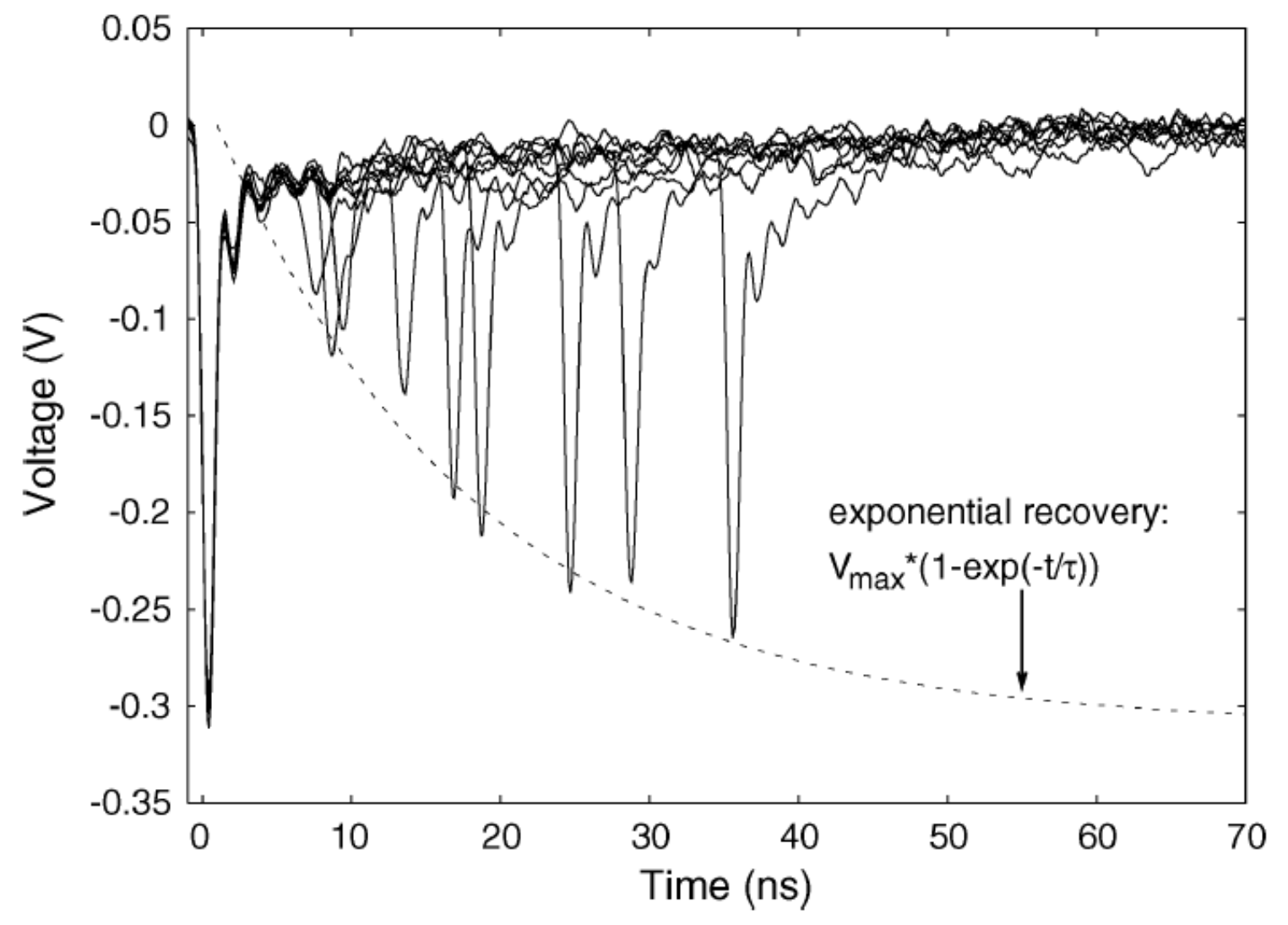}
    \caption{ }
    \label{fig:PulseRecover}
   \end{subfigure}%
   \caption{SiPM current transients from Ref.\,\cite{Piemonte:2007} illustrating the different pulse categories mentioned in the text.
   (a) Single Geiger discharges "s", prompt cross-talk "d", and single discharge with two after-pulses "s+a+a".
   (b) Transients of after-pulses as a function of time after the initial Geiger discharge.
   In the figure $\tau $ is used for the recovery time $\tau _r$, and $V_{max}$ for the amplitude of the signal from a single Geiger discharge. }
  \label{fig:Nuisence}
 \end{figure}

 An ideal photon-detector produces signals with identical shapes linearly scaled with the number photons which have initiated Geiger discharges, and
 the charge spectrum will consist of $\delta $- functions at 0, 1, 2, ... pe.
 SiPMs however show a number of differences from an ideal detector, frequently called  nuisance parameters.
 These are:
 \begin{enumerate}
   \item Dark counts produce background signals at the primary dark count rate, $DCR$.
   \item Secondary photons produced during  Geiger discharges can generate an electron-hole pair in an adjacent pixel and cause a Geiger discharge there, which results in a double-size signal (d in Fig.\,\ref{fig:PulseMultiple}). This effect is called prompt cross-talk, and its probability is $P_{pCT}$.
   \item Secondary photons produced in  Geiger discharges can generate an electron-hole pair in the non-depleted Si and charge carriers can diffuse into the amplification region of a neighbouring pixel, where they cause a Geiger discharge. This effect is called delayed cross-talk, and its probability is $P_{dCT}$.
   \item During the Geiger discharge, charge carriers can be trapped in defect states and released after some time causing a Geiger discharge in the same pixel as the primary discharge. This effect is called after-pulsing (s+a+a in Fig.\,\ref{fig:PulseMultiple}). The trapping probability for a state $i$ is called $P_{trap,\,i}$ and the corresponding time constant $\tau _{trap,\,i}$. As can be seen from Fig.\,\ref{fig:PulseRecover} the signal strength of after-pulses depends on the recovery state of the pixel, and increases proportional to $1 - e^{-t/\tau _r}$.
       In addition, secondary photons generating electron-hole pairs in the non-depleted Si with charge carriers diffusing into the same pixel as the primary Geiger discharge, contribute to after-pulses.
       This is called optically-induced after-pulsing.
 \end{enumerate}

 In addition, pixel-to-pixel gain variations and read-out noise will result in signal fluctuations.

 The effects discussed above can also be observed in the charge ($Q$) spectra recorded with a QDC (Charge-to-Digital-Convertor).
 Fig.\,\ref{fig:Dark33} shows the charge spectrum for a KETEK SiPM with a pixel size of 15\,$\upmu $m at $V_{bias} = 33$\,V measured in the dark using a CAEN QDC.
 The peak around 380\,QDC channels corresponds to zero, and the peak at 550 QDC channels to a single Geiger discharge.
 Double and triple Geiger discharges are also visible.
 The width of the zero discharge peak is caused by the electronics noise.
 The single Geiger discharge peak is caused by dark pulses which significantly overlap with the 100\,ns gate used for the measurement.
 The tail to the left of the single discharge peak and the flat part between single and zero peak are due to dark pulses which only partially overlap with the gate.
 The curve Fit\,2 is the result of the fit to the data by a model which includes the nuisance effects enumerated above (Ref.\,\cite{Chmill:2017}).
 Fit\,1 considers only single dark counts without correlated noise.

 Fig.\,\ref{fig:Light33} shows the $Q$ spectrum for the same KETEK SiPM illuminated by a sub-nanosecond laser pulse.
 The laser intensity was tuned to result in approximately 1.3 primary Geiger discharges per pulse.
 As discussed in Ref.\,\cite{Vinogradov:2012}, and also observed in the $Q$ spectrum shown, the number of events in the peaks does not follow a Poisson distribution, which is expected for an ideal photon-detector if the incoming photons are also Poisson distributed.
 The observed number of pulses at high $Q$ significantly exceeds the Poisson expectation, which is ascribed to prompt and delayed cross-talk.
 The statistics of cross-talk, which can be described by a Generalised Poisson distribution, is discussed in detail in Ref.\,\cite{Vinogradov:2012}.
 The events in-between the $N_G = 0$ and the $N_G = 1$\,pe peak are  again ascribed to dark counts.
 The events in-between the following peaks and the background below the peaks are ascribed to after-pulses and delayed cross-talk.
 The shape of the $Q$\,spectrum depends on the integration time of the readout electronics, in particular, if only a fraction of the signal is integrated.
 The curve in Fig.\,\ref{fig:Light33} is the model fit described in Ref.\,\cite{Chmill:2017}, which includes effects 2--4, but not 1, which explains the disagreement around channel 450.
 Note, that fitting separately the peaks with Gauss functions and ignoring the background in-between, which is frequently done, does not give the correct number of Geiger discharges, in particular for high $N_G$\,values.

 The effect of the nuisance parameters is to change the measured distribution with respect to the distribution of converting photons, which would be the response of the ideal photon detector.
 Two parameters, the excess charge factor, $ECF$, and the excess noise factor, $ENF$, are frequently used to describe the worse performance of a non-ideal detector\,\cite{Piemonte:2012}.
 They are discussed next.
 The distribution of photons and the number of primary Geiger discharges, $N_{pG}$, are assumed to follow a Poisson distribution with the mean $ \langle N_{pG} \rangle $ and the root-mean-square (rms) deviation $ \sqrt{\langle N_{pG} \rangle }$.
 The response of the ideal photon-detector will just be the Poisson distribution multiplied with $q_0 \cdot G^\ast$ resulting in the mean $\langle Q_{P} \rangle = q_0 \cdot G^\ast \cdot \langle N_{pG} \rangle $ and the rms deviation $\sigma_{P} = q_0 \cdot  G^\ast \cdot \sqrt{\langle N_{pG} \rangle }$.
 If the measured charge distribution of the real photon-detector has the mean $\langle Q \rangle$ and the rms deviation $\sigma _Q$ for the same number of primary Geiger discharges as the ideal detector, then by definition
  \begin{equation}\label{equ:ECF}
     ECF = \frac{ \langle Q \rangle} {\langle Q_{P} \rangle},
  \end{equation}
 and
  \begin{equation}\label{equ:ENF}
   ENF = \frac{(\sigma _Q / \langle Q \rangle)^2} {(\sigma_{P} / \langle Q_{P} \rangle)^2 }.
  \end{equation}

 As the contribution of the nuisance effects to the measured signal depends on the effective integration time, also $G^\ast $, $ECF$ and $ENF$ depend on the readout and the analysis method used, which presents a significant complication.
 It should also be noted that assuming a Poisson distribution for the photons producing primary Geiger discharges is not necessarily correct for all light sources.

 Non-linearity and saturation are other limitations of SiPMs.
 As the charge from a single pixel is approximately the same for one and more than one simultaneous Geiger discharge, the signal is expected to saturate at $Q_{sat} = N_{pix} \cdot q_0 \cdot G^\ast$ for high number of photons, $N_\gamma $.
 The saturation can be described by a decrease of the photon-detection efficiency, $PDE$, because of the already busy pixels.
 Well below saturation, the mean number of Geiger discharges is approximately given by $ PDE_0 \cdot ECF \cdot N_\gamma $, with the photon-detection efficiency without saturation effects $PDE_0$.
 For high numbers of simultaneous photons
 \begin{equation}\label{equ:NGsat}
    N_G \approx N_{pix} \cdot \big(1 - e^{-(PDE_0 \cdot ECF \cdot N_\gamma)/N_{pix}} \big)
 \end{equation}
 is expected because of multiple Geiger discharges in individual pixels.
 This relation is only valid if the photons are uniformly distributed over the SiPM.
 If this is not the case, the non-linearity sets in already at lower $N_ \gamma$\,values and the functional form is different.
 If the arrival time of the photons is spread over time, some of the pixels will have already partially recovered when the next photon arrives, and signals  exceeding $Q_{sat}$ are expected, and actually observed.
 The situation is quite complex, however phenomenological parametrisations are available, which describe detailed measurements\,\cite{Bretz:2016}.
 High dark count rates, e.\,g. due to radiation damage, also cause a decrease of $PDE$ due to pixels in the recharging state after Geiger discharges.
 This topic is addressed in Sect.\,\ref{sect:non-linearity} and in the contribution on radiation damage of this Special Issue\,\cite{Garutti:2018}.

 For the description of the nonlinearity, the terms \emph{Linearity}, \emph{Nonlinearity} and \emph{Dynamic Range} are frequently used.
 Note that different definitions are found in the literature.
 For the linearity, $Lin$,  a minimum and a maximum value of the number of photons to be detected, $N_{\gamma ,\, min}$ and $N_{\gamma ,\, max}$, have to be defined.
 Then $Lin (N_{\gamma ,\, min},N_{\gamma ,\, max}) = Res(N_ {\gamma ,\, max}) / Res(N_{\gamma ,\, min})$, with the \emph{Responsivity } $Res(N_ \gamma) = \langle Q(N_ \gamma) \rangle /N_ \gamma$.
 The non-linearity is just $ NLin = 1 - Lin$.
 For the dynamic range, values for $NLin$ and for $N_{\gamma ,\, min}$ have to be specified.
 The ratio of $N_\gamma $ where the specified $NLin$ is reached to $N_{\gamma ,\, min}$ is defined as the dynamic range.
 In the situation where zero and one Geiger discharges can be distinguished, $N_{\gamma ,\, min} = 1 / PDE_ 0$ appears to be a reasonable convention.


 \begin{table}
  \caption{Parameters and symbols used for the characterisation of SiPMs.
   The measurement methods are $I_f-V$ and $I_r-V$ for the forward and reverse $I-V$ measurement, $Trans$ for the current-transient measurement, and $Q$ for the  spectra obtained either by integrating the transients or from the maximum of the pulse of the transient, or from the charge recorded with a charge-to-digital convertor. }
   \centering
    \begin{tabular}{c c c }
   Symbol & Parameter & Measurement \\
  \hline   \hline
   $V_{bias}$ & Bias voltage & -- \\
   $V_{bd}$ & Breakdown voltage & $I_r-V$ \\
   $V_{off}$ & Turnoff voltage & $Q$ \\
   $V_{OV}=V_{bias}-V_{bd}$ & Overvoltage & $I_r-V$ \\
   $I_{dark}$ & Dark current & $I_r-V$ \\
   $I_{light}$ & Current with illumination & $I_r-V$  \\
   $I_{photo}=I_{light}-I_{dark}$ & Photo current & $I_r-V$  \\
   $Q$ & Measured charge (amplitude)& $Q$\\
   $\langle Q \rangle$ & Mean $Q$ & $Q$\\
   $\sigma _Q ^2$ &Variance of $Q$ & $Q$\\
  \hline
   $N_{pix},\,(N_{total})$ & Number of pixels & -- \\
   $R_{q}$ & Quenching resistance & $I_f-V,\, C-V$ \\
   $C_{q}$ & Quenching capacitance & $C-V$ \\
   $C_{d}$, ($C_{pix}$) & Pixel capacitance & $C-V, Q$ \\
   $R_{s}$ & Shunt resistor readout & -- \\
   $I_{disc}$ & Pixel discharge current & -- \\
   $R_{d}$ & Pixel discharge resistor & -- \\
   $V_{d}$ & Voltage drop over pixel & -- \\
   $C_{eq}=N_{pix} (1/C_d +1/C_q)^{-1}$ & Capacitance seen by readout & -- \\
   $\tau _{in}=R_s \cdot C_{eq}$ & Time const. fast component & $Trans$ \\
   $\tau _{r}=R_q \cdot (C_d + C_q)$ & Recharging time const. & $Trans$ \\
   $G = (V_{bias}-V_{off})(C_d + C_q)/q_0 $ & SiPM overall gain & $Trans,\,Q$ \\
   $f_Q $& Fraction SiPM signal recorded & $Trans,\,Q$ \\
   $G^\ast = G \cdot f_Q$ & Measured gain & $Q$ \\
  \hline
   $N_{\gamma}$ & Number of photons on SiPM & -- \\
   $PDE=FF \cdot QE \cdot P_T$ & Photon-detection efficiency & $Q,\,Trans$ \\
   $PDE_0$ & $PDE$ in linear range (low $N _\gamma $) & $Q,\,Trans$ \\
   $FF$ & Fill factor & -- \\
   $QE$ & Quantum efficiency & -- \\
   $P_T$ & Geiger breakdown probability & -- \\
   $P_{T, \, photo} $ & $P_T$ for photons & -- \\
   $N_{G}$ & Number Geiger discharges & $Q$\\
   $N_{pG}$ & Number primary Geiger discharges & $Q$\\
   $N_{pG, \, photo}$ & $N_{pG}$ due to photons & $Q$\\
   pe & Unit \emph{Geiger discharges}, (\emph{photo-electrons}) & $Trans,\,Q$\\
   $f_0$ & Fraction events in $N_{G} = 0$ peak & $Q$\\
   $f _{0,\,dark}$ & $f_0$ in the dark & $Q$\\
   $f _{0,\,light}$ & $f_0$ with light & $Q$\\
   $f_{0.5}$ & Fraction events above 0.5 pe & $Q$\\
   $f_{1.5}$ & Fraction events above 1.5 pe & $Q$\\
     \hline
   $DCR$ & Dark count rate & $I_r - V,\,Q,\,Trans$ \\
   $DCR_p$ & Primary $DCR$ & $I_r - V,\,Q,\,Trans$ \\
   $P_{pCT}$ & Probability prompt cross-talk & $Q,\,Trans $ \\
   $P_{dCT}$ & Probability delayed cross-talk & $Q,\,Trans$ \\
   $P_{AP}$ & After-pulse probability  & $Q,\,Trans$ \\
   $ECF$ & Excess charge factor & $Q$ \\
   $ENF$ & Excess noise factor & $Q$ \\
   $Lin$ & Linearity & $Q$ \\
   $NLin = 1 - Lin$ & Non-linearity & $Q$ \\
   $DR$ & Dynamic range & $Q$ \\
   $Res = \langle Q (N_\gamma)\rangle /N_\gamma $ & Responsivity & $Q$ \\
  \hline
   \end{tabular}
  \label{table:parameters}
 \end{table}

  \section{Measurement setups}
  \label{sect:Setups}

 In this section an overview of different setups used for characterising SiPMs is presented, and some recommendations given.

 \subsection{$I-V$ and $C-V$ setup}
  \label{sect:IVCV-setup}

\begin{figure}[!ht]
   \centering
    \includegraphics[width=0.4\textwidth]{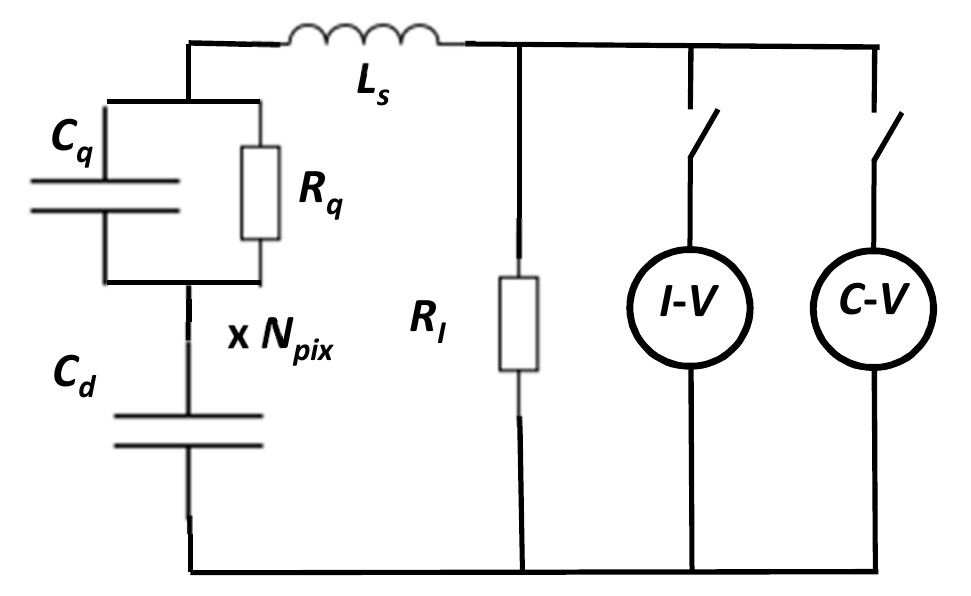}
   \caption{ Generic setup for the $I-V$ and $C-V$ measurements. The elements used to analyse the measurements in addition to the ones already shown in Fig.\,\ref{fig:Emodel} are the parasitic elements $L_s$, and $R_{I}$ for the dark current.
   Their meaning is described in the text.
   The grid capacitance, $C_{grid}$, shown in Fig.\,\ref{fig:Emodel} has not been implemented in the analysis.}
     \label{fig:IVCV-setup}
 \end{figure}

 Fig.\,\ref{fig:IVCV-setup} shows a schematic layout of the measurement setup used for the $I-V$ and $C-V$ measurements.
 They are best performed on a temperature-controlled chuck in a light tight and electrically shielded box.
 As it has been observed that SiPM parameters can be influenced by humidity, a humidity measurement and control of the atmosphere in the box is recommended.

 For the $I-V$ measurements the ramping of the voltage should be sufficiently slow so that stable conditions at the individual voltage steps are reached.
 This can be verified  by taking $I-V$\,data  ramping the voltage up and down.
 For the precise ($ \lesssim 10$\,mV) determination of the breakdown voltage $V_{bd}$, a voltage step around $V_{bd}$ of 100\,mV is recommended.
 This small step size should already be used well below $V_{bd}$ (e.\,g. 3\,V), to avoid problems with fitting the data or numerically calculating derivatives.
 In addition, note that the Keithley voltage source, which is typically used for the measurements, has a setting accuracy of $\pm 10$\,mV with a saw-tooth deviation as a function of voltage.
 This can cause problems for a precise determination of $V_{bd}$.
 Last but not least, the possibility to illuminate the SiPM with DC light is highly recommended.
 For low dark currents, (e.g. at low operating temperatures), this is needed for a precise determination of $V_{bd}$, and for highly irradiated SiPMs with high pixel occupancies, the comparison of the difference of the current with and without illumination for different radiation fluences can give a first idea on the degradation of the SiPM as photon-detector due to radiation damage (Ref.\,\cite{Garutti:2018} and Sect.\,\ref{sect:non-linearity}).

 The measurement of the admittance $Y(f)$ as a function of frequency, $f$, can be used to determine the SiPM electrical parameters.
 In addition to those already described in Fig.\,\ref{fig:Emodel}, these are
 $L_s$ an effective inductance for the biasing lines, and
 $R_I$ to parameterise the SiPM dark current.
 Note that in this model the capacitance of the voltage distribution grid in parallel to $R_I$, discussed e.g. in Refs.\,\cite{Seifert:2009, Marano:2014}, is not included.
 As will be shown in Sect.\,\ref{sect:ElParameters} this model gives a fair description of the measured data.
 For the measurements a large frequency range, e.\,g. $f = 100$\,Hz to 2\,MHz should be chosen with about 3 $f$-values per decade.
 High frequencies are in particular relevant for the determination of $C_q$.
 Only at high frequencies a significant fraction of the AC-current flows through $C_q$ and its effect can be seen in the $Y-f$ measurements.
 Experience has shown that a value of $V_{bias}$ between 0.5 and 1\,V below $V_{bd}$ gives reliable results, even if the dark-count rate  is very high (>\,1\,GHz).
 As discussed in Sect.\,\ref{sect:Efield}, $C-V$ measurements can  be used to estimate the doping profile and the electric field of the avalanche region.

 \subsection{Current-transient setup}
   \label{sect:Transient-setup}

\begin{figure}[!ht]
   \centering
    \includegraphics[width=\textwidth]{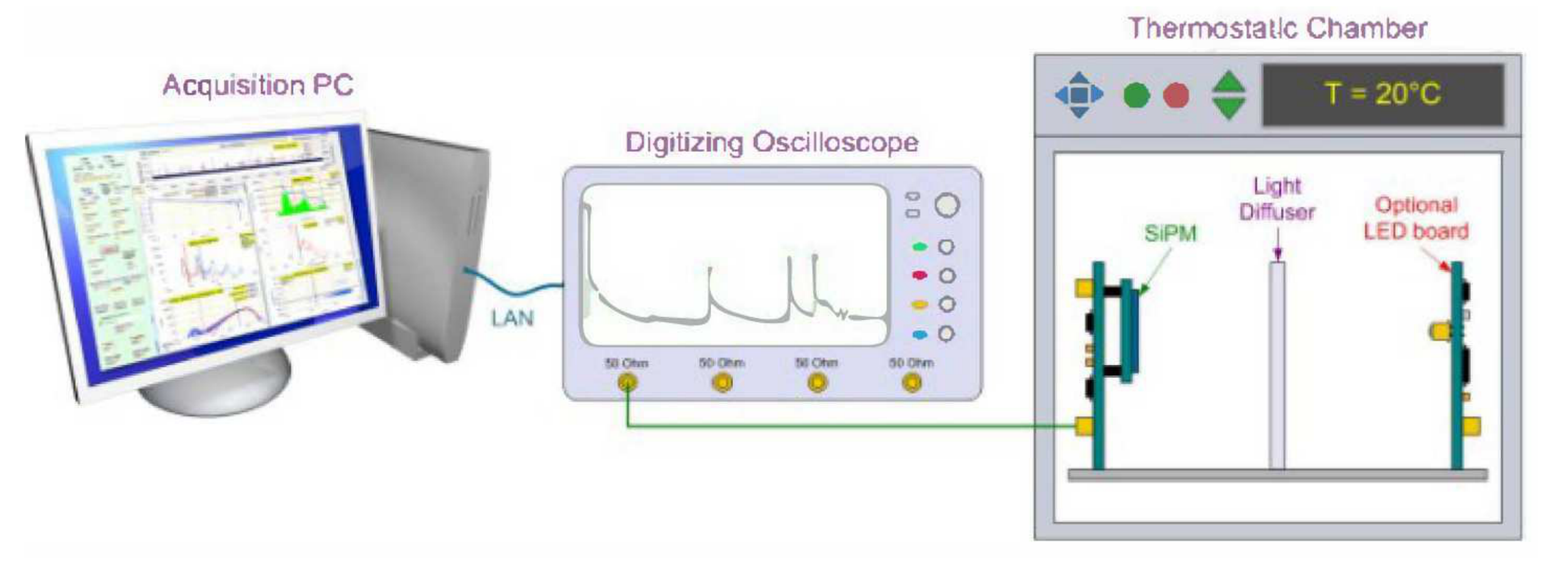}
   \caption{Setup from Ref.\,\cite{Piemonte:2012} for the characterisation of SiPMs using a digital scope for transient recording.
    It consists of a climate-controlled chamber,  the SiPM with its amplifier, a digital scope with a sampling rate of 10\,GS/s and a bandwidth set to 500\,MHz and a PC for data acquistion.}
  \label{fig:Trans-setup}
 \end{figure}

 A number of groups (e.g.\,\cite{Piemonte:2012, Rosado:2015, Otte:2016}) are using setups to characterise SiPMs by recording the current transients.
 They all follow a similar design:
 The SiPM is mounted in a temperature-controlled chamber, where it can be uniformly illuminated by a sub-nanosecond pulsed light source.
 The SiPM signal is amplified by a low-noise high-bandwidth amplifier and the waveform digitised by a digital oscilloscope or digitiser.
 A PC is used for steering the measurements, for storing the data and for performing a first on-line analysis.
 Fig.\,\ref{fig:Trans-setup} shows the setup at FBK as an example.
 Details can be found in Ref.\,\cite{Piemonte:2012}.
 Together with this setup a complete analysis chain has been developed which allows a fast and reliable characterisation of large samples of SiPMs.
 It should be noted that, if such a setup is used to investigate highly-irradiated SiPMs where the dark current can exceed tens of mA, the heating of the SiPM is significant and the exact knowledge of the SiPM temperature is quite a challenge.
 In addition, the voltage drop over the protection, filter and readout resistors has to be taken into account.
 Such effects can be investigated using a non-irradiated SiPM and simulating the high DCR by an additional DC light source.
 To the author's knowledge, such a study has so far not been reported.

 The recording of the transient allows for a most complete characterisation of SiPMs:
 In the off-line data analysis, pulse amplitudes and time delays of pulses correlated with the primary discharges can be investigated, as well as charge and amplitude distributions for different pulse integration times and digital pulse shaping evaluated.
 However, the effort to set up a system with low noise,  high performance and precise temperature control is significant and requires quite some expertise.

 \subsection{Charge-measurement setup}
   \label{sect:PH-setup}

\begin{figure}[!ht]
   \centering
    \includegraphics[width=0.6\textwidth]{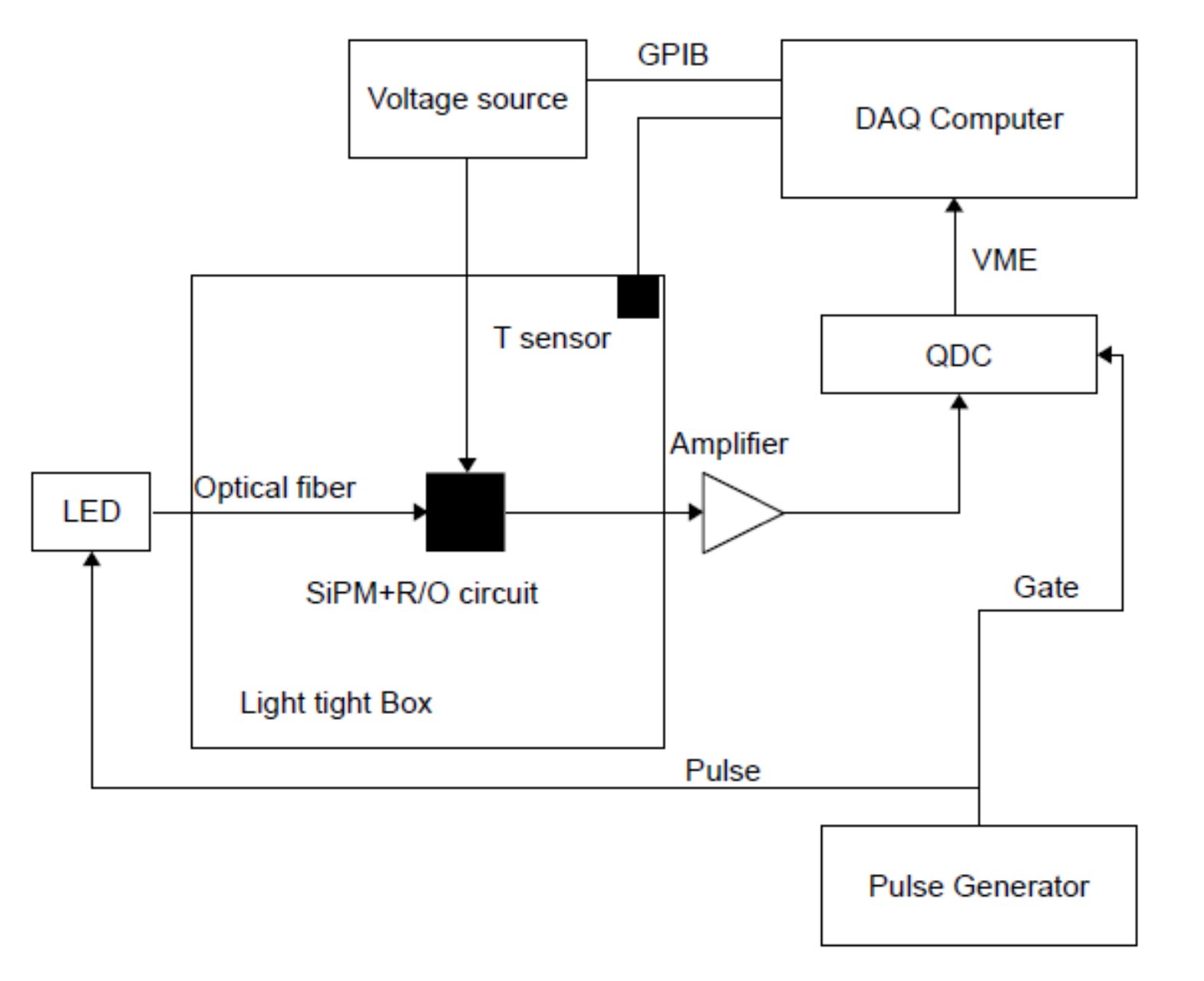}
   \caption{Schematic diagram of the setup from Ref.\,\cite{Xu:2014} for the characterisation of SiPMs recording charge spectra. }
  \label{fig:PH-setup}
 \end{figure}

 Recording charge spectra from SiPMs is significantly simpler than recording and analysing current transients.
 However, the time information, required for a detailed understanding of the nuisance parameters, is not available.
 Again a number of groups (e.g.\,\cite{Dziewiecki:2009, Eckert:2010, Barbosa:2012, Xu:2014, Arosio:2014}) have set up such systems.
 An example from Ref.\,\cite{Xu:2014} is shown in Fig.\,\ref{fig:PH-setup}.
 A pulse generator triggers a LED, which illuminates the SiPM.
 The SiPM signal is amplified by a factor 50 (for a $50\,\Omega $ load) by a high-bandwidth amplifier and recorded by a Charge-to-Digital-Convertor (QDC) with the gate generated by the pulse generator.

\begin{figure}[!ht]
   \centering
    \includegraphics[width=0.6\textwidth]{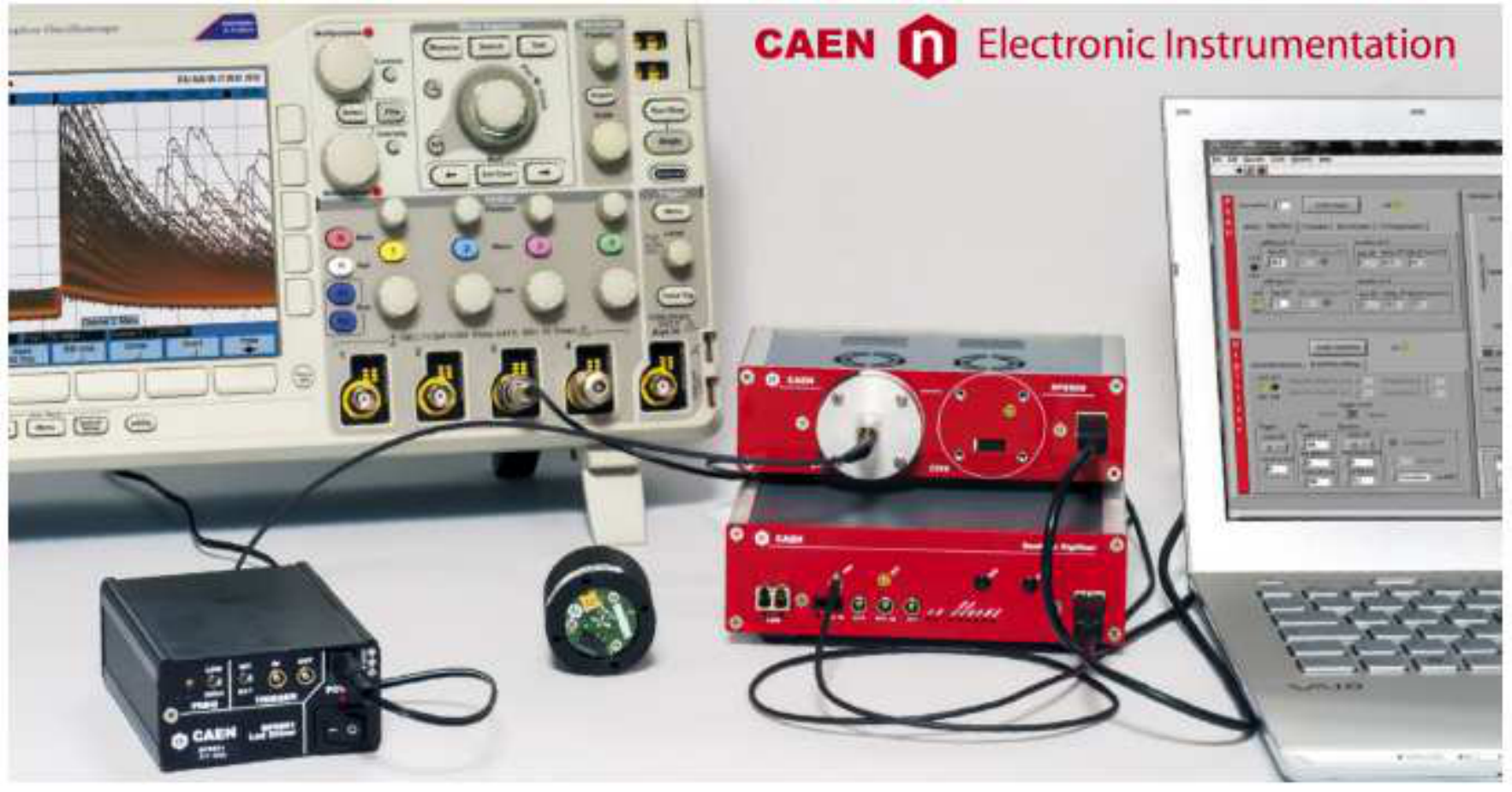}
   \caption{CAEN setup (Ref.\,\cite{Arosio:2014}) for the characterisation of SiPMs. It is a modular plug-and-play system which is simple to set up and allows characterising many properties of SiPMs. A suite of analysis software comes with the system. Similar systems are also available from other vendors. These systems are ideal for a first step towards characterising SiPMs and also well suited for laboratories for pupils and students.}
  \label{fig:Photo_CAEN}
 \end{figure}

 In addition to home-built systems, several firms offer SiPM evaluation kits.
 An example is the \emph{SiPM Educational Kit} from CAEN\,\cite{Arosio:2014}.
 A photo of such a setup is shown in Fig.\,\ref{fig:Photo_CAEN}.
 It consists of a LED emitting light of 400\,nm with sub-nanosecond rise time and 5\,ns decay time, a two-channel power supply-amplifier unit and a two-channel 250\,MS/s digitiser with 12\,bit dynamic range.
 The firmware allows charge integration, pulse-shape discrimination and triggering.
 In this way high-speed recording of charge spectra is possible.
 Commercial and custom built systems, which record charge spectra, are particularly well suited for the high-throughput characterisation of SiPMs.

  \subsection{Absolute \emph{PDE} setup}
   \label{sect:PDE-setup}

\begin{figure}[!ht]
   \centering
   \begin{subfigure}[a]{0.5\textwidth}
    \includegraphics[width=0.8\textwidth]{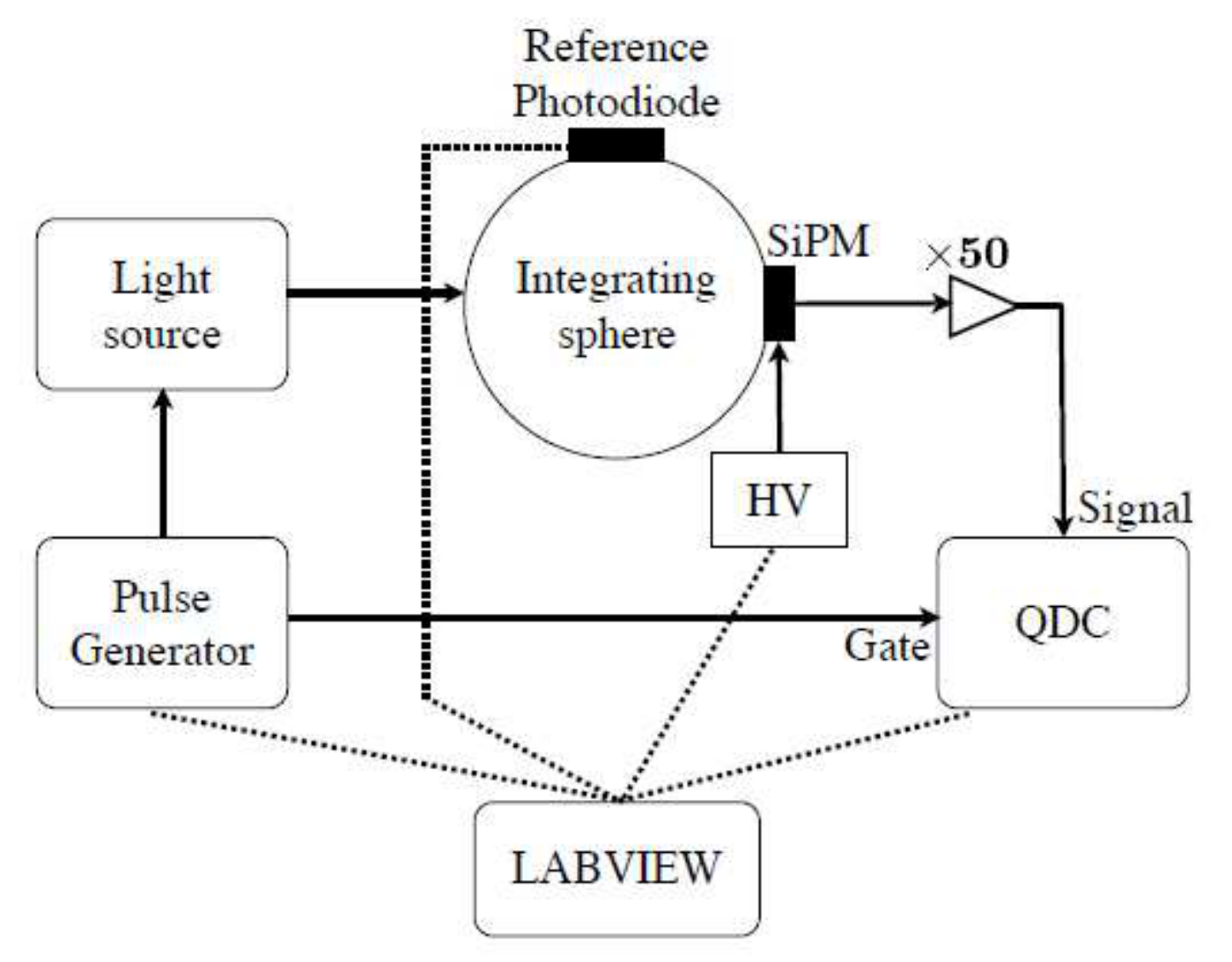}
    \caption{ }
    \label{fig:LayoutPDE}
   \end{subfigure}%
    ~
   \begin{subfigure}[a]{0.5\textwidth}
    \includegraphics[width=0.8\textwidth]{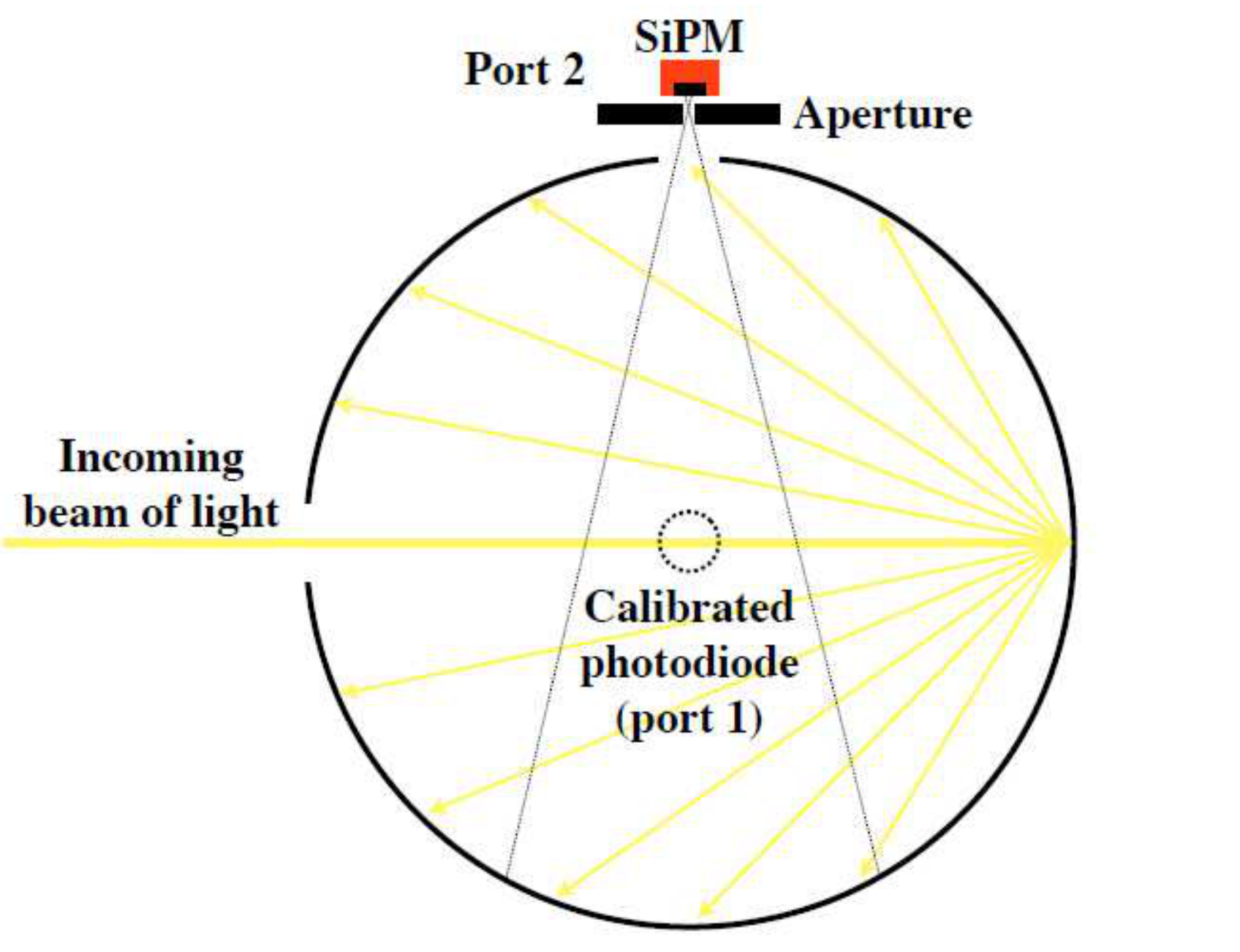}
    \caption{ }
    \label{fig:LSpherePDE}
   \end{subfigure}%
   \caption{(a) Schematic layout of the $PDE$ measurement from Ref.\,\cite{Eckert:2010}. The absolute normalisation is obtained by measuring the power of the light source with a calibrated photo-diode. (b) Sketch of the integrating sphere and the positions of the SiPM and the calibrated photo-diode. The angles between the individual openings are $90^\circ $.}
  \label{fig:PDE-setup}
 \end{figure}

 For measuring the photon-detection efficiency, $PDE$, the response of the SiPM is compared to the response of a calibrated photo-detector.
 Both pulsed and DC measurements, or a combination of both are used.
 Again, several setups (e.g.\,\cite{Piemonte:2012, Otte:2016, Eckert:2010, Otte:2006, Bonanno:2009, Yue:2015}) following similar concepts are in use.
 As an example, the layout from Ref.\,\cite{Eckert:2010} is shown in Fig.\,\ref{fig:PDE-setup}.

 As light sources pulsed laser diodes and LEDs with pulse widths below 2\,ns are used.
 The wavelength spectra have a FWHM of typically 5\,nm for the laser and 10 to 20\,nm for the LED.
 The SiPM output signal is amplified by a fast amplifier and digitised by a QDC with an integration gate of 50 to a few 100\,ns depending on the SiPM pulse shape.
 Dark spectra and spectra with pulsed light are recorded.
 The light intensity is adjusted so that the fraction of events without a SiPM pulse, $f_{0,\,light}$, can be determined precisely.
 Assuming Poisson statistics for the number of dark counts and of primary Geiger discharges, the  mean number of primary Geiger discharges per pulse from the photons of the light source is
 \begin{equation}\label{equ:NpGphoto}
  \langle N_{pG, \, photo} \rangle = \ln (f_{0,\,dark} / f_{0,\,light}) ,
 \end{equation}
 with $f_{0,\,dark}$ the fraction of events without a SiPM pulse under dark conditions.
 When deriving this formula the fact is used that in the absence of a Geiger discharge, there are  no correlated pulses, and the mean number of primary discharges $\langle N_{pG} \rangle$ for both light and dark condition is obtained from the zero probability of the Poisson distribution: $P(0, \langle N_{pG} \rangle) = e^{- \langle N_{pG} \rangle}$.
 Finally, the absolute $PDE$ is obtained by normalising to the power $P_{ref}$ measured by the calibrated reference diode and $PR_{1/2}$, the measured power ratio of port\,1 to port\,2 using

 \begin{equation}\label{equ:Abspde}
   PDE = \frac{\langle N_{pG, \, photo} \rangle\cdot PR_{1/2} \cdot f _{Laser} } {P_{ref} / (h \nu)},
 \end{equation}
 with the laser repetition rate $f _{Laser}$ and the photon energy $h \nu$.

 The $PDE$ for typically four wavelengths is determined as described above.
 In order to extend the measurements to wavelengths in the range between 300 and 1000\,nm, a Xe lamp with a monochromator is used as light source and the current from the SiPM and the reference diode is measured.
 As the current includes cross-talk and after-pulses, the measurements have to be normalised to the $PDE$\,measurements described above.
 With a careful control of different systematic effects, absolute $PDE$\,values with an $\approx 3$\,\% uncertainty for wavelengths between 350 and 800\,nm have been determined\,\cite{Otte:2016}.
 For lower wavelengths the uncertainties are dominated by stray light, and above 800\,nm by the knowledge of the quantum efficiency of the Si reference diode.
 In Ref.\,\cite{Yang:2014} a precision method with two integrating light spheres is presented.
 In Ref.\,\cite{Kueck:2015} a double attenuator techniques is described which achieves an absolute uncertainty below 0.5\,\% at a wavelength of 770\,nm.

  \subsection{Counting methods}
   \label{sect:counting}

\begin{figure}[!ht]
   \centering
   \begin{subfigure}[a]{0.5\textwidth}
    \includegraphics[width=\textwidth]{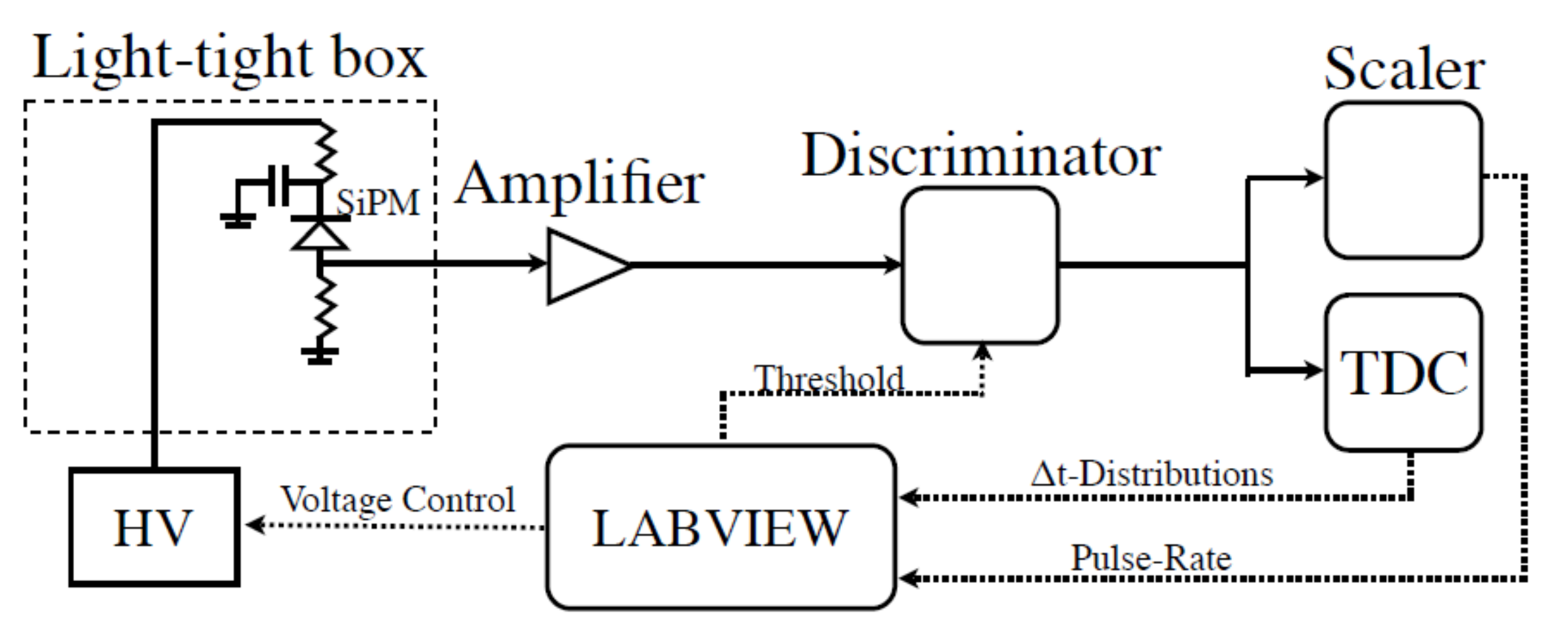}
    \caption{ }
    \label{fig:LayoutCounting}
   \end{subfigure}%
    ~
   \begin{subfigure}[a]{0.5\textwidth}
    \includegraphics[width=\textwidth]{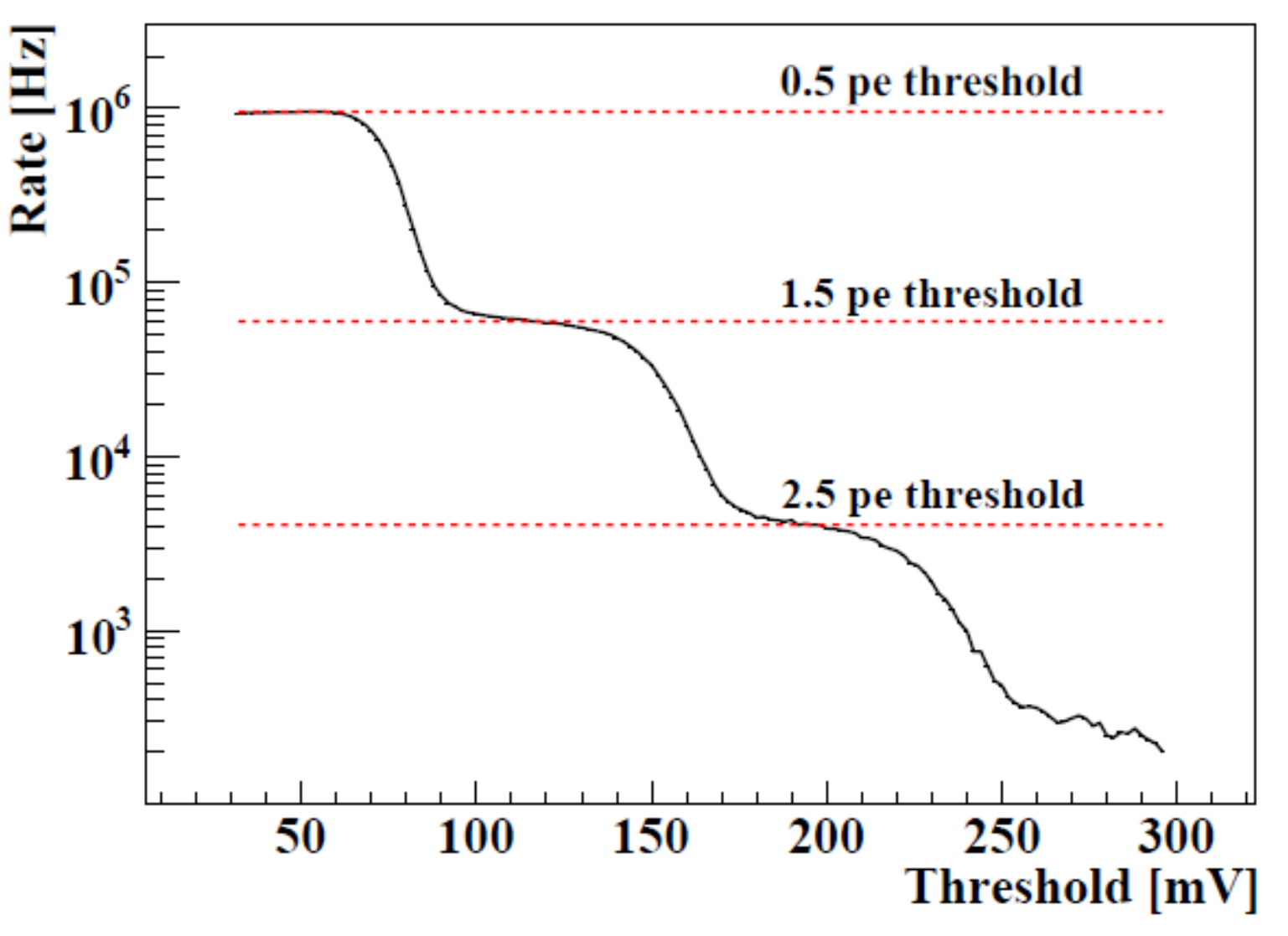}
    \caption{ }
    \label{fig:CountDCR}
   \end{subfigure}%
   \caption{(a) Setup from Ref.\,\cite{Eckert:2010} for the $DCR$, cross-talk and after-pulse measurements. For the $DCR$ and cross-talk measurement the count rate as function of discriminator threshold is measured; for the after-pulse measurement the discriminator output is connected to the TDC.
   (b) Measured count rate as function of discriminator threshold. The unit pe corresponds to the amplitude of a single Geiger discharge.}
  \label{fig:Counting}
 \end{figure}

  An elegant method for a quick determination of the nuisance parameters $DCR$ and correlated noise, is described in Ref.\,\cite{Eckert:2010}.
  The schematic layout is shown in Fig.\,\ref{fig:LayoutCounting}.

 The measurements are performed in the dark.
 Fig.\,\ref{fig:CountDCR} shows the count rate as function of the discriminator threshold in units of pe, the amplitude of a single Geiger discharge.
 The curve, which corresponds to the cumulative pulse-amplitude distribution, shows characteristic plateaus at 0.5, 1.5, and 2.5 pe.
 The rate $Rate_{0.5}$ for 0.5\,pe gives the $DCR$, and the ratio $Rate_{1.5} / Rate_{0.5}$ approximately the overall cross-talk probability.

 For the measurement of the time dependence of the delayed correlated pulses, the discriminator threshold is set to a value well above the electronics noise and the time between triggers is measured using the TDC.
 The measured time distribution can be fitted by the sum of delayed pulses with two time constants and the dark-count contribution.
 More details are given in \cite{Eckert:2010} noting that the functions used for the fits (Eq.\,5 and 6 in the Ref.) are only approximately correct.
 A similar analysis with an improved formula is given in Ref.\,\cite{Garutti:2014}.

 It should be noted that this and more information can be obtained from the the $\Delta t$\,method using  current transients as described in Sect.\,\ref{sect:Transient-setup}, which is probably the reason why the counting method is not widely used.

  \subsection{Optical observation of Geiger discharges}
   \label{sect:Optical}

\begin{figure}[!ht]
   \centering
   \begin{subfigure}[a]{0.5\textwidth}
    \includegraphics[width=\textwidth]{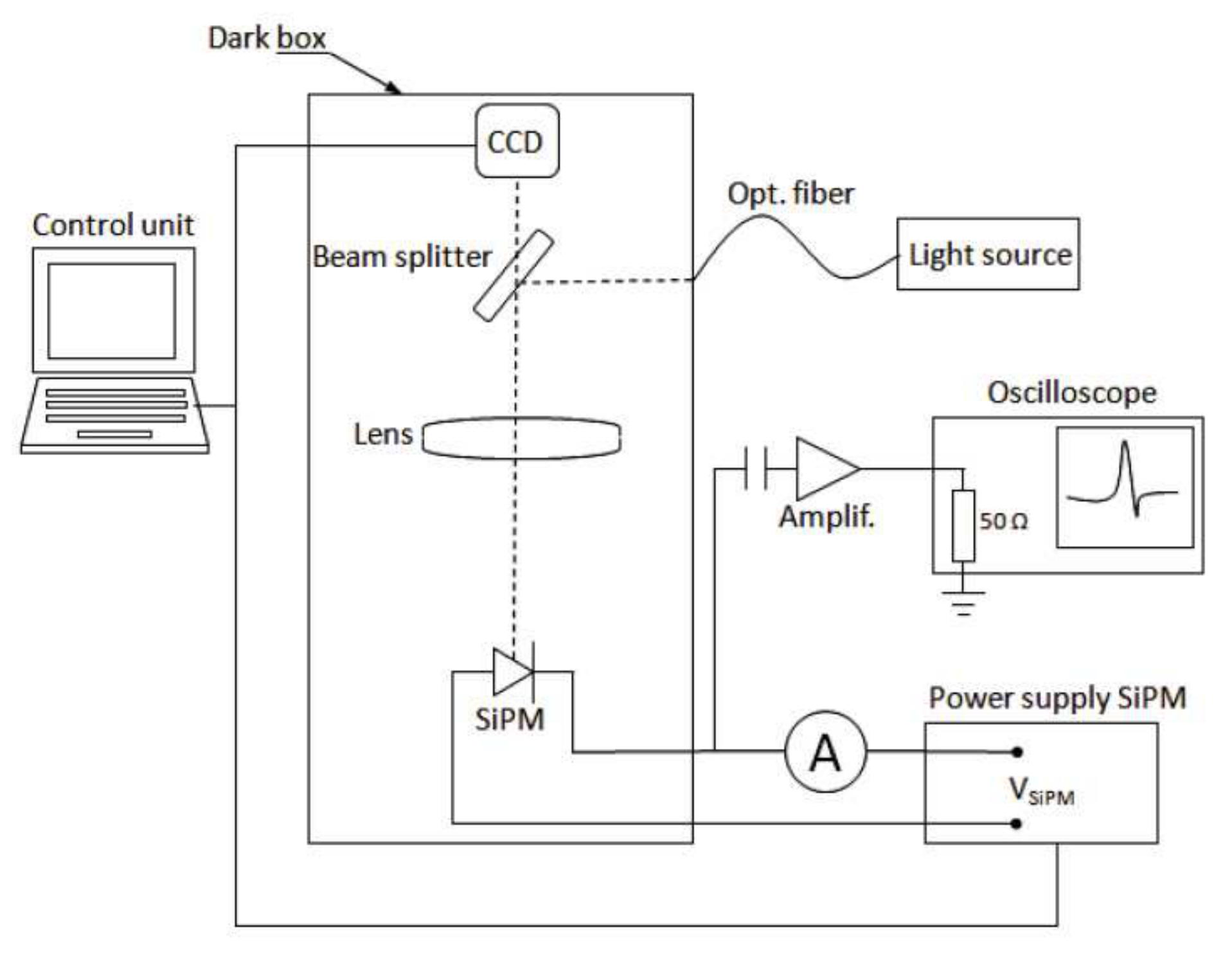}
    \caption{ }
    \label{fig:LightSetup}
   \end{subfigure}%
    ~
   \begin{subfigure}[a]{0.5\textwidth}
    \includegraphics[width=\textwidth]{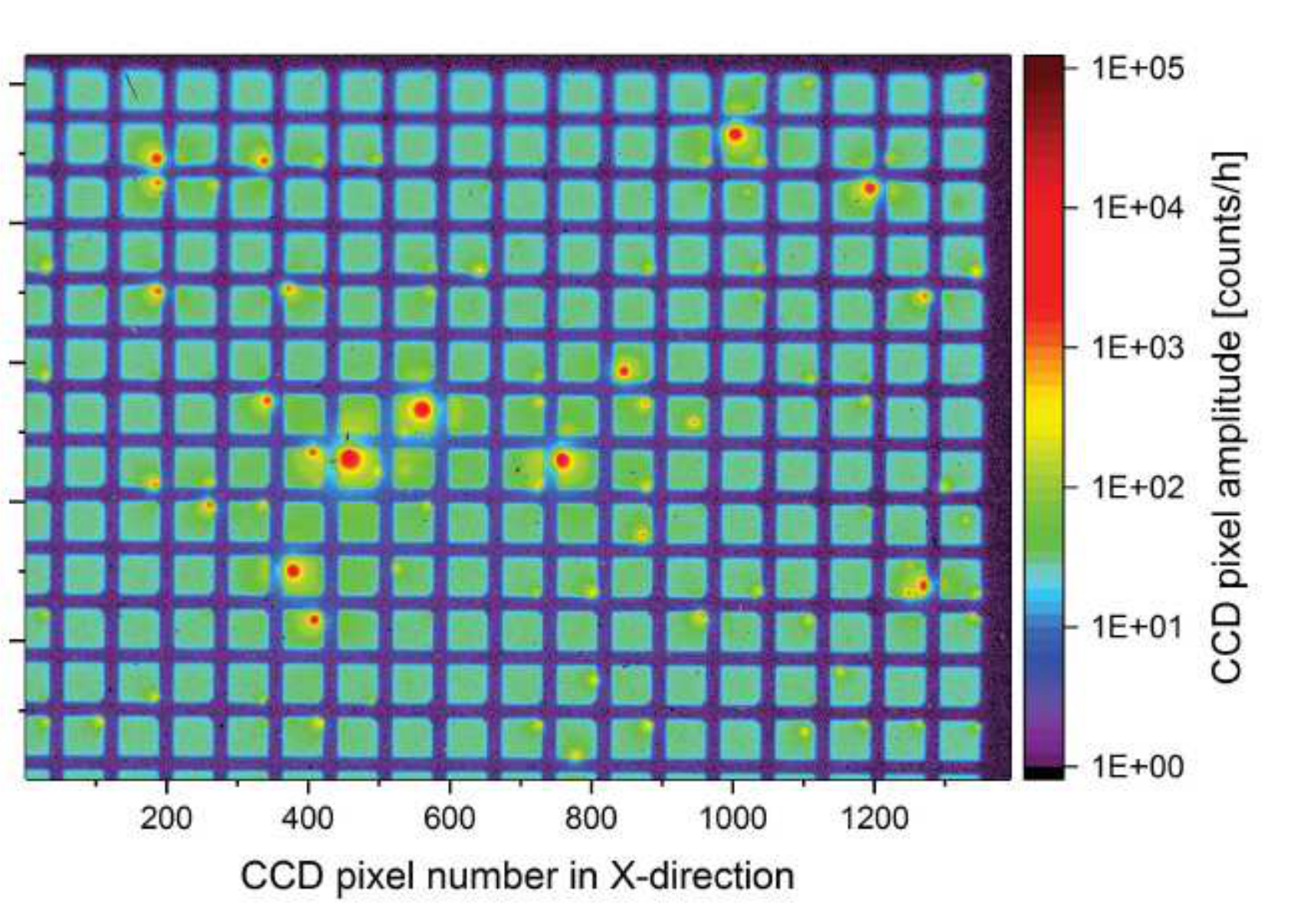}
    \caption{ }
    \label{fig:LightResult}
   \end{subfigure}%
   \caption{(a) Setup from Ref.\,\cite{Engelmann:2018} to study the light emission from Geiger discharges in SiPMs.
   (b) Distribution of the observed light intensity from Geiger discharges for a KETEK SiPM with a pixel size of 50\,$\upmu$m.}
  \label{fig:Light}
 \end{figure}

 To study the spatial distribution and extension of Geiger discharges, the author of Ref.\,\cite{Engelmann:2018} uses the setup shown in Fig.\,\ref{fig:LightSetup}.
 The method is based on the observation that  Geiger discharges emit optical and near-infrared photons, as first shown in Ref.\,\cite{Newman:1955} and studied quantitatively in Ref.\,\cite{Lacaita:1993}.
 In Ref.\,\cite{Mirzoyan:2009} the light spectrum from a Hamamatsu SiPM has been measured in the wavelength range between 450 and 1600\,nm.

 In Ref.\,\cite{Engelmann:2018} the SiPM is imaged by a high resolution CCD camera with a sensitivity for photons between 300 and 900\,nm.
 Fig.\,\ref{fig:LightResult} shows the image of a KETEK PM3350T SiPM (50\,$\upmu$m pixels) in the dark at $20\,^\circ$C for $V_{OV} = 5.4$\,V and an exposure time of 4\,h.
 Assuming that on average every Geiger discharge produces the same amount of light, the observed light intensity is  proportional to  $DCR$.
 Hot-spots are observed with a light intensity approximately 20 times higher than the average.
 These high generation rates are explained by the presence of point defects, either of the starting material or generated during the fabrication process.
 A similar observation has been made in Ref.\,\cite{Frach:2009} using a digital SiPM, where individual pixels can be disabled and the $DCR$ of individual pixels measured.
 Thus, the frequently made assumption, that the distribution of dark counts can be described by a Poisson distribution with the same mean for every pixel, is in strong disagreement with this observation.
 It may  be a better approximation for highly radiation-damaged SiPMs.
 This however has not yet been demonstrated.

 Fig.\,\ref{fig:LightResult} also shows that the hot-spots are fixed in space and that the light spots have a diameter of about 10\,$\upmu$m, much smaller than the 50\,$\upmu$m pixel size, which allows estimating the diameter of the micro-discharge channels.
 It is also reported that the diameter of the light spots does not depend on $V_{OV}$.

  \section{Determination of the SiPM parameters}
   \label{sect:Determination}

 In the following, it is described how the different parameters discussed in Sect.\,\ref{sect:Parameters} can be determined using the setups presented in Sect.\,\ref{sect:Setups}.
 Most of the parameters can be determined in several ways.
 Some comments will be given, which way the author considers to be the most trustworthy.
 As discussed in Ref.\,\cite{Garutti:2018}, most of the methods cannot be applied if the $DCR$ or the noise is so high that 0, 1,  and more Geiger discharges cannot be distinguished.
 Ideas on how to characterise SiPMs in these situations will be presented.

  \subsection{Electrical parameters}
   \label{sect:ElParameters}

  To illustrate the determination of the electrical parameters, results are presented for 4 different KETEK SiPMs studied in Ref.\,\cite{Chmill1:2017}. Their names and parameters are given in Table\,\ref{table:ElPar}.
  They all have an area of 1\,mm$^2$,
  PNCV is a special, single pixel produced by KETEK for testing purposes.
  It cannot be used as photo-detector for voltages above $V_{bd}$, because the value of $R_q$ is too low to quench the Geiger discharge.

 \begin{table}[!ht]
  \centering
   \begin{tabular}{c|c|c|c|c|c}
      & PM15 & PM25  & PM50 & PM100 & PNCV \\
  \hline
     $N_{pix}$ & 4384 & 1600 & 400 & 100 & 1 \\
     $pitch$ & 15\,$\upmu $m & 25\,$\upmu $m  & 50\,$\upmu $m  & 100\,$\upmu $m  & 1\,mm \\
   \hline
     $C_{d}$ & 18 fF & 69 fF & 330 fF & 1.5 pF & 110 pF \\
     $R_q$ & 750 k$\Omega $ & 500 k$\Omega $  & 340 k$\Omega $  & 410 k$\Omega $  & 130 $\Omega $  \\
     $C_q$ & < 5 fF & < 10 fF & 25 fF & 155 fF & -- \\
     $R_I$ & 85 G$ \Omega $  & 80 G$\Omega $  & 70 G$ \Omega $  & 50 G$\Omega $  & 85 G$ \Omega $  \\
     $\tau _r $ & 14 ns & 25 ns & 100 ns & 620 ns & 14 ns \\
     \hline
   \end{tabular}
  \caption{Geometrical parameters (top) and electrical parameters as determined from the admittance-frequency ($Y-f$) measurements (bottom) of the KETEK SiPMs investigated. }
  \label{table:ElPar}
 \end{table}

\begin{figure}[!ht]
   \centering
   \begin{subfigure}[a]{0.5\textwidth}
    \includegraphics[width=\textwidth]{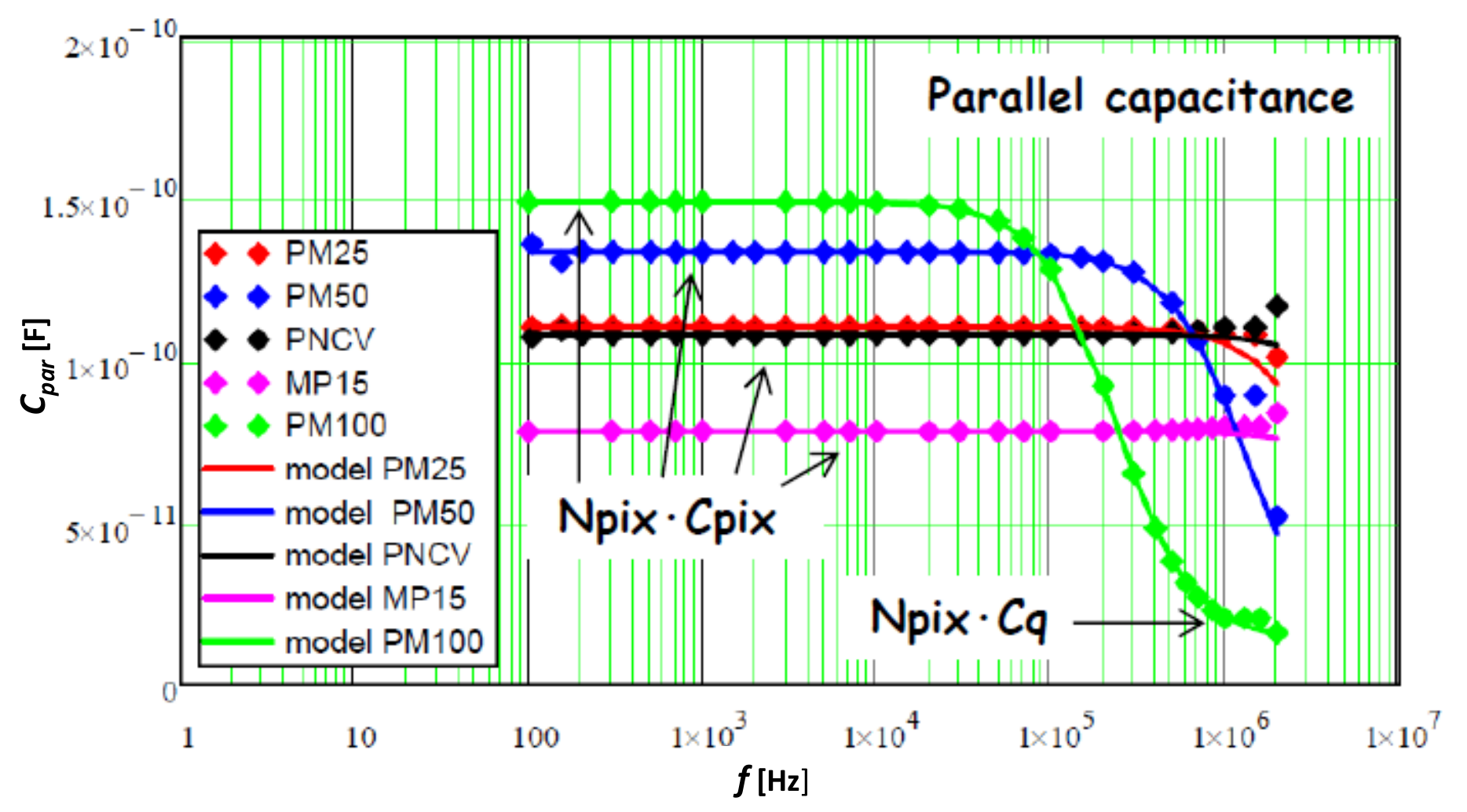}
    \caption{ }
    \label{fig:Cpar}
   \end{subfigure}%
    ~
   \begin{subfigure}[a]{0.5\textwidth}
    \includegraphics[width=\textwidth]{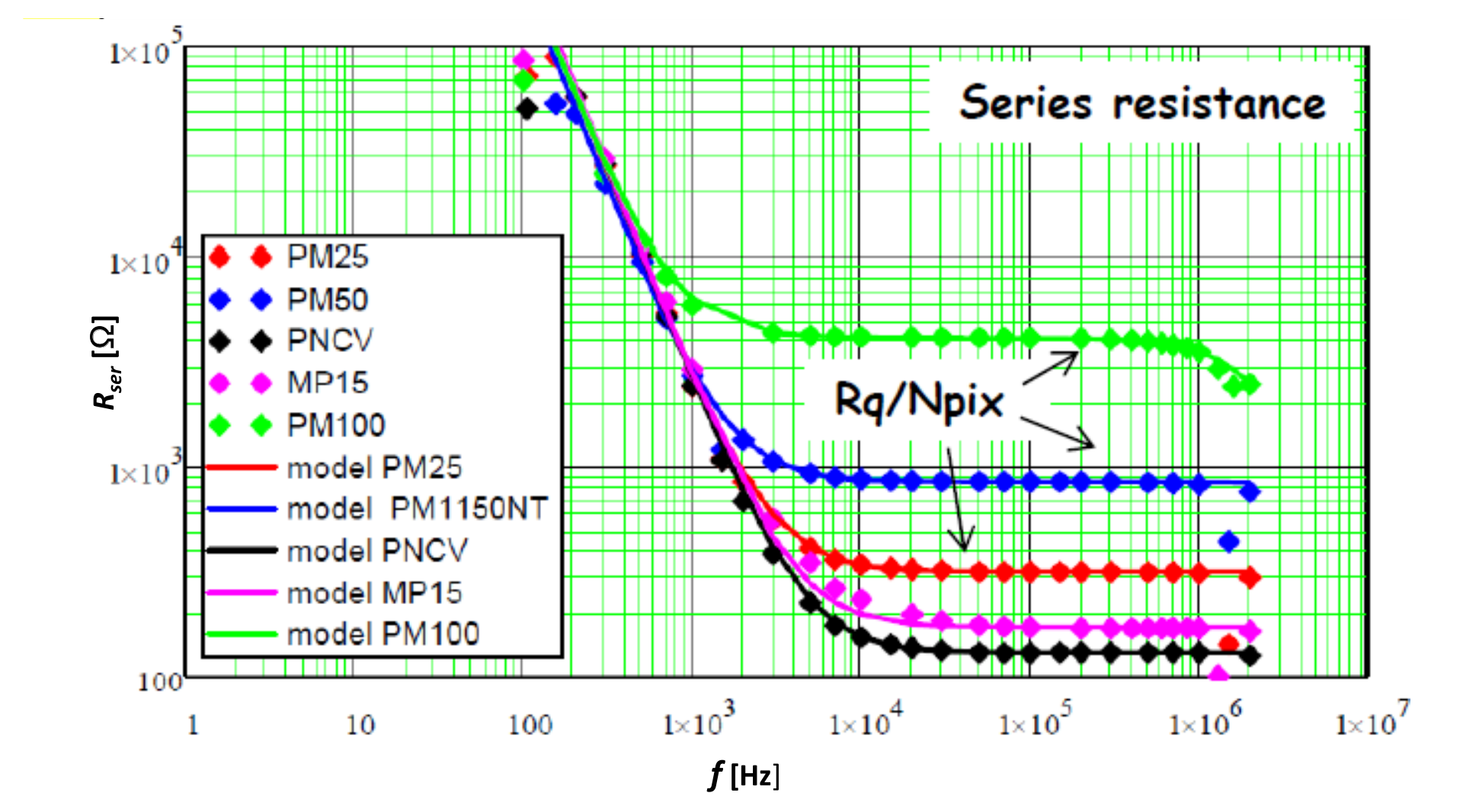}
    \caption{ }
    \label{fig:Rse}
   \end{subfigure}%
   \caption{Analysis of the admittance-frequency ($Y-f$) measurements for KETEK SiPMs with different pixel sizes measured 0.5\,V below the breakdown voltage, $V_{bd} \approx 27.5\,V$, and $ 20\,^\circ $C.
    From the $Y-f$ data
   (a) the parallel capacitance, $C_{par}$, and
   (b) the series resistance, $R_{ser}$, as a function of frequency are shown.
   As discussed in the text, approximate values of the electrical parameters $C_d$ ($C_{pix}$ in the figure) and $R_q$ can be obtained directly from the values of constant $C_{par}$ and $R_{ser}$.}
  \label{fig:Yf}
 \end{figure}

 The admittance-frequency, $Y-f$, measurements were performed at 0.5 and 1\,V below the breakdown voltage for 27 frequencies between 100\,Hz and 2\,GHz.
 The LCR-meter used records
 \begin{equation}\label{equ:YLCR}
   Y(f) = 1/R_{par}(f) + i\,\omega \cdot C_{par}(f),
 \end{equation}
 with the parallel resistance, $R_{par}$, and the parallel capacitance, $C_{par}$.
 The series capacitance, $C_{ser}$, and the series resistance, $R_{ser}$, are obtained from Eq.\,\ref{equ:YLCR} using $Z(f) = 1/Y(f) = R_{ser} +1/(i\,\omega \cdot C_{ser})$.
 For the analysis, the electrical model shown in Fig.\,\ref{fig:IVCV-setup} with the $C-V$\,switch closed, is used.
 The admittance of a single pixel is given by
 \begin{equation}\label{Ypix}
   Y_{pix} = \Big( \big(\frac {1} {R_q} + i\,\omega \cdot C_{q} \big) ^{-1} +\frac{1} {i\,\omega \cdot C_{d}} \Big) ^{-1},
 \end{equation}
 and the total admittance by
 \begin{equation}\label{equ:Ytot}
 Y_{tot} = \Big( (N_{pix} \cdot Y_{pix})^{-1} + i\,\omega \cdot L_s \Big)^{-1} + \frac{1} {R_I}.
 \end{equation}
 Fig.\,\ref{fig:Yf} shows as a function of frequency the measured $C_{par}$ and $R_{ser}$.
 From Eq.\,\ref{equ:Ytot} follows that for intermediate frequencies $C_{par} \approx N_{pix} \cdot C_d$, and at high frequencies, for $\omega \cdot C_q \gg  1/R_q$, $C_{par} \approx N_{pix} \cdot (1/C_q + 1/C_d)^{-1}$.
 For the SiPMs PM50 and PM100, where a significant fast component is observed in the current transient (see Fig.\,\ref{fig:Pulse}), the decrease of $C_{par}$ at high frequencies can be seen in Fig.\,\ref{fig:Cpar}.
 At high frequencies, the dominant contribution to $Z_{tot} = 1/Y_{tot}$ is $R_q /N_{pix}$ in series with $N_{pix} \cdot C_d$.
 Thus in Fig.\,\ref{fig:Rse} at high frequencies the constant value of $R_{ser} \approx R_q/N_{pix}$ gives an approximate value of $R_q$.
 With these initial values for $C_d$, $C_q$, and $R_q$, all 5 parameters of the model ($C_d$, $C_q$, $R_q$, $L_s$, $R_I$) are adjusted until the data are well described.
 The results are shown as solid lines in Fig.\,\ref{fig:Yf}.

 It is concluded that the electrical SiPM parameters can be approximately determined from $Y-f$ measurements and  that with this method the change of these parameters with irradiation can be determined for highly irradiated SiPMs, where dark-count rates exceed GHz.
 Ref.\,\cite{Xu1:2014} reports such a study for radiation damage by X-rays, and Ref.\,\cite{Vignali:2017} by reactor neutrons up to fluences of $5 \times 10^{14}$\,cm$^{-2}$.
 A detailed study of the accuracy of this method and its dependence on the SiPM design has so far not been published.
 However, it is surprising that this method of determining the electrical SiPM parameters is hardly used.

\begin{figure}[!ht]
   \centering
   \begin{subfigure}[a]{0.5\textwidth}
    \includegraphics[width=\textwidth]{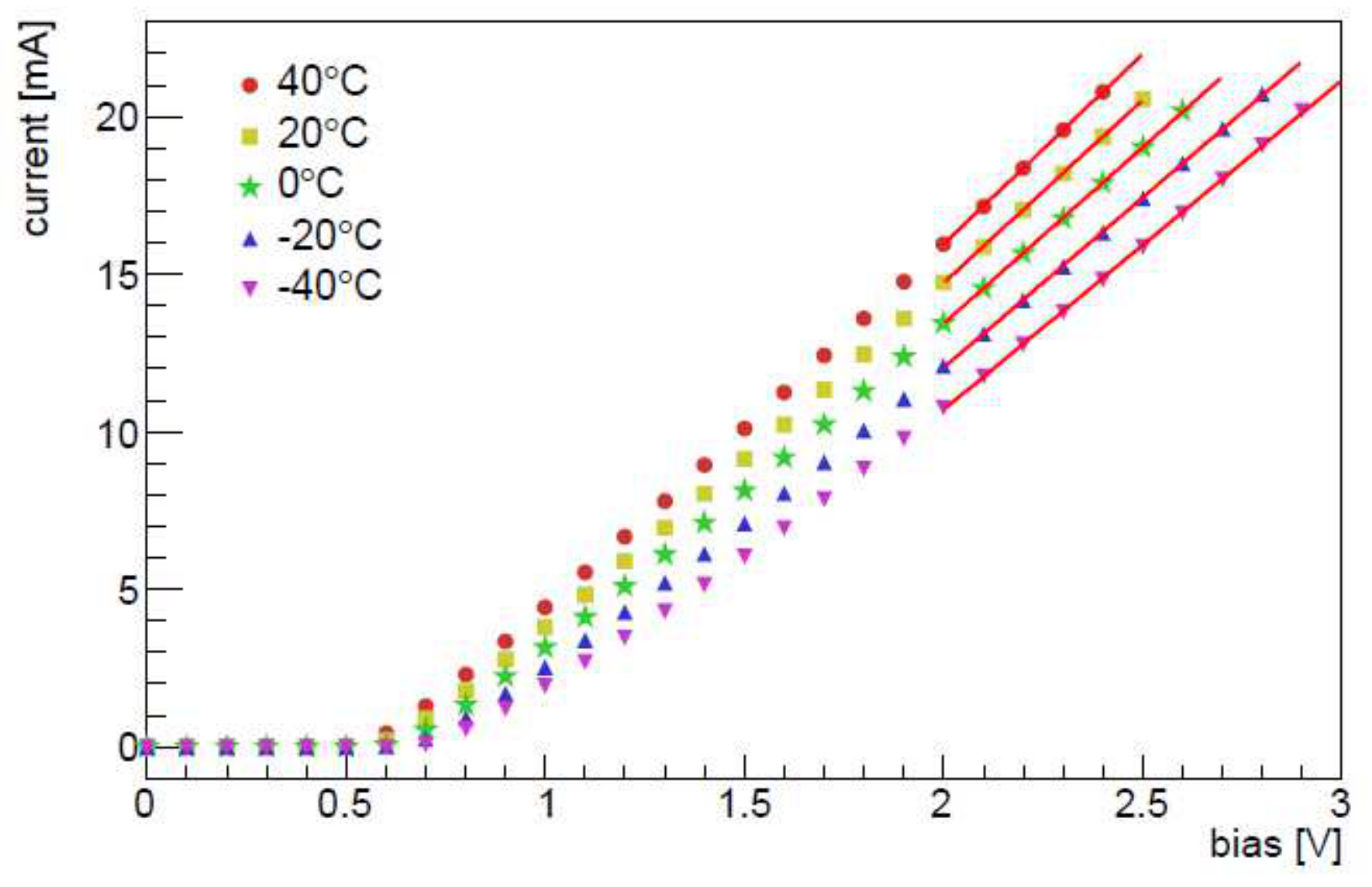}
    \caption{ }
    \label{fig:IRq}
   \end{subfigure}%
    ~
   \begin{subfigure}[a]{0.5\textwidth}
    \includegraphics[width=\textwidth]{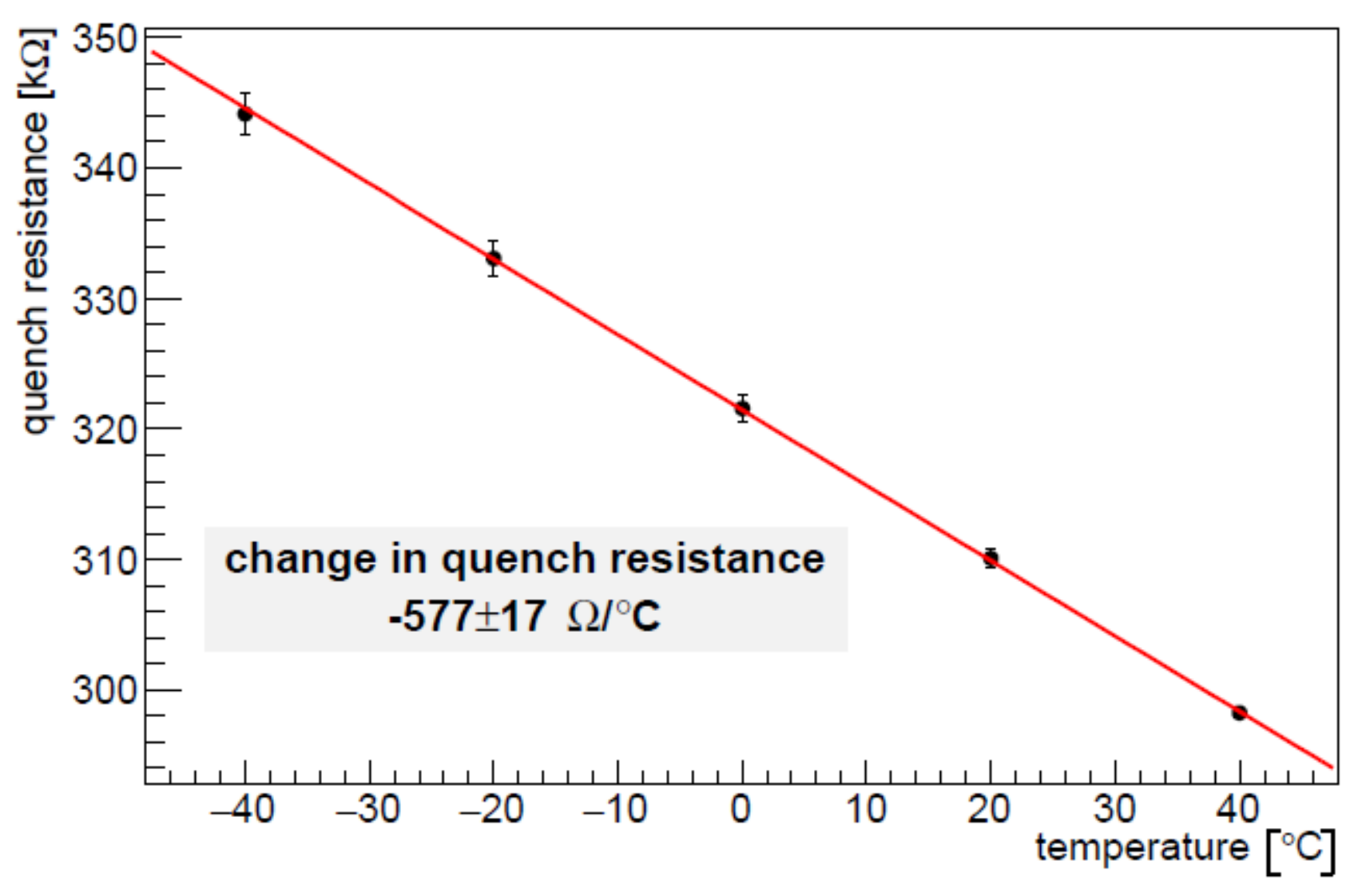}
    \caption{ }
    \label{fig:TRQ}
   \end{subfigure}%
   \caption{Determination of the quenching resistance, $R_q$, from the $I_f -V$ measurement for different temperatures with forward bias for the Hamamatsu S13360-3050CS SiPM from Ref.\,\cite{Otte:2016}.
   (a) The value of $ R_q$ is obtained from a linear fit to the $I_f-V$ measurement for $V_{bias} > 2$\,V.
   (b) The value of $R_q$ as function of temperature.}
  \label{fig:Rq}
 \end{figure}

 The standard way of determining $R_q$ is to measure the current for forward bias  with a setup as shown in Fig.\,\ref{fig:IVCV-setup} with the $I-V$\,switch closed.
 For sufficiently high $V_{bias} $\,values the diode becomes conductive and the differential resistance is $1/(\mathrm{d} I_f / \mathrm{d}V_{bias}) \approx R_q/N_{pix}$.
 Examples for such measurements from Ref.\,\cite{Otte:2016} are shown in Fig.\,\ref{fig:Rq}.

 Fig.\,\ref{fig:IRq} shows the $I_f -V$ results for temperatures between $-40\,^\circ $ and $+40\,^\circ $C with straight-line fits for $V_{bias} > 2$\,V.
 The inverse of the slope gives $R_q / N_{pix}$.
 Fig.\,\ref{fig:TRQ} shows the temperature dependence of $R_q$.
 As expected for a poly-Si resistor, the resistance increases with decreasing temperature.

 From the author's experience, the value obtained for $R_q$ depends on the fit range, and the derivative $\mathrm{d}I_f /\mathrm{d}V_{bias}$ approaches, but does not reach a constant value.
 For the KETEK SiPMs studied by the author, the value of $R_q$ from the $I_f-V$ measurement is typically 5\,\% higher than the one found from the $Y-f$ measurements, which is assumed to be more accurate.
 However, for the SiPM characterisation the precise knowledge of $R_q$ is not so important.


  \subsection{Electric field}
   \label{sect:Efield}

\begin{figure}[!ht]
   \centering
    \includegraphics[width=0.5\textwidth]{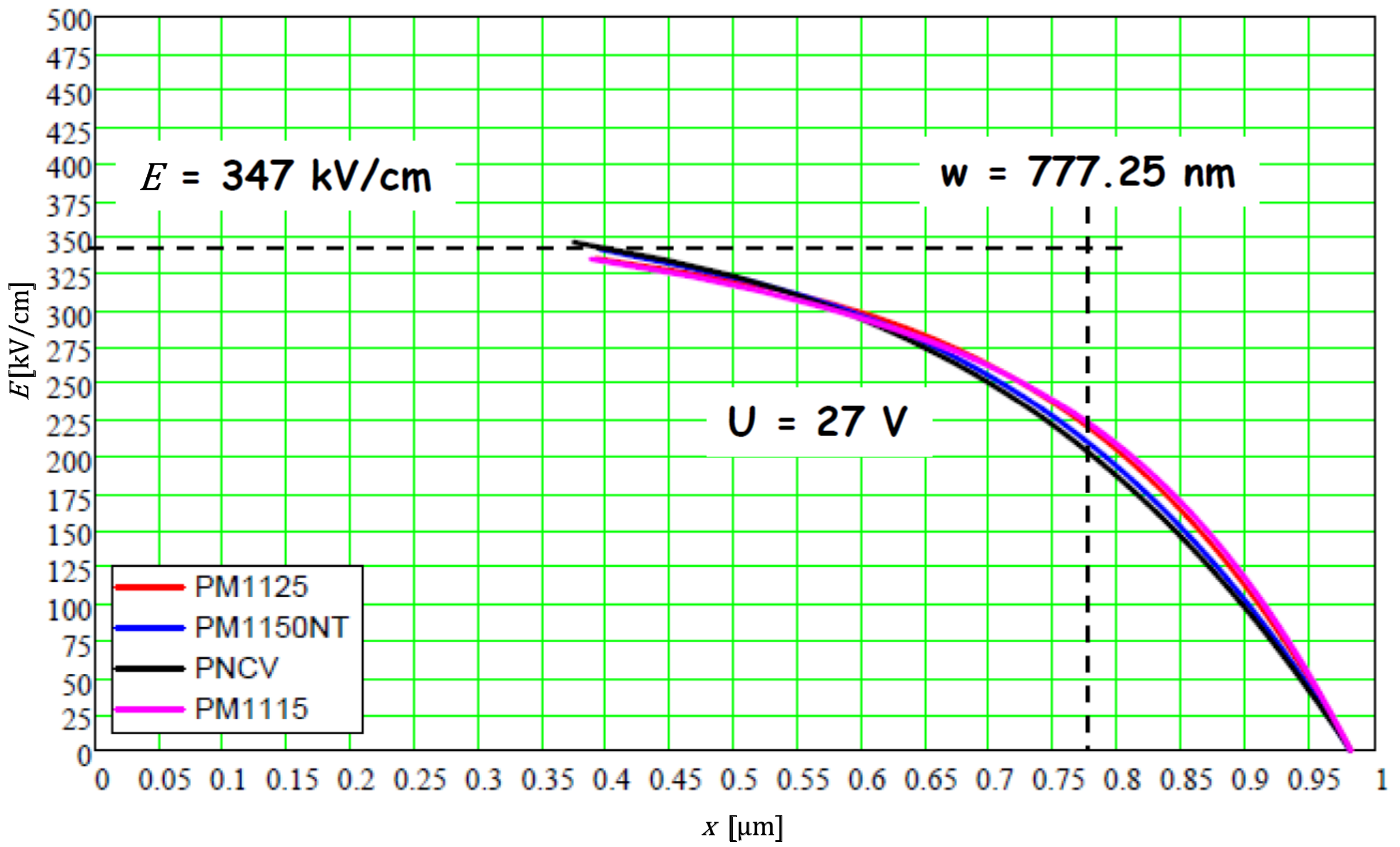}
   \caption{Electric field as a function of the distance from the $pn$\,junction determined from $C-V$\,measurements in the range $V_{bias} = 1$ to 27\,V for the SiPMs listed in Table\,\ref{table:ElPar}.}
  \label{fig:Efield}
 \end{figure}

 From  $C-V$ measurements it is  possible to estimate the doping density and the electric field in the amplification region.
 Such information, which is only rarely communicated by the vendor to the user, is required to simulate the Geiger breakdown probability as a function of position, which can be done using the formulae given in Ref.\,\cite{McIntyre:1973}.
 For the determination of the electric field the standard 1-D textbook formulae for an asymmetric $pn$ junction given e.\,g. in Ref.\,\cite{Grove:1967} can be used:
 \begin{equation}\label{eq:xV}
   x(V_{bias}) = \frac{\varepsilon _0\, \varepsilon _{Si} A } {C(V_{bias})} \hspace{5mm} \mathrm{and} \hspace{5mm} N_d(x)=\frac{2} {q_0 \,\varepsilon _0\, \varepsilon _{Si} \,A^2} \cdot \frac{1} {\mathrm{d}(1/C)^2/\mathrm{d} V_{bias}}
 \end{equation}
 with the distance from the $pn$\,junction $x$, and the doping density $N_d(x)$, and
 \begin{equation}\label{eq:Efield}
   E(x) = \int _{x_{max}} ^x \frac{q_0 \, N_d(x) } {\varepsilon _0\, \varepsilon _{Si} } \mathrm{d}x
 \end{equation}
 for the electric field $E(x)$.
 The SiPM area is denoted by $A = N_{pix} \cdot pitch^2$, the elementary charge by $q_0$ and the dielectric constant of Si by $\varepsilon _0\, \varepsilon _{Si}$.
 The maximal depletion depth reached is $x_{max} = x(V_{bias,\,max})$, where $V_{bias,\,max} = 27$\,V is the maximum bias voltage used in the measurements.
 Fig.\,\ref{fig:Efield} shows the results for the KETEK SiPMs of Table\,\ref{table:ElPar}.
 For these SiPMs the $pn$\,junction is close to the entrance window, the built-in depletion depth is about $0.35\,\upmu$m, and the maximal electric field  $\approx 350$\,kV/cm for $V_{bias}$ approximately 0.5\,V below the breakdown voltage $V_{bd}$.
 The full depletion depth is about $1\,\upmu$m, which is quite shallow, and results in a relatively narrow amplification region.
 The observation, that the electric field obtained for the PNCV, a single 1\,mm$^2$ diode, and for the SiPMs with different pixel sizes are approximately the same, confirms that assuming  the 1-D model and taking the entire SiPM area for $A$ in the analysis, are reasonable for the SiPM investigated.
 The electric field above $V_{bd}$ can be estimated, by adding $(V_{bias}-V_{bias,\,max})/x_{max}$ to the field determined below $V_{bd}$.

  \subsection{Breakdown and turnoff voltage}
   \label{sect:Vbd}

 Consistent with the discussion in Sect.\,\ref{sect:Parameters}, a distinction is made in this paper between $V_{bd}$, the voltage at which Geiger discharges start to occur, and $V_{off}$, the voltage at which the Geiger discharges are quenched.
 In Ref.\,\cite{Chmill1:2017} a difference $V_{bd} - V_{off}$ of about 1\,V is reported for a specific SiPM, and in Ref.\,\cite{Marinov:2007} a model calculation for $V_{bd} - V_{off}$  is presented.
 However, in most of the literature only $V_{bd}$ is used, and this issue still has to be clarified.
 If the SiPM is operated well above $V_{bd}$, a small difference $V_{bd} - V_{off}$ has only a small effect.
 However, if the SiPM is operated close to $V_{bd}$, which may be required at high dark count rates due to background light or radiation damage, the effect could be significant.

 Two types of measurements are used to determine the breakdown voltage $V_{bd}$:
 Analysis of the $I-V$ characteristics and extrapolation of $PDE (V_{bias}) $ to $PDE(V_{bd}) = 0$.
 For the determination of the turn-off voltage $V_{off}$, the linear gain-voltage dependence, $G^\ast (V_{bias})$ is extrapolated to $G^\ast (V_{off}) = 1$.

\begin{figure}[!ht]
   \centering
    \includegraphics[width=0.75\textwidth]{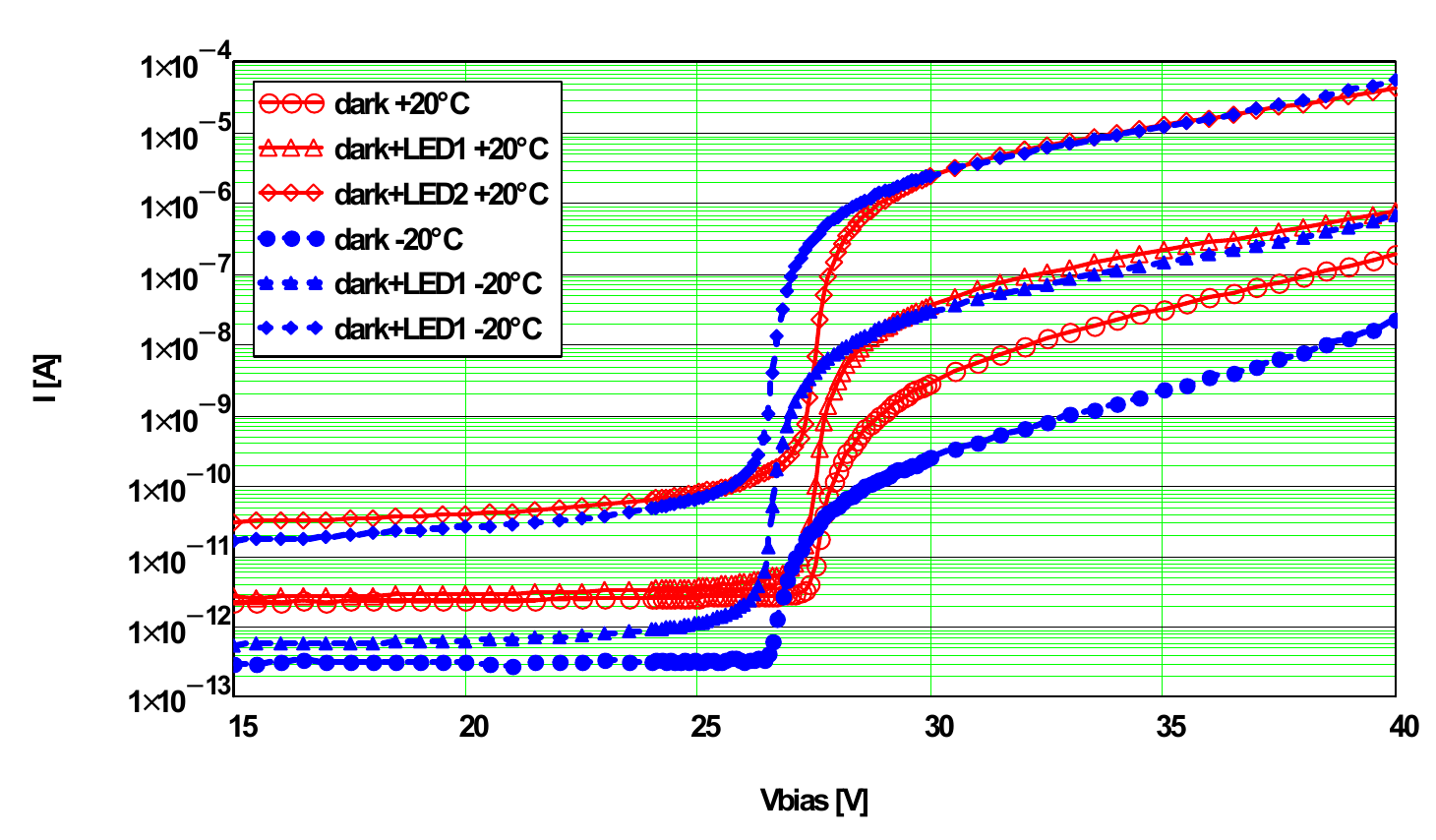}
   \caption{Current versus $V_{bias}$ for the KETEK SiPM measured at $+20\,^\circ $C and $-20\,^\circ $C in the dark ("dark") and with DC-illumination by a blue LED with low ("dark+LED1") and high ("dark+LED2") intensity.
   After subtracting the dark current, the photo-currents for LED1 and LED2 scale (not shown). }
  \label{fig:IVrev}
 \end{figure}

 Fig.\,\ref{fig:IVrev} shows $I-V$\, measurements for the KETEK SiPM MP15 at $+20\,^\circ$C and $-20\,^\circ $C in the dark and with DC illumination by a blue LED with low and high photon intensity.
 The current scale extends over 9 orders of magnitude.
 At $V_{bd}$ the currents with and without illumination increase rapidly due to the onset of Geiger discharges.
 As will be shown later quantitatively, at a given temperature the same $V_{bd}$ value is observed with and without illumination.
 Between  $+20\,^\circ $C and $-20\,^\circ $C $V_{bd}$ decreases by $\approx 900$\,mV, because of the increase of the charge-carrier ionisation coefficients with decreasing temperature.

 Below $V_{bd}$ the $I-V$\,characteristics are very different for the data with and without illumination:
 With illumination ($I_{light}$) the expected increase in current due to avalanche multiplication -- the regime in which Avalanche Photo-Diodes (APDs) are operated -- is observed, whereas without illumination ($I_{dark}$) the current is constant up to $V_{bd}$.
 The reason is that $I_{dark}$ below $V_{bd}$ is dominated by surface-generation current from the depleted Si-SiO$_2$ interface, which misses the amplification region and is therefore not amplified.

 At $V_{bd}$ the current rises rapidly, with an increase which is higher for the illuminated SiPM.
 The reason is again that charge carriers from the Si-SiO$_2$ interface bypass the amplification region.
 It is also seen that the relative slope of  $I_{dark}$ is steeper than of $I_{light}$ above $V_{bd}$.
 The reason is the position dependence of the Geiger trigger probability, $P_T$.
 It is highest close to the $pn$ junction, which for this SiPM is located  near to the SiPM entrance window.
 Whereas the thermally generated  $eh$ pairs are approximately uniformly generated in the depletion region, the blue light has an absorption length  of $\approx 0.1 \upmu $m and generates $eh$ pairs in the region of highest $P_T$ only.
 With increasing $V_{bias}$ the region of high $P_T$ extends further and further into the amplification region, thus increasing $\langle P_T \rangle $ for the uniformly generated $eh$ pairs from dark counts.

 At voltages above $V_{bias} \approx 37$\,V,   $I_{dark}$ and $I_{light}$ at $- 20\,^\circ$C show an increase of the slope of $\ln (I)$ compared to the $+ 20\,^\circ$C data.
 It still has to be investigated if this increase is due to Geiger discharges for which the quenching is delayed and the current through $R_q$ during the discharge contributes significantly to the signal, or to an increased correlated noise (e.\,g. after-pulses) at high electric fields and low temperature.

\begin{figure}[!ht]
   \centering
    \includegraphics[width=0.75\textwidth]{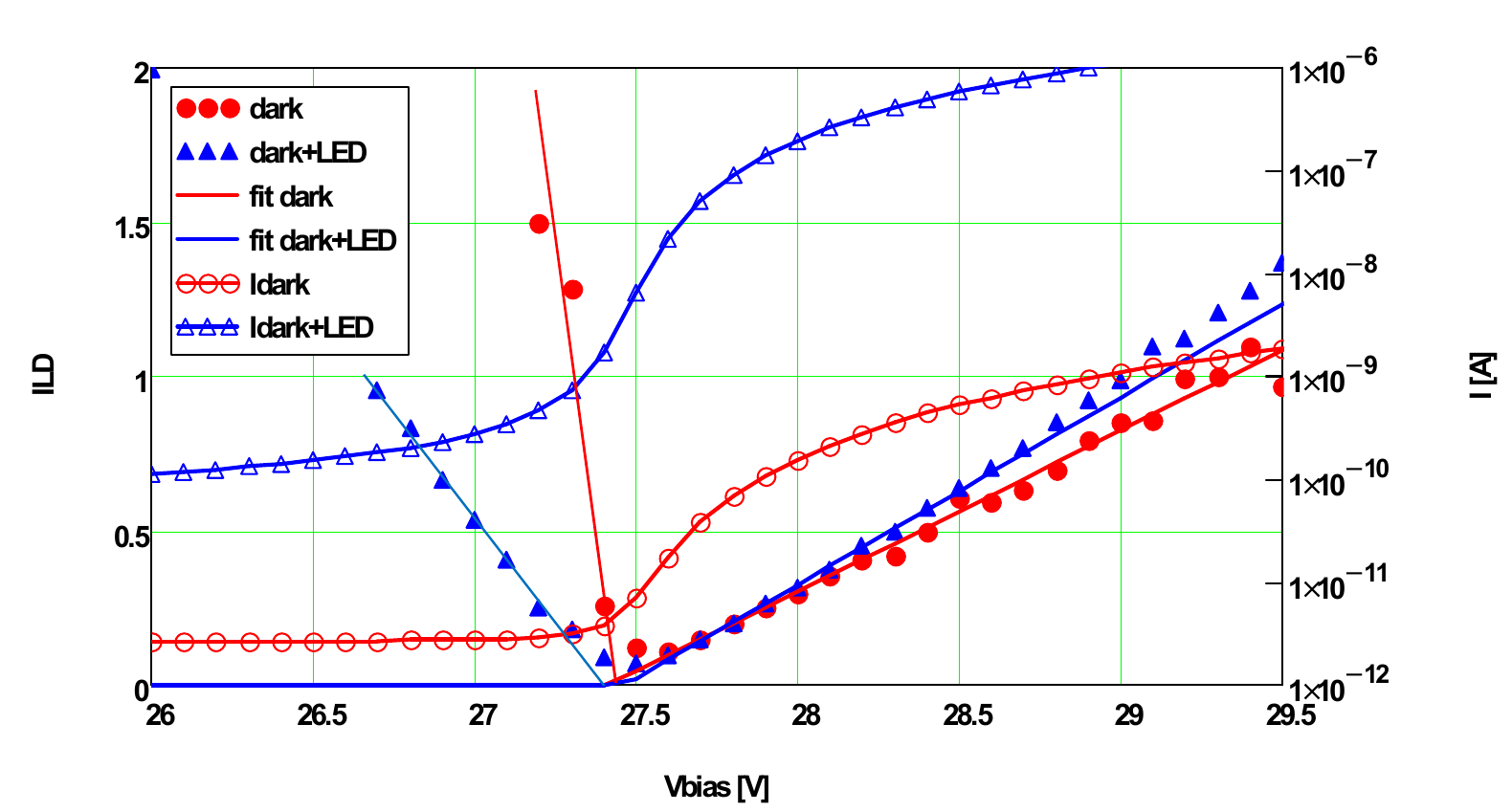}
   \caption{For the KETEK SiPM with 15\,$\upmu $m pitch measured at $+20\,^\circ $C: $I_{dark}$ and $I_{light}$ ($I_{dark+LED}$ in the figure) with the scale on the right, and the corresponding $ILD$ values with straight-line fits below and above $V_{bd}$ with the scale on the left. }
  \label{fig:ILD}
 \end{figure}

 Several methods are used to determine $V_{bd}$ from the $I(V_{bias})$ measurements giving all similar results, and it is a matter of taste which one to use.
 Most of them use either the logarithmic derivative, $LD = \mathrm{d} \ln(I)/ \mathrm{d} V_{bias}$, or its inverse $ILD = 1/LD$.
 The advantage of $LD$ and $ILD$ is that they are only sensitive to the shape and not to the  value of $I(V_{bias})$,  and $I(V_{bias})$ measurements can be easily compared, even if the current values are vastly different.
 This is seen in Fig.\,\ref{fig:ILD}, where $I_{dark}$ and $I_{light}$ differ by three orders of magnitude, and the $ILD$s are quite similar.

 The breakdown voltage $V_{bd}$ for the different methods is  determined as the voltage at which
  \begin{enumerate}
    \item $LD$ has its maximum.
    \item The parabola through the 3 points around the minimum of $ILD$ has its minimum.
    \item The extrapolation of a straight-line (or parabola) fit of $ILD$ for $V_{bias} > V_{bd}$ is zero.
    \item The extrapolation of a straight-line fit of $ILD$ for $V_{bias} < V_{bd}$ is zero
    \item The second derivative of $\ln I(V_{bias})$ with respect to $V_{bias}$ has its maximum.
    \item A second order polynomial, fitted to $I(V_{bias})$ above $V_{bd}$ after surface-current subtraction, crosses the $V_{bias}$ axis.
  \end{enumerate}

 Fig.\,\ref{fig:ILD} and Table\,\ref{table:Vbd} show the results of methods $1 - 4$ for the KETEK SiPM PM15.
 Shown in the figure are $I_{dark}$ and $I_{light}$ with the scale  on the right, and the corresponding $ILD(V_{bias})$ results with straight-line fits below and above $V_{bd}$, with the scale on the left.
 It is found that the results for $V_{bd}$ from $I_{dark}$ and $I_{light}$  for $V_{bias} > V_{bd}$ agree within $\pm 20$\,mV.
 The values found from $I_{dark}$ for methods 1 and 2 are systematically higher by $\approx 100$\,mV, which is related to the nearly constant $I_{dark}$ for $V_{bias} < V_{bd}$, which results in a very high $ILD$\,value and shifts the $ILD$\,minimum to somewhat higher values.

 In Ref.\,\cite{Otte:2016} method\,5 is compared to methods\,1 and 2, and agreement at the 100\,mV is reported.
 To the knowledge of the author, a comparison at the 20\,mV level is not available.
 In addition it is noted, that obtaining reliably second derivatives from experimental data can by quite tricky.

 As discussed below, the second order polynomial of method\,6, which is proposed in Ref.\,\cite{Piemonte:2007} and also recommended in Ref.\,\cite{Otte:2016}, describes only the $I_{dark}$ but not the $I_{light}$ data for the KETEK SiPMs.
 Therefore it was not used.
 The method assumes the functional form for the Geiger breakdown probability $P_T \propto \big[1 - \exp\ \big(- \alpha \cdot \big(V_{bias} - V_{bd})\big)\big]$ for both $I_{dark}$ and $I_{light}$.
 Apparently the functional form of $P_T(V_{bias})$ depends on the SiPM design and is also position dependent.
 As a result, a power-law fit $I(V_{bias}) \propto (V_{bias} - V_{bd})^n$ with the free parameter $n$, which is equivalent to method\,3, is the safer approach.

 For a quick and reliable determination at the 50\,mV level it is recommended to use method\,2 with the SiPM illuminated with DC light.
 In particular at low temperatures $I_{dark}$ is so low that the measurement errors are significant, which makes the $V_{bd}$ results unreliable.
 An idea about the dominant systematic uncertainties and a more precise determination can be obtained by varying the fit range in method 3 and by using a second order polynomial to fit $ILD$.

\begin{table}
  \centering
   \begin{tabular}{c|c|c|c|c}
     Method & 1 & 2 & 3 & 4 \\
    \hline
     $V_{bd}\, \mathrm{from}\, I_{dark}$ [V]   & 27.6 & 27.57 & 27.42 & 27.44 \\
     $V_{bd}\, \mathrm{from}\, I_{light}$ [V]  & 27.5 & 27.51 & 27.49 & 27.41 \\
    \hline
   \end{tabular}
  \caption{Results of the different methods for determining $V_{bd}$ for the KETEK SiPM with $15\,\upmu$m pitch.}
 \label{table:Vbd}
\end{table}

 From the fit using method 3 the inverse slope $1/n $ of $ILD$ is obtained: $n_{dark}$ from $I_{dark}$, and $n_{photo}$ from $I_{photo}$.
 The values found for $n _{dark}$ and for $n _{photo}$ are 1.96 and 1.43, respectively.
 For an $I(V) = (V-V_{bd})^{n} $ dependence, $ILD(V_{bias}) = (V_{bias} - V_{bd}) / n$.
 Thus a straight line of $ILD$ means that above $V_{bd}$ the current obeys the power law $I(V_{bias}) \propto (V_{bias}-V_{bd})^n.$

 From $n_{dark}$ and $n_{photo}$  information on the position- and $V_{bias}$-dependence of the Geiger-discharge probability, $P_T$, can be obtained:
 Assuming a uniform,  voltage-independent thermal volume-generation rate $U_{gen}$ in the depletion region, i.\,e. ignoring high-field effects, the primary dark count rate $DCR_p$ = $U_{gen} \cdot \langle P_T \rangle _{dep}$, and $I_{dark} = q_0 \cdot G \cdot DCR \cdot ECF \propto G \cdot \langle P_T \rangle _{dep} \cdot ECF$, where $P_T$ is averaged over the entire depletion region.
 In the approximation $V_{bd} \approx V_{off}$, $G \propto (V_{bias} - V_{bd})^1 $, and taking into account that $ECF \approx 1$ for small $ V_{bias} - V_{bd}$ values, $\langle P_T \rangle _{dep} \propto (V_{bias} - V_{bd})^{n_{dark} - 1}$.
 The corresponding relation for the photo-current is: $I_{photo} = q_0 \cdot G \cdot N_\gamma \cdot PDE \cdot ECF \propto G \cdot \langle P_T \rangle _{photo} \cdot ECF$, from which follows $\langle P_T \rangle _{light} \propto (V_{bias} - V_{bd})^{n_{light} - 1}$.
 Here the average of $P_T$ is taken over the region in which the photons are absorbed, which extends only to $\approx 0.1\,\upmu$m from the entrance window for the blue light used.
 Thus $n_{photo} - 1$ is related to $P_T(V_{bias})$ at the SiPM entrance window, and $n_{dark}-1$ to $P_T(V_{bias})$ in the entire depletion region.
 This information can be used to validate simulations of the position dependence of $P_T$ for different $V_{bias}$ values.

 Another approach of determining $V_{bd}$ is presented in Refs.\,\cite{Otte:2016, Otte:2018}.
 The measured voltage dependence of $PDE$ is fitted by the phenomenological function
 \begin{equation}\label{eq:pdeOtte}
   PDE(V_{bias}) = PDE_{max} \big(1 - e^{- \mathfrak{O} \cdot V_{rel}}\big) \hspace{5mm} \mathrm{with} \hspace{5mm} V_{rel} = \frac{V_{bias} - V_{bd}} {V_{bd}},
 \end{equation}
 with the phenomenological parameter $\mathfrak{O}$.
 The measurement of the $PDE$ will be described in Sect.\,\ref{sect:Gain}.
  The values found for $V_{bd}$ agree with the values found  using the methods described above, however the uncertainty is significantly larger.
 The authors point out that in first approximation $\mathfrak{O}$ does not depend on the width of the multiplication region and conclude that the wavelength dependence $\mathfrak{O} (\lambda )$ reflects the position dependence of $P_T$.
 These results still have to be compared to simulations using the formulae given in Ref.\,\cite{McIntyre:1973} with realistic electric fields, or TCAD or Monte Carlo programs.
 In Ref.\,\cite{Chmill1:2017} a similar approach is followed: $PDE(V_{bias})$ is fitted with the dependence derived assuming that all electron-hole pairs are generated at the SiPM entrance window and a constant electric field in the depletion region of effective width $w_{eff}$.
 The values found for $V_{bd}$ are again compatible with the values using the $I(V_{bias})$ methods.

\begin{figure}[!ht]
   \centering
   \begin{subfigure}[a]{0.52\textwidth}
    \includegraphics[width=\textwidth]{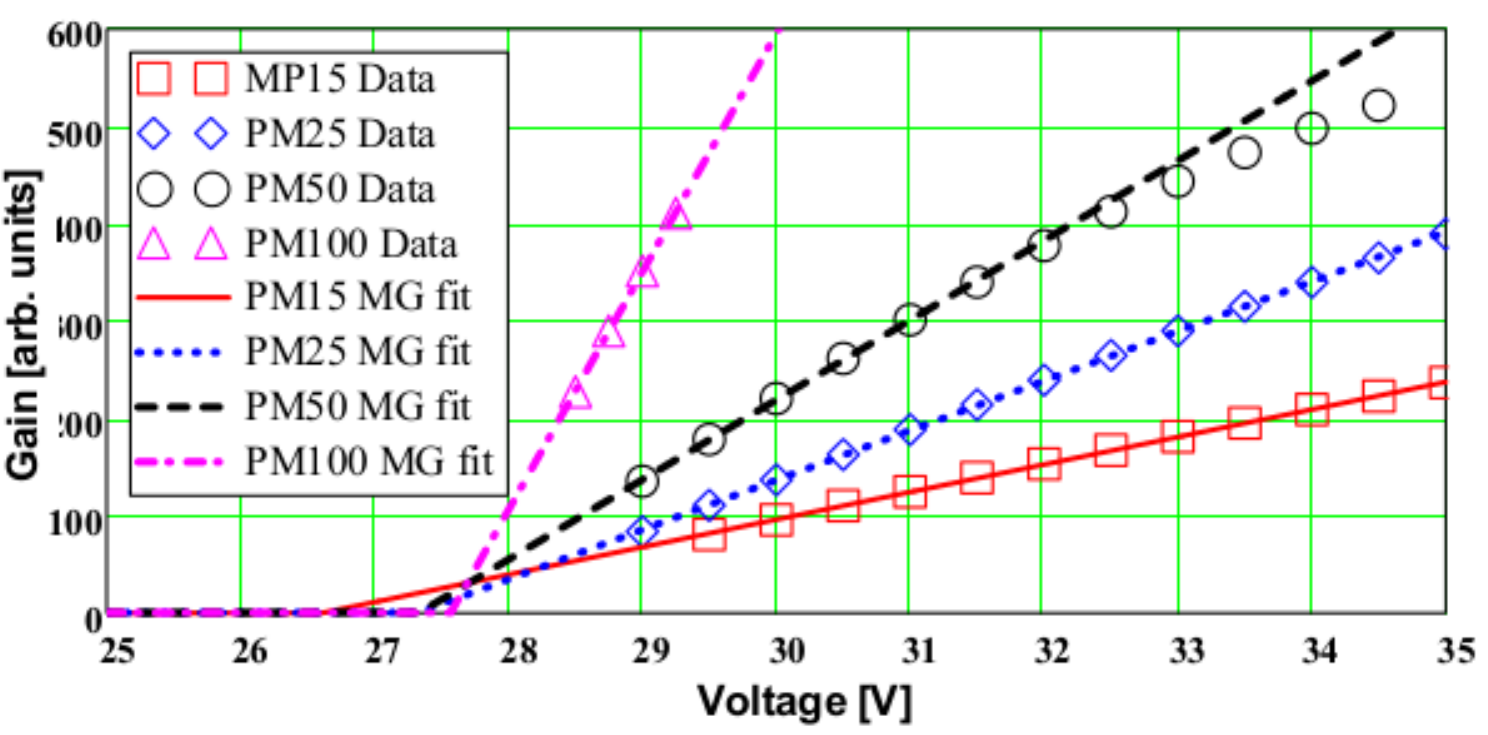}
    \caption{ }
    \label{fig:VbdGain}
   \end{subfigure}%
    ~
   \begin{subfigure}[a]{0.48\textwidth}
    \includegraphics[width=\textwidth]{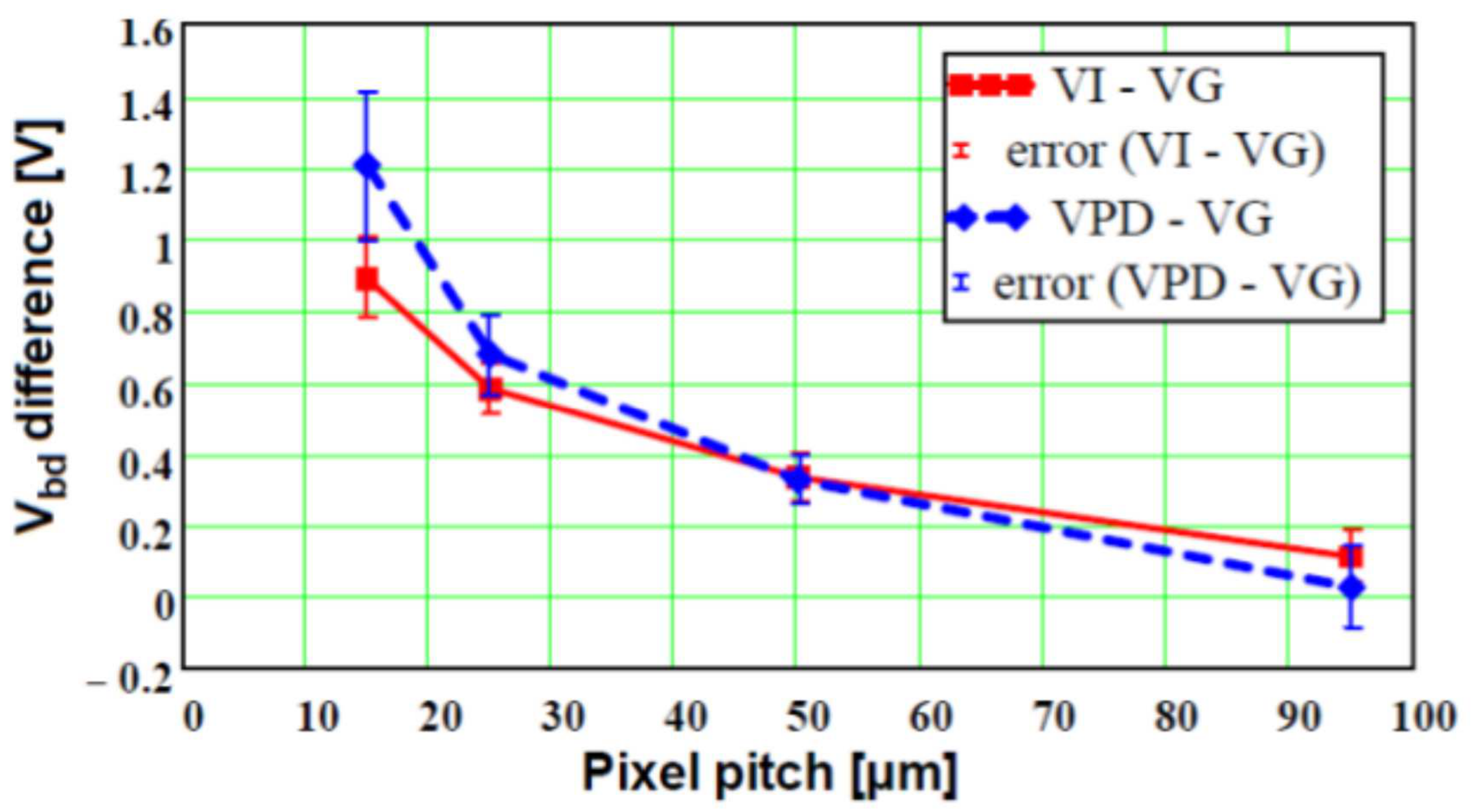}
    \caption{ }
    \label{fig:DeltaVbd}
   \end{subfigure}%
   \caption{ Results for the determination of $V_{bd}$ and $V_{off}$ for four KETEK SiPMs with different pitch from Ref.\,\cite{Chmill1:2017}.
    (a) Gain versus bias voltage, $G(V_{bias}) $, and straight-line fits to determine the turn-off voltage $V_{off}$.
    (b) Difference $V_{bd} - V_{off}$ as a function of the pixel pitch.
     $V_{off}$ from the $G(V_{bias})$ measurement is denoted VG, $V_{bd}$ from $I(V_{bias})$ VI, and $V_{bd}$  from $PDE(V_{bias})$ VPD.}
  \label{fig:Vbd}
 \end{figure}

 The turn-off voltage, $V_{off}$, is obtained from the voltage dependence of the SiPM gain, $G(V_{bias}) \approx (C_d + C_q) \cdot (V_{bias} - V_{off})$, by fitting a straight line to the data and extrapolating to $G(V_{off}) = 1$.
 Examples for $G(V_{bias})$ for the KETEK SiPM with pitch values between 15$\,\upmu$m and 100$\,\upmu$m and the corresponding straight-line fits are shown in Fig.\,\ref{fig:VbdGain}.
 The determination of  $G$ and of the fluctuations of $V_{off}$ is discussed in Sect.\,\ref{sect:Gain}.

 In Fig.\,\ref{fig:DeltaVbd} the differences $V_{bd} - V_{off}$ for the different pitch values of the KETEK SiPMs are shown.
 The $V_{bd}$ value from $I(V_{bias})$ is labeled VI, and the one from $PDE$, VPD.
 The symbol used for $V_{off}$ from $G(V_{bias})$ is labeled VG.
 The values found from VI and VPD are compatible, confirming that they both determine $V_{bd}$.
 However they differ from the values from VG, which determines $V_{off}$.
 The difference is approximately 1\,V for the SiPM with $15\,\upmu$m pixels and decreases with increasing pixel size.
 The reason for this dependence is not understood, however to the author's knowledge no simulations with realistic 3-D electric fields have  been performed so far.

 For the determination of $V_{off}$, instead of $G$ derived from charge spectra, $G$ from the pulse amplitude can also be used.
 The results obtained are compatible.
 Given the sensitivity of the amplitude to the band-width of the readout, in particular in the presence of a fast component, this method is not recommended.

 In summary: A  difference of up to 1\,V between $V_{bd}$ and $V_{off}$ has been observed for a KETEK SiPM with a pitch of 15$\,\upmu$m.
 For larger pitch values, the difference decreases.
 For the gain the relevant voltage is $V_{off}$, i.\,e. $G \propto (V_{bias} - V_{off})$.
 To avoid confusion, in publications it should be  clearly stated, which voltage, $V_{bd}$ or $V_{off}$, is used.

  \subsection{Photon-detection efficiency, number of primary Geiger discharges and gain}
   \label{sect:Gain}

\begin{figure}[!ht]
   \centering
   \begin{subfigure}[a]{0.5\textwidth}
    \includegraphics[width=\textwidth]{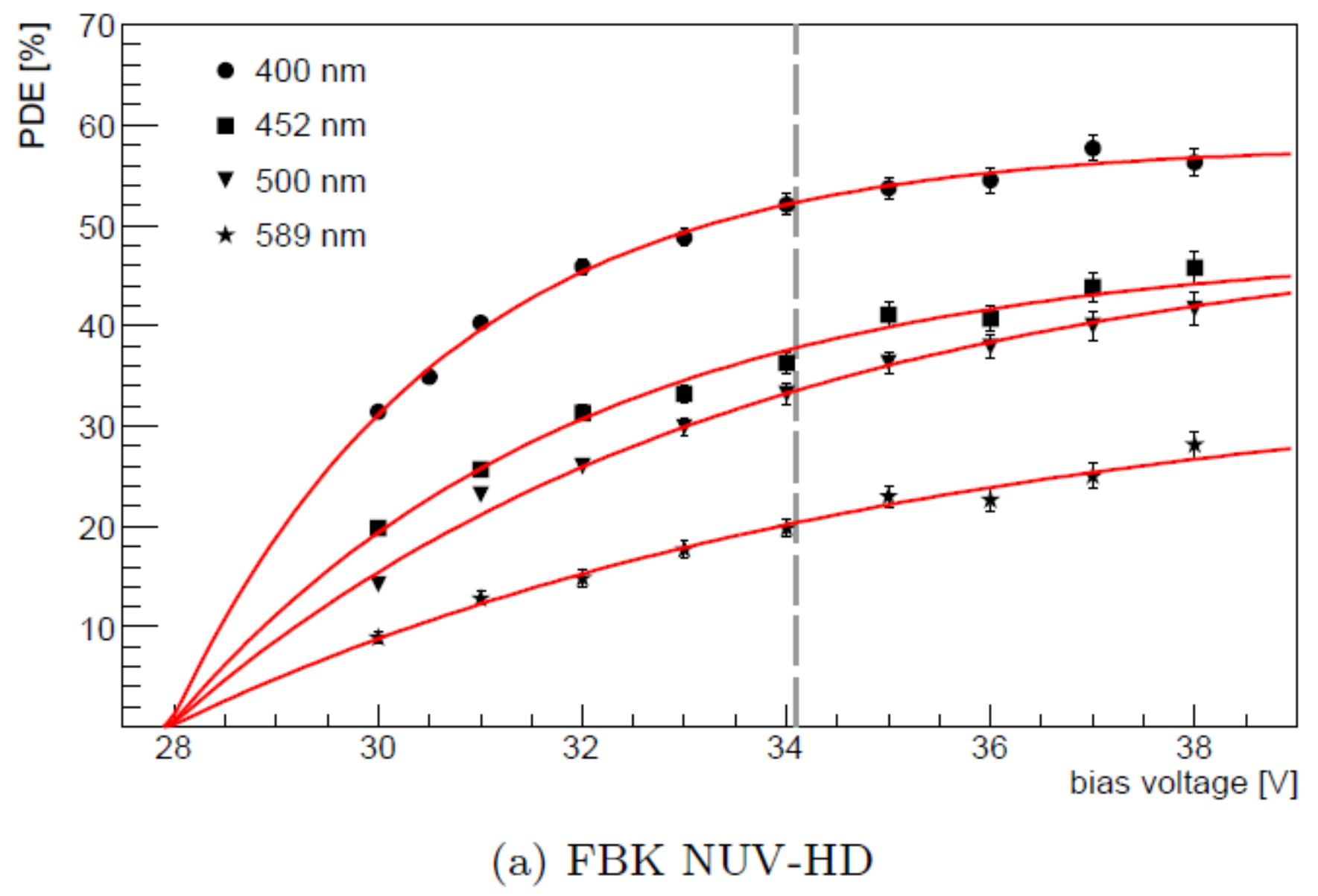}
    \caption{ }
    \label{fig:pdeFBK}
   \end{subfigure}%
    ~
   \begin{subfigure}[a]{0.5\textwidth}
    \includegraphics[width=\textwidth]{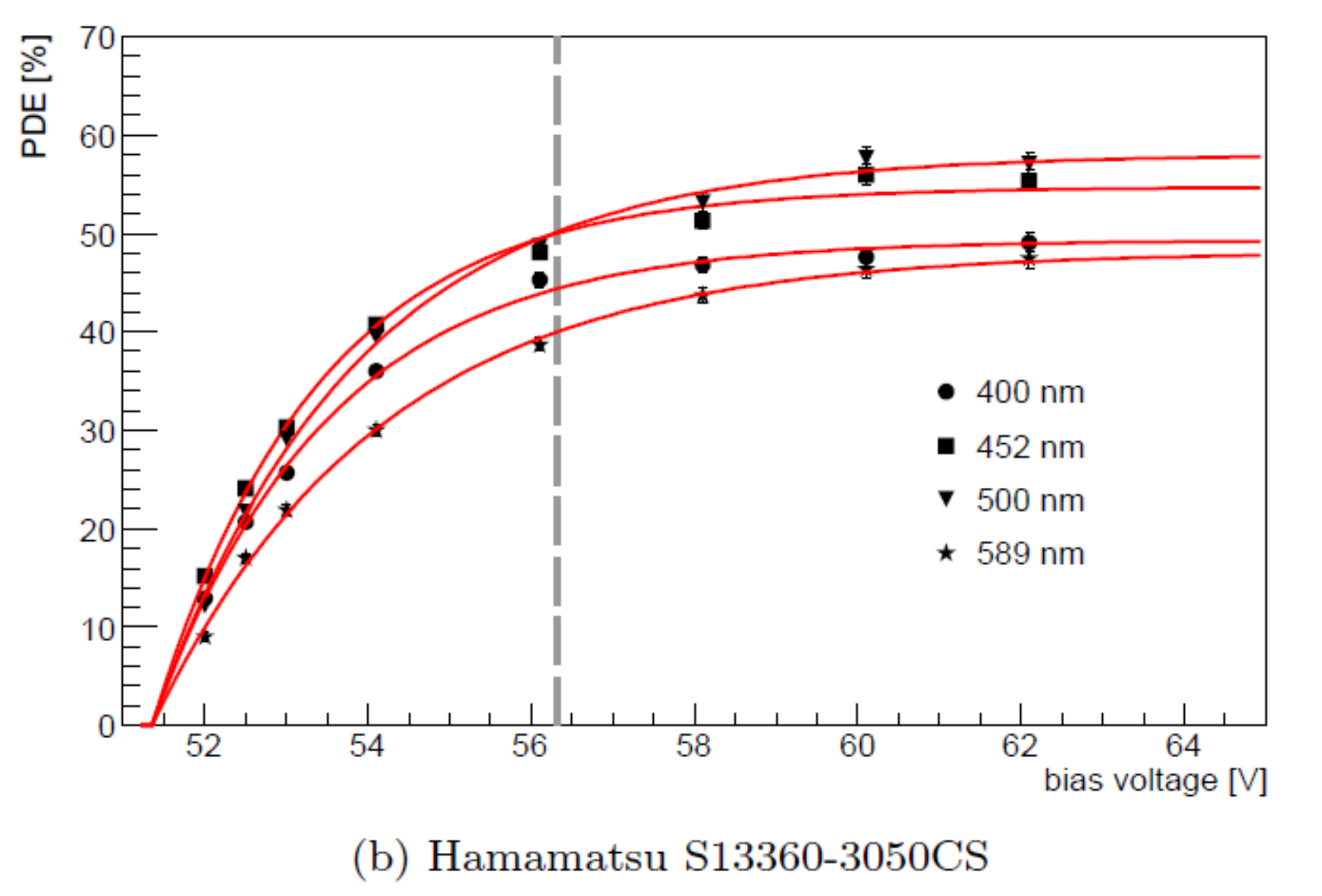}
    \caption{ }
    \label{fig:pdeHPK}
   \end{subfigure}%
   \caption{ Photon-detection efficiencies $PDE(V_{bias}, \lambda ) $ as a function of $V_{bias}$ and the wavelength $\lambda $ of the light, for two SiPMs from Ref.\,\cite{Otte:2016}.
   The vertical lines denote the bias voltage at which 90\,\% of the maximum $PDE$ is reached.
   The pixel pitch is $30 \,\upmu$m for the FBK SiPM, and $50 \,\upmu$m for the Hamamatsu SiPM.
   In addition, the design of the two SiPMs is different and optimised for different wavelengths, which has to be taken into account when judging the PDE dependencies.}
  \label{fig:pde}
 \end{figure}

 If peaks corresponding to different numbers of Geiger discharges, $N_G$, can be separated in the charge or amplitude spectra, the $V_{bias}$ dependence of the relative $PDE$ can be obtained from $f_{0,\,light}$ and from  $f_{0,\,dark}$  using Eq.\,\ref{equ:NpGphoto}.
 Fig\,\ref{fig:pde}, taken from Ref.\,\cite{Otte:2016}, shows the $V_{bias}$ dependence of $PDE$ for a number of wavelengths for two SiPMs.
 The relative values are obtained with the method described above.

\begin{figure}[!ht]
   \centering
    \includegraphics[width=0.5\textwidth]{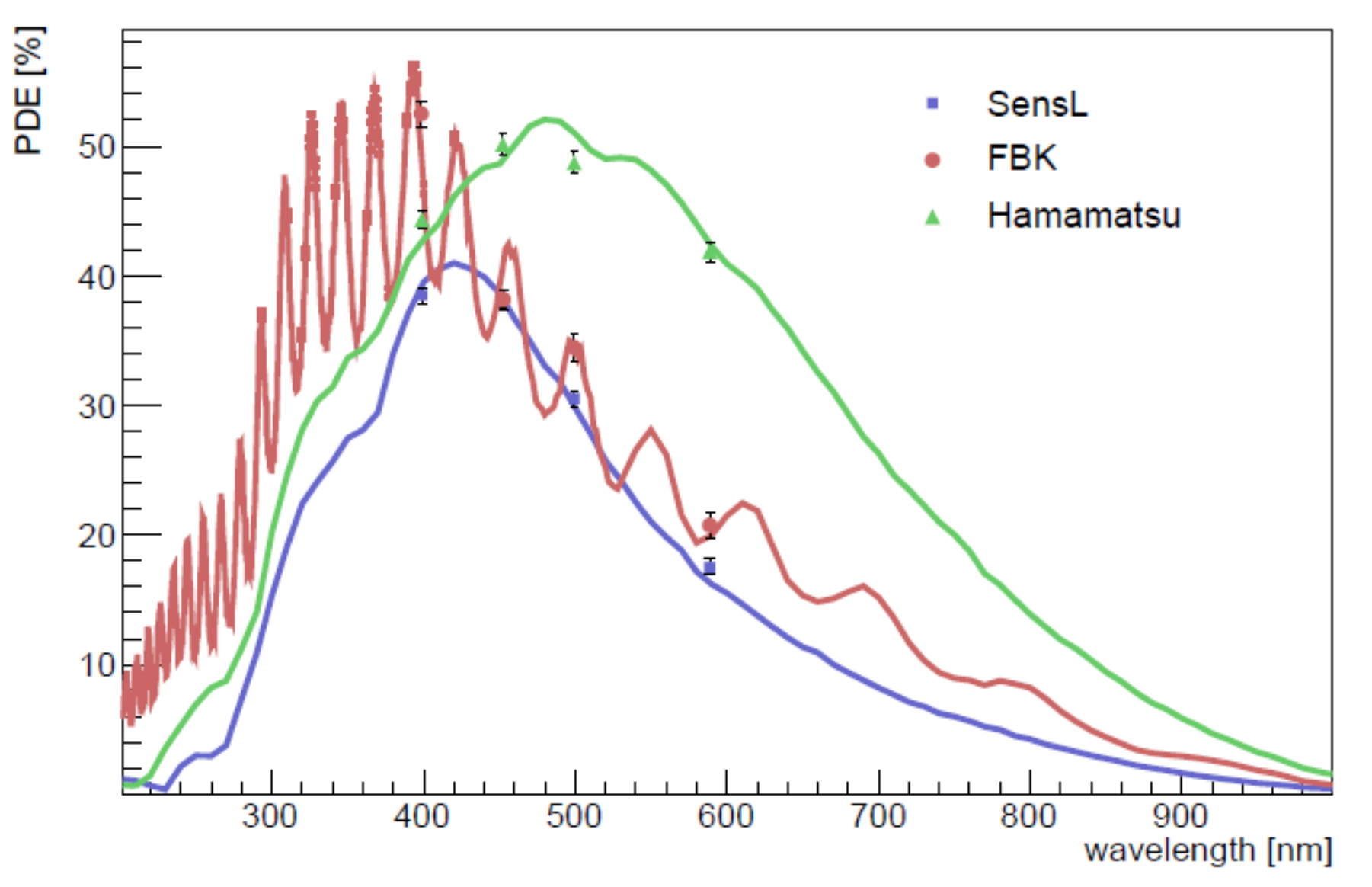}
     \caption{$PDE$ as a function of wavelength between 200\,nm and 1000\,nm for three SiPMs from Ref.\,\cite{Otte:2016}.
     $V_{bias}$ has been adjusted to give a Geiger-breakdown probability $P_T \approx 90$\,\%.}
   \label{fig:pde-lambda}
  \end{figure}

  The determination of the absolute $PDE$ uses calibrated photo-diodes, as already discussed in Sect.\,\ref{sect:PDE-setup}.
 There are several setups, both at producers and at research laboratories, which measure the absolute $PDE$ of SiPMs as function of wavelength and $V_{bias}$ with an accuracy of a few \%.
 Examples from Ref.\,\cite{Otte:2016} of the $V_{bias}$ dependence of the absolute $PDE$ are shown in Fig.\,\ref{fig:pde}, and of the wavelength dependence at a Geiger-breakdown probability $P_T \approx 90$\,\%, in Fig.\,\ref{fig:pde-lambda}.
 The accuracy achieved in these measurements is impressive, and so is the increase in $PDE$ achieved by the producers in recent years.

 Next, different methods of determining the SiPM gain are discussed.
 The most straight-forward method of measuring the gain, $G^\ast (V_{bias} )$, of the combined system SiPM--readout, is to record charge spectra, as shown in Fig.\,\ref{fig:PHspectra}, and determine the distance between the peaks corresponding to different number of Geiger discharges.
 Several methods are used:
 \begin{enumerate}
   \item Fit individual peaks by Gauss\,functions and determine the distance between them.
   \item Determine the distance using the Fourier transformed spectrum (Ref.\,\cite{Xu:2014}).
   \item Perform a complete fit of the spectrum with $G^\ast$ as one of the free parameter of the fit (Ref.\,\cite{Chmill:2017}).
 \end{enumerate}
 All three methods give very precise and compatible results, and it is matter of taste and convenience which one to use.
 However, they all require that 0, 1, 2, etc. are well separated.

 In addition to $G$, the rms width $\sigma _{N_{G} } $ of the peaks corresponding to different number of Geiger discharges, $N_G$, can be obtained from the spectra.
 As expected and observed, the data can be described by $\sigma _{N_{G}} ^2(V_{bias}) = \sigma _0 ^2 + N_G \cdot \sigma _1 ^2 (V_{bias})$, with $\sigma _0$ the contribution from the electronics noise and $\sigma _1(V_{bias})$ the contribution from the fluctuations of $G$ for single Geiger discharges.
 For the KETEK SiPM investigated in Ref.\,\cite{Chmill:2017} it is found that $\sigma _1$ has only a weak $V_{bias}$ dependence:
 Between $V_{bias} = 30$\,V and 35\,V it increases by $\approx 20$\,\% only.
 For this SiPM $V_{bd} = 26.64$\,V at $20\,^\circ$C.
 As $G  = (C_d + C_q) \cdot (V_{bias} - V_{off})$, there are  two contributions to $\sigma _1$.
 One from the pixel-to-pixel variations of $C_d + C_q$, called $\delta C$, and one
 from the fluctuations of $V_{off}$, called $\delta V_{off}$.
 The two terms can be distinguished using the $V_{bias}$ dependence of $\sigma _1$.
 Using the measured slope $\mathrm{d} G / \mathrm{d} V_{bias}$ and the definition of $G$, one finds:
 \begin{equation}\label{equ:sig1}
   \sigma _1^2 = \Big(\frac{\mathrm{d} G} {\mathrm{d} V_{bias}}\Big)^2 \cdot  \delta V_{off} ^2 + \Big(\frac{\mathrm{d} G} {\mathrm{d} (C_d + C_q)}\Big)^2 \cdot  \delta C ^2 = \big(C_d + C_q\big)^2 \cdot  \delta V_{off} ^2 + \big(V_{bias} - V_{off}\big)^2 \cdot \delta C ^2.
 \end{equation}
 As $\sigma _1 $ is approximately independent of $V_{bias}$, the second term is small and: $\delta V_{off} \approx \sigma_1 /(C_d + C_q)$, giving $\delta V_{off} \approx 175$\,mV for the data from Ref.\,\cite{Chmill:2017}.
 It is concluded that the increase of the width of the peaks in the SiPM charge (or amplitude) spectra is caused by differences in $V_{off}$ and not by differences in pixel capacitances.
 The reason for the rather large value of $\delta V_{off}$ could be differences of the 3-D electric field distribution within a pixel.
 To the author's knowledge, no realistic simulations of $V_{off}$ and $\delta V_{off}$  have been performed so far.

 If $N_G = 0$, 1, 2, ... peaks can not be separated, the gain, $G^\ast$, and the mean number of primary Geiger discharges, $\langle N_{pG} \rangle$, can nevertheless be determined from the mean, $\langle Q \rangle$, and the root-mean square, $\sigma _Q $, of the measured charge (or amplitude) distribution if the excess charge factor, $ECF$, and the excess noise factor, $ENF$, are known.
 The method is an extension of the well known method used to determine the gain, $G ^\ast$, and the mean number of photo-electrons, $\langle N_{pe} \rangle$, for vacuum photomultipliers, which are (incorrectly) assumed to be ideal photon-detectors with $ECF = ENF = 1$.
 For the ideal photomultiplier the distribution of the photo-electrons generated by the pulsed light is assumed to follow a Poisson distribution, for which both mean and variance are equal to $\langle N_{pe} \rangle$.
 With the gain $G^\ast $, the mean of the measured charge distribution becomes $\langle Q_{P} \rangle = q_0 \cdot G^\ast \cdot \langle N_{pe} \rangle$, and the square of the rms spread $\sigma _P ^2 = q_0^2 \cdot G^{\ast\,2} \cdot \langle N_{pe} \rangle$.
 The subscript $P$, which stands for \emph{Poisson}, refers to the ideal detector.
 From these two equations follows:
 \begin{equation}\label{equ:GNPoisson}
   G^\ast = \frac{\sigma _P ^2 } {q_0 \cdot \langle Q_{P} \rangle} \hspace{5mm} \mathrm{and} \hspace{5mm}
     \langle N_{pe} \rangle = \frac{\langle Q_{P} \rangle ^2} {\sigma _P^2 }.
 \end{equation}

 For a non-ideal SiPM, $ECF$ and $ENF$, defined in Eq.\,\ref{equ:ECF} and \ref{equ:ENF}, have to be taken into account, which results in
 \begin{equation}\label{equ:meanQ}
   \langle Q \rangle = ECF \cdot \langle Q_P \rangle = ECF \cdot q_0 \cdot G^\ast \cdot \langle N_{pG} \rangle , \hspace{5mm} \mathrm{and}
 \end{equation}
 \begin{equation}\label{equ:sigQ}
   \sigma _Q ^2 = ECF^2 \cdot ENF \cdot \langle Q_P \rangle = ECF^2 \cdot ENF \cdot q_0^2 \cdot G^{\ast \, 2} \cdot \langle N_{pG} \rangle,
 \end{equation}
 where for the SiPM the mean number of photo-electrons of the vacuum photomultiplier,  $\langle N_{pe} \rangle$, has been replaced by the mean number of primary Geiger discharges $\langle N_{pG} \rangle$.
 Solving the two equations for $G^\ast$ and $\langle N_{pG} \rangle$ gives
 \begin{equation}\label{equ:GN}
   G^\ast = \frac{\sigma _Q ^2 } {q_0 \cdot  \langle Q \rangle \cdot ECF \cdot ENC} \hspace{5mm} \mathrm{and} \hspace{5mm}
     \langle N_{pG} \rangle = \frac{\langle Q \rangle ^2 \cdot ENF} {\sigma _Q ^2 }.
 \end{equation}
 A method to determine $ECF$ and $ECN$ is presented in Sect.\,\ref{sect:nuisance}.

\begin{figure}[!ht]
   \centering
   \begin{subfigure}[a]{0.5\textwidth}
    \includegraphics[width=\textwidth]{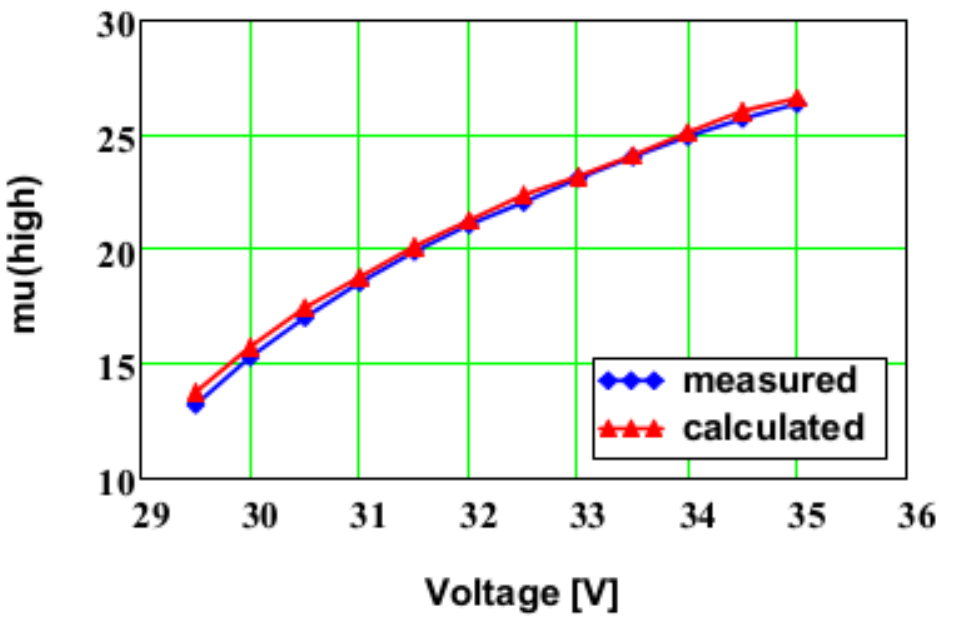}
    \caption{ }
    \label{fig:muhi}
   \end{subfigure}%
    ~
   \begin{subfigure}[a]{0.5\textwidth}
    \includegraphics[width=\textwidth]{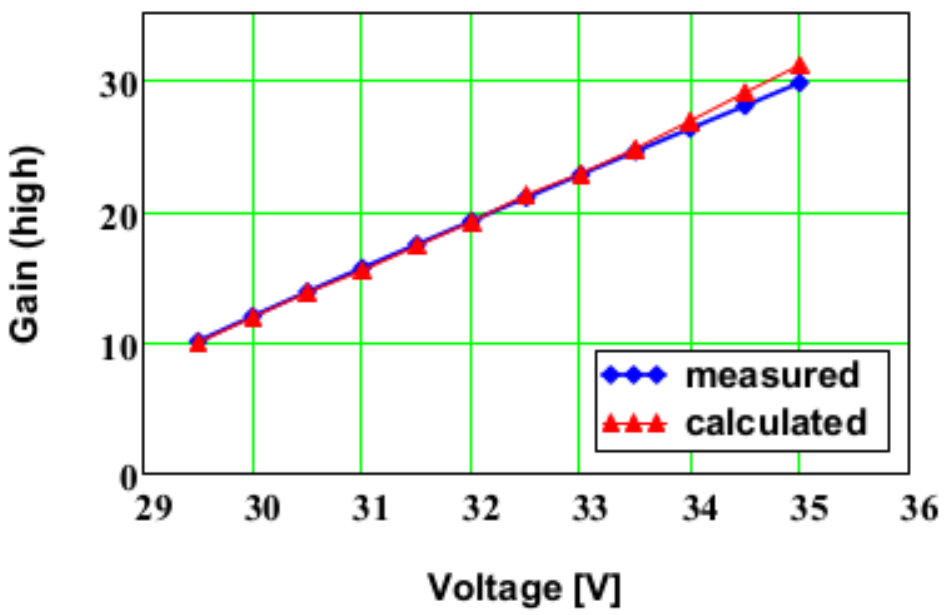}
    \caption{ }
    \label{fig:gainhi}
   \end{subfigure}%
   \caption{ Comparison of (a) the average number of primary Geiger discharges, $\mu \equiv \langle N_{pG} \rangle $,  and  (b) the gain $G ^\ast$ from a fit to the measured charge spectrum  (measured) with the method of the moments of the charge distribution (calculated) using Eq.\,\ref{equ:GN}.
   The figure is from Ref.\,\cite{Chmill:2017}, which also gives the corresponding values of $ECF$ and $ENF$. }
  \label{fig:MuGain}
 \end{figure}

 In Ref.\,\cite{Chmill:2017} the results for $\langle N_{pG} \rangle $ and $G ^\ast$ determined by a fit to the charge distribution are compared to the ones from $\langle Q \rangle $ and $\sigma _Q$ for the KETEK SiPM with 15\,$\upmu$m pitch illuminated with a pulsed LED.
 Fig.\,\ref{fig:MuGain} shows the results.
 For both $\langle N_{pG} \rangle $ and $G ^\ast$ the agreement is within a few percent, demonstrating the validity of the method.
 This method is straight-forward to use and suitable for the in-situ calibration and the monitoring of large numbers of SiPMs.
 It is used routinely in Ref.\,\cite{Gaug:2005}.
 It should be noted that in the case of significant noise, the width of the $N_G = 0$ peak has to be subtracted quadratically from $\sigma _Q$.
 In addition, it should be mentioned that the method as described does not work if the response of the system SiPM--readout is non-linear.
 But it is straight-forward to extend the method to non-linear regions, which, however, to the author's knowledge, has not yet been reported.

  \subsection{Nuisance parameters: Dark-count rate and correlated noise}
   \label{sect:nuisance}

 Compared to an ideal photon-detector, the SiPM performance is affected by a number of nuisance sources, in particular random dark counts and pulses correlated with  primary discharges.
 The different types of nuisance parameters have been discussed in Sect.\,\ref{sect:Parameters}.
 The best way to study them in detail, is to record current transients without or with low-intensity illumination.
 The analysis and results presented in Ref.\,\cite{Otte:2016} will be discussed next.
 The analysis procedure used follows closely the one reported in Ref.\,\cite{Piemonte:2012}.
 Similar analyses are reported in Refs.\,\cite{Du:2008, Acerbi:2015, Acerbi:2017}.

\begin{figure}[!ht]
   \centering
   \begin{subfigure}[a]{0.5\textwidth}
    \includegraphics[width=\textwidth]{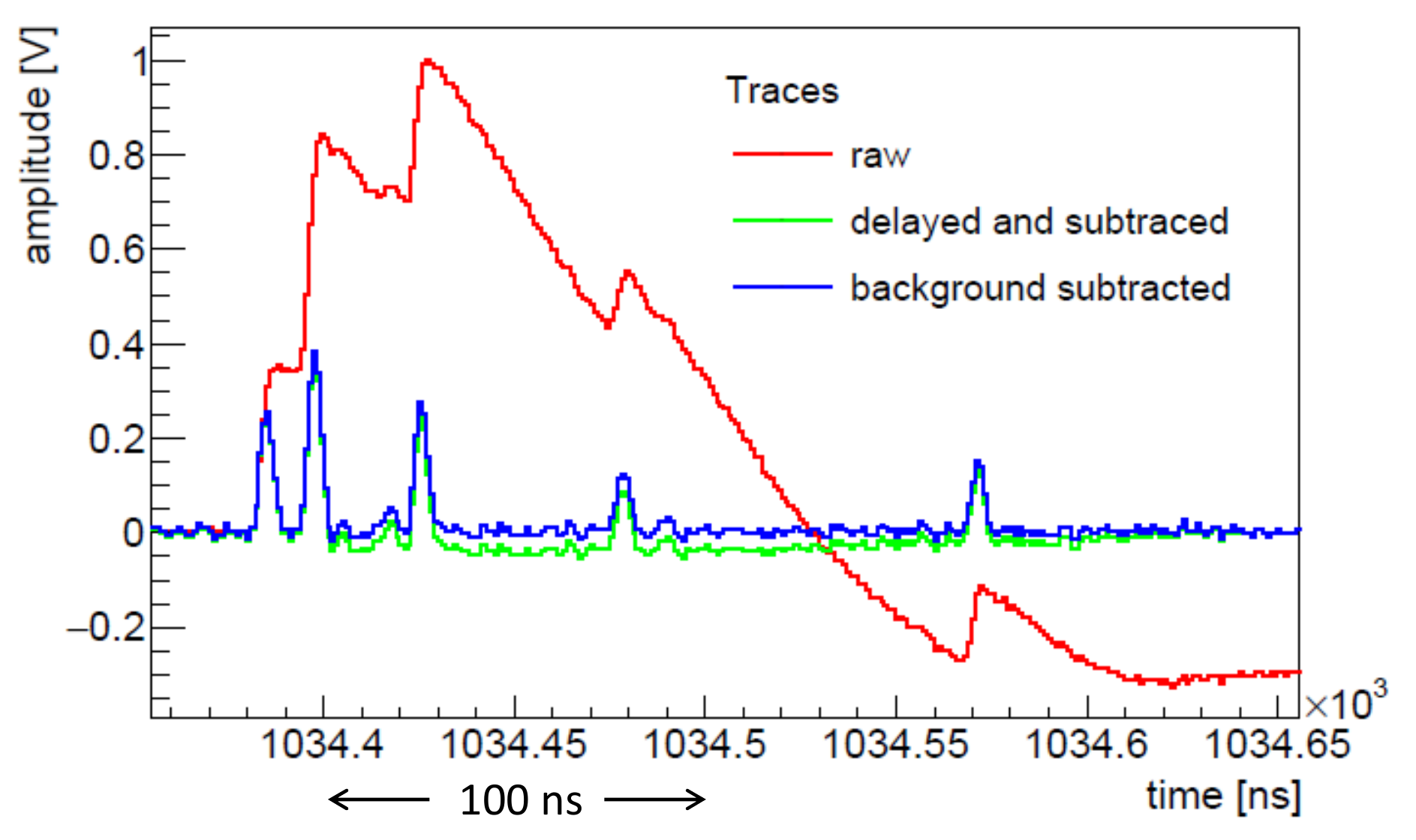}
    \caption{ }
    \label{fig:TransCN}
   \end{subfigure}%
    ~
   \begin{subfigure}[a]{0.5\textwidth}
    \includegraphics[width=\textwidth]{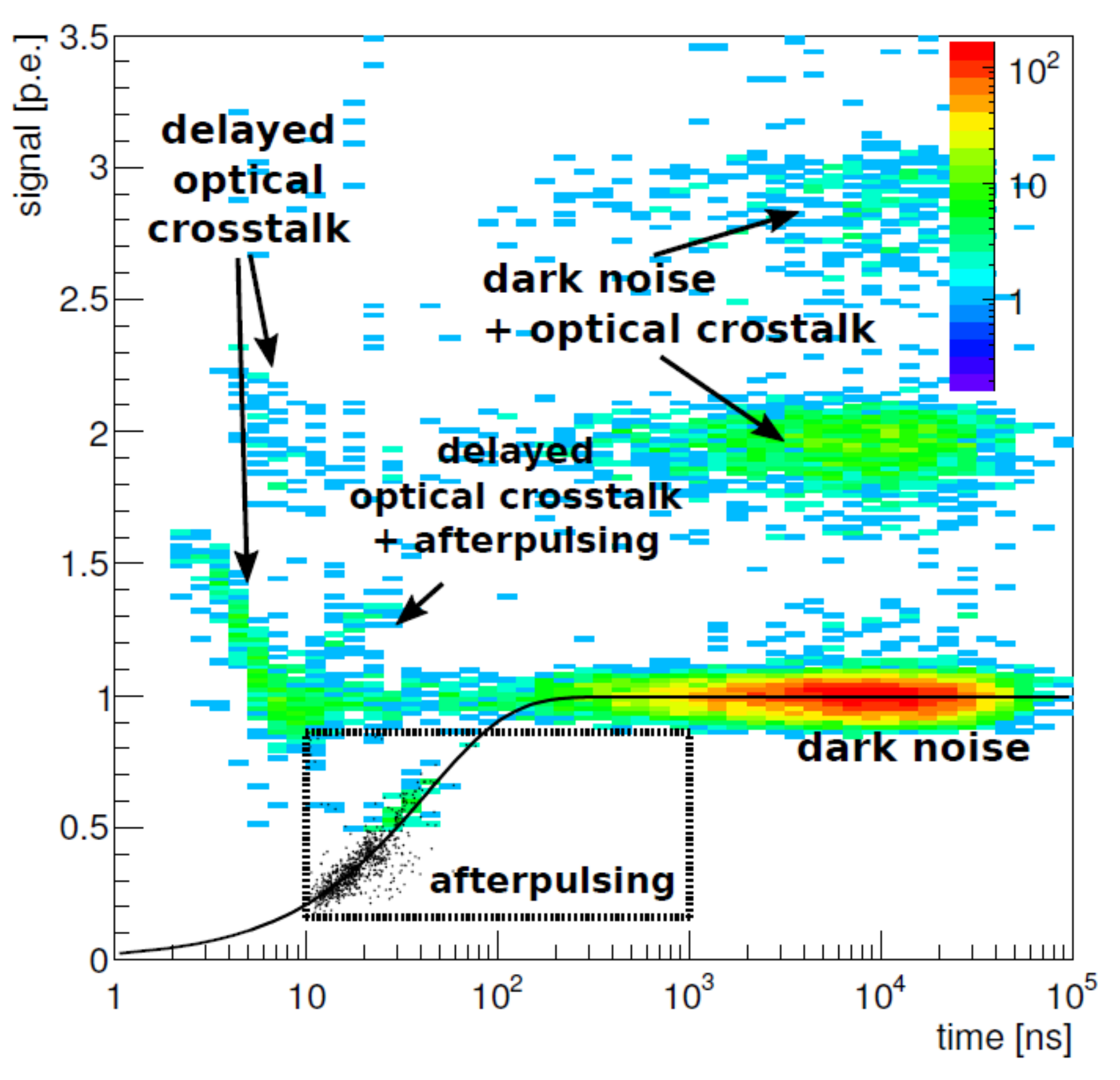}
    \caption{ }
    \label{fig:2dCN}
   \end{subfigure}%
   \caption{Analysis and results for the determination of the nuisance parameters using current transient measurements from Ref.\,\cite{Otte:2016}.
   (a) A SiPM transient recorded with 1\,GS/s and 8 bit resolution (raw), after subtraction of the transient shifted by 3\,ns (delayed and subtracted), and after the undershoot correction (background subtracted).
   (b) 2-D plot of the number of events in bins of the logarithm of the time difference between two consecutive SiPM pulses (x\,axis) and the amplitude of the second pulse (y axis), after selecting pulses corresponding to single Geiger discharges (1 pe) for the first pulse.
   As explained in the text and indicated on the figure, dark counts and the different types of correlated noise events can be classified and identified. }
  \label{fig:AnalysisCN}
 \end{figure}

 In Ref.\,\cite{Otte:2016} the transients are differentiated by subtracting a copy of the transient shifted by 3\,ns (see Fig.\,\ref{fig:TransCN}).
 In this way the pulse tails are removed and the pulses have a full width of about 9\,ns.
 Next the undershoot is removed by applying a background-subtraction algorithm, and pulses with an amplitude exceeding 0.5 pe (pe = the average amplitude of a single Geiger discharge) are marked.
 Finally pulses corresponding to single Geiger discharges are selected, and the time difference $\Delta t$ to the following pulse versus its amplitude plotted, as shown in Fig.\,\ref{fig:2dCN}.
 Note that the $\Delta t$ scale and the $\Delta t$ bin widths are logarithmic.

 Dark counts without correlated noise have an average amplitude corresponding to one Geiger discharge.
 They appear as horizontal line around pe = 1.
 Dark-count pulses with one or two prompt optical cross-talk pulses, each with an amplitude of 1 pe, appear as horizontal lines at pe = 2 and 3, respectively.
 After-pulses have an amplitude which increases with time due to the recharging of the pixel (see also Fig.\,\ref{fig:PulseRecover}).
 Pulses with delayed optical cross-talk are the  sum of the decaying pulse of the primary Geiger discharge and one or two optical cross-talk pulses, each with an amplitude of 1 pe.
 Their amplitudes decrease towards pe = 1 or 2 with increasing $\Delta t$.
 By selecting events in the different regions, all nuisance parameters can be determined in a quantitative way.
 In the following a few examples are given.


 The total dark-count rate , $DCR$, can be approximately determined by counting all pulses and dividing the number by the total duration of all analysed transients.
 $DCR$ is given by the primary dark count rate plus the effects of after-pulses and delayed optical cross-talk.
 A more precise procedure is to analyse the $\Delta t$ distribution, $\mathrm{d} N/\mathrm{d} \Delta t$,  which for random dark pulses at the rate $DCR$ is expected to have the form  $\mathrm{d} N/\mathrm{d} (\Delta t ) \propto e^{- (\Delta t \cdot DCR) }$.
 This dependence follows from the properties of the Poisson distribution:
 The mean number of dark counts ($DC$) in the time interval $\Delta t$ is $\langle N(\Delta t) \rangle = DCR \cdot \Delta t$, and the probability of zero $DC$s in $\Delta t$ is $P(0,\Delta t) = e^{-(\Delta t \cdot \ DCR)}$.
 The absolute value of the derivative $|\mathrm{d} P(0,\Delta t) / \mathrm{d} (\Delta t)| = DCR \cdot e^{-(\Delta t \cdot DCR)}$ is proportional to the probability of the change from 0 to $\geq 1$ $DC$s, thus the occurrence of a $DC$ at $\Delta t$.
 The $\Delta t$ distribution for random dark counts,  when plotted in bins of $\ln (\Delta t)$, is $\propto \Delta t \cdot e^{-(\Delta t \cdot DCR)}$ with the maximum at $\Delta t _{max} =1/DCR$.
 This can be clearly seen in Fig.\,\ref{fig:2dCN}.

\begin{figure}[!ht]
   \centering
   \begin{subfigure}[a]{0.4\textwidth}
    \includegraphics[width=\textwidth]{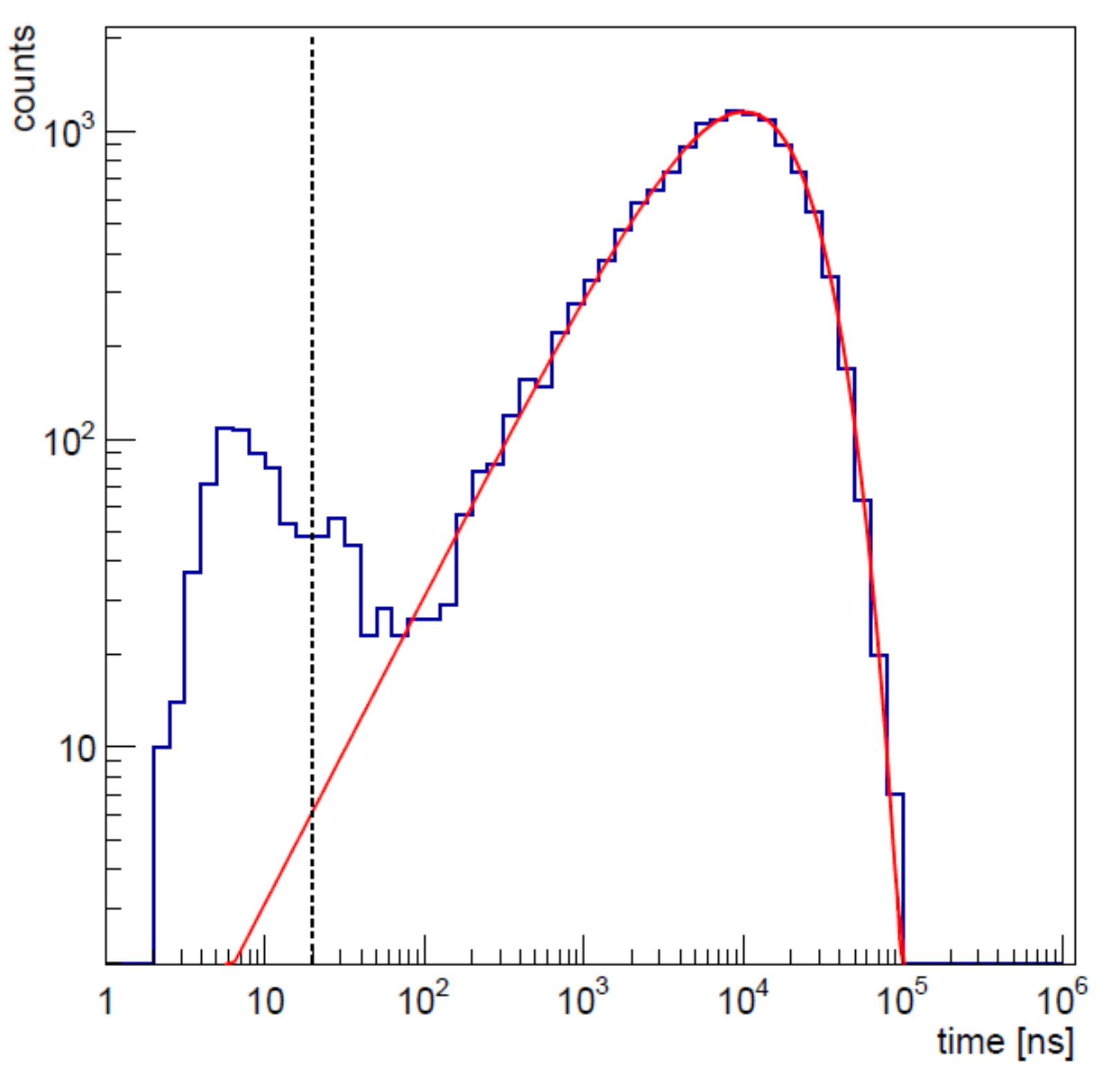}
    \caption{ }
    \label{fig:lnDeltat}
   \end{subfigure}%
    ~
   \begin{subfigure}[a]{0.5\textwidth}
    \includegraphics[width=\textwidth]{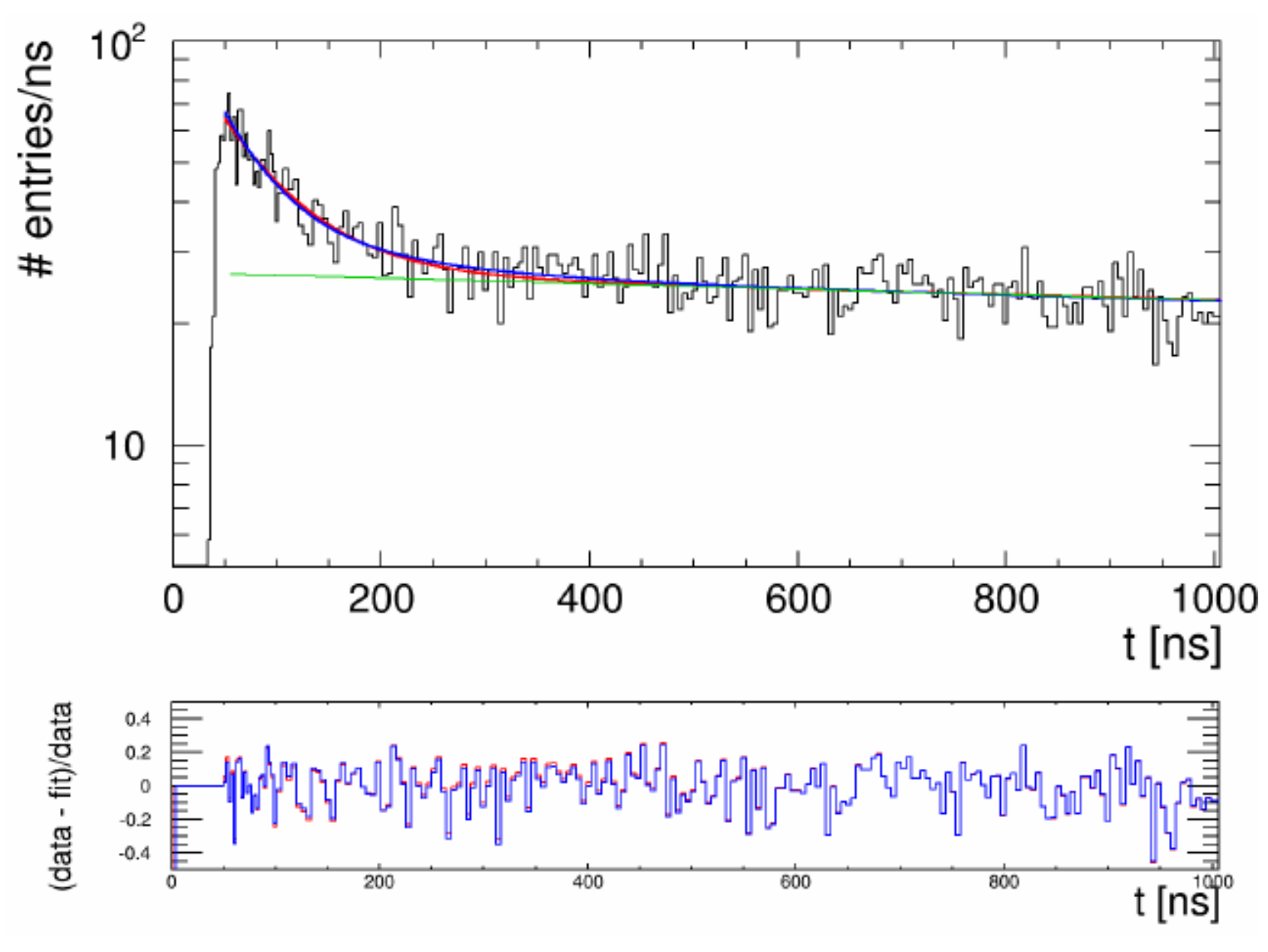}
    \caption{ }
    \label{fig:LinDeltat}
   \end{subfigure}%
   \caption{ (a) Time difference $\Delta t$  in logarithmic $\Delta t$ scale from Ref.\,\cite{Otte:2016}.
    The solid line is a fit for $\Delta t > 200$\,ns by $\Delta t \cdot e^{-(\Delta t \cdot DCR)}$, the expected shape for random dark-count pulses in $\log (\Delta t)$\,bins.
    The maximum of the peak is at $1/DCR$.
   (b) Top: Time-difference distribution in linear $\Delta t$ scale for a Hamamatsu SiPM from Ref.\,\cite{Garutti:2014}.
    The solid lines are fits of the sum of dark counts and after-pulses for $\Delta t > 50$\,ns, where the algorithm for the identification of after-pulses is fully efficient.
    The essentially straight line is the contribution of dark counts.
    Bottom: Difference (data -- fit) /data. }
  \label{fig:Deltat}
 \end{figure}

 Fig.\,\ref{fig:lnDeltat} shows an example of a $\mathrm{d} N/\mathrm{d} (\ln (\Delta t ))$ distribution with a fit of the expected $DC$ dependence for  $\Delta t > 200$\,ns.
 At lower $\Delta t $ values the effects of correlated pulses are clearly visible.
 In Refs.\,\cite{Eckert:2010, Garutti:2014} the $\Delta t $ distribution in linear $\Delta t$ scale is fitted to the sum of $DC$s and after-pulses with exponential time distributions.
 If only $DC$s and after-pulses are considered, the expected $\Delta t $ distribution can be derived by replacing $\langle N(\Delta t) \rangle = DCR \cdot \Delta t$ valid in the absence of after-pulses and delayed cross-talk, by  $\langle N(\Delta t) \rangle = DCR \cdot \Delta t + \varepsilon _{AP} \cdot (1 - e^{- \Delta t/\tau _{AP}})$ for one state, with the probability of after-pulses, $\varepsilon _{AP}$, and the time constant $\tau _{AP}$.
 Differentiation of $P(0,\Delta t) = e^{-\langle N(\Delta t) \rangle} $ with respect to $\Delta t$ gives the $\Delta t $ dependence.
 Fig.\,\ref{fig:LinDeltat} shows an example of such an analysis from Ref.\,\cite{Garutti:2014}, which shows that the data are well described by the model and that $DCR, \, \varepsilon _{AP} $ and $\tau _{AP}$ are determined with good accuracy.
 Delayed cross-talk can be implemented in a similar way, if a parametrisation for its time dependence is available.

 The $DCR$ can also be obtained from charge or amplitude spectra measured without illumination, as the one shown in Fig.\,\ref{fig:Dark33}, using the relation
 \begin{equation}\label{equ:fhalf}
   DCR = - \frac {\ln (f_{0.5, \, dark})} {t_{gate}},
 \end{equation}
 with $f_{0.5, \, dark}$, the fraction of events with a charge exceeding half the signal of a single Geiger discharge (1/2 pe), and $t_{gate}$ the gate width used for the current integration.
 As  gate and dark pulses are uncorrelated in time, the charge spectrum contains pulses with different overlaps with the gate, resulting in signals between the $N_G =0 $ and the $N_G = 1$ peak.
 In Ref.\,\cite{Chmill:2017} it is shown, that only for $f_{0.5, \, dark}$ Eq.\,\ref{equ:fhalf} is exact.
 If a lower or a higher threshold than 0.5 pe is chosen, the value for $t_{gate} $ in Eq.\,\ref{equ:fhalf} has to be decreased or increased with respect to the actual $t_{gate}$.
 Thus fitting  the $N_G = 0$ peak and using the fraction of events in the peak instead of $f_{0.5, \, dark}$, which is frequently done, is only approximate and should be avoided.
 In cases where the tail of the zero-Geiger discharge peak results in a significant fraction of events above  0.5 pe, these events have to be subtracted when determining $f_{0.5, \, dark}$.
 It should be noted that after-pulses and delayed cross-talk result in a systematic bias of this $DCR$ determination.
 It is estimated that the effect is small, but a systematic study is not known to the author.

 To summarise: The $f_{0.5, \, dark}$ method is straight-forward and recommended for determining the $DCR$, but has a bias, which however in most practical cases will be small.
 For a more precise determination, the $\Delta t$\,method described above should be used.

\begin{figure}[!ht]
   \centering
   \begin{subfigure}[a]{0.5\textwidth}
    \includegraphics[width=\textwidth]{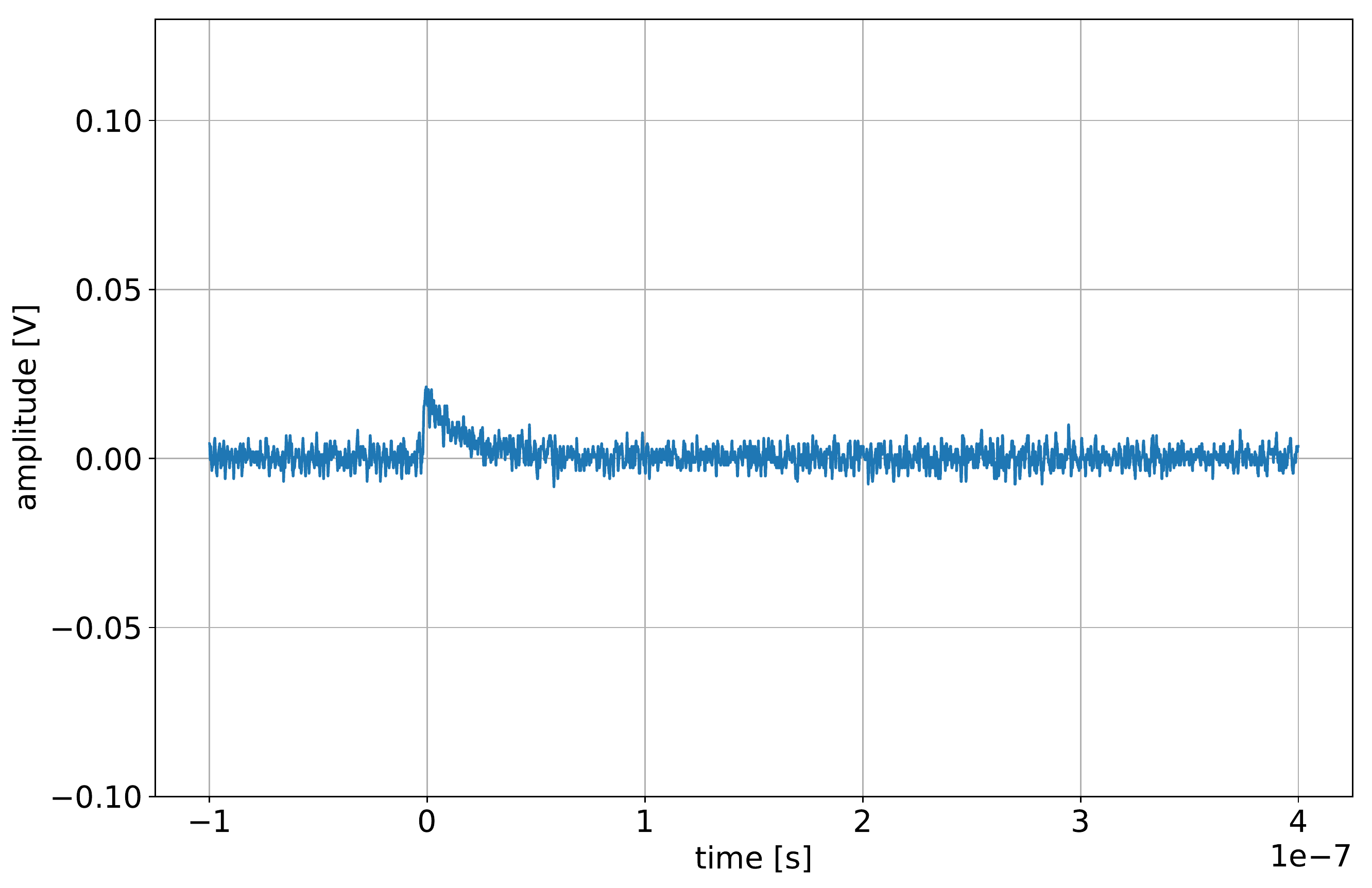}
    \caption{ }
    \label{fig:non-irr}
   \end{subfigure}%
    ~
   \begin{subfigure}[a]{0.5\textwidth}
    \includegraphics[width=\textwidth]{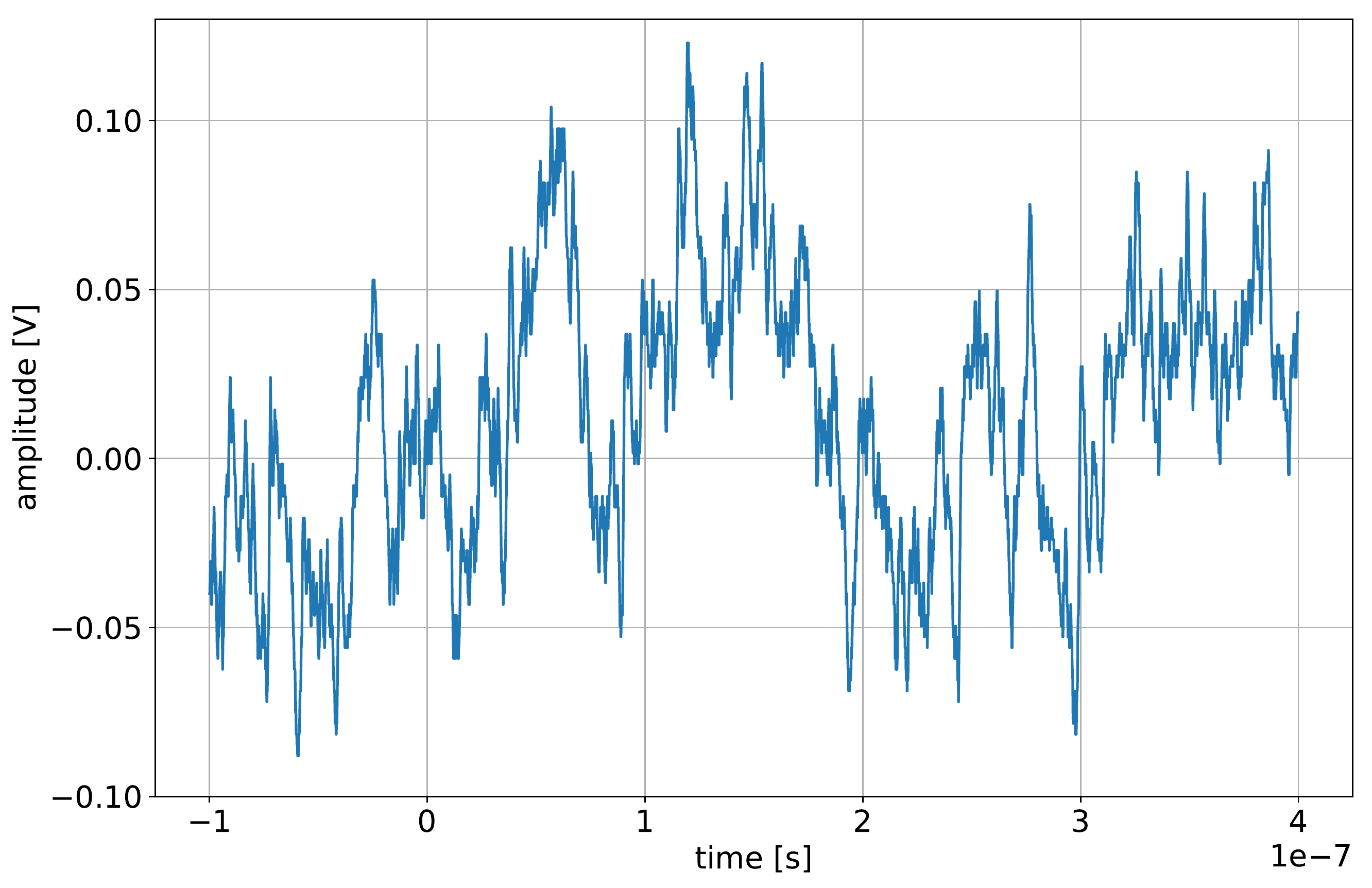}
    \caption{ }
    \label{fig:irr}
   \end{subfigure}%
   \caption{Transients recorded with a KETEK PM15 SiPM at 29.4\,V and $- 30 \,^\circ $C for
    (a) before irradiation, and
    (b) after irradiation by neutrons to a fluence of $5 \times 10^{13}$\,cm$^{-2}$ causing a $DCR \approx 3$\,GHz.}
  \label{fig:Transirr}
 \end{figure}

 The methods described so far can only be applied if peaks corresponding to different number of Geiger discharges can be distinguished.
 Determining $DCR$ when this is not the case, is significantly more complex and a number of assumptions have to be made in the analysis.
 Fig.\,\ref{fig:Transirr}, which shows current transients with low light for a SiPM before irradiation (a), and after irradiation (b), shows the problem.
 Whereas in (a) it is straight-forward to analyse the single Geiger discharge pulse, this is impossible for (b), which shows wild fluctuations with  amplitudes, which are larger by one order of magnitude.
 Transients, as shown in Fig.\,\ref{fig:irr}, can be reproduced by a simple Monte Carlo simulation by adding $DCR \cdot \Delta t_{trans}$ pulses as shown in Fig.\,\ref{fig:non-irr} randomly distributed in the time interval of the transient, $\Delta t_{trans}$.
 For estimating $DCR$ in such a situation, two methods will be described. One uses the measured dark current, $I_{dark}$, the other $\sigma _{dark}$, the \emph{rms} of the charge distribution measured without illumination.
 These methods are discussed in Ref.\,\cite{Klanner:2018} and used in Ref.\,\cite{Garutti:2018} to characterise radiation-damaged SiPMs.
 At  high $DCR$ values, $I_{dark}$ can exceed several mA and an AC-coupled readout is typically used, so that the average current is zero and contains no information.
 In addition, the high current results in a significant power dissipation causing an uncertainty in the knowledge of the SiPM temperature.

 $I_{dark}$ is related to to the primary dark count rate  $DCR_p$ by
 \begin{equation}\label{equ:Idark}
   I_{dark} = q_0 \cdot G \cdot ECF \cdot DCR_p = q_0 \cdot (C_d + C_q) \cdot (V_{bias} - V_{off})\cdot ECF \cdot DCR_p,
 \end{equation}
 from which follows
 \begin{equation}\label{equ:DCR-Idark}
   DCR_p (V_{bias})= \frac {I_{dark}} {ECF (V_{bias}) \cdot q_0 \cdot (C_d + C_q) \cdot (V_{bias}-V_{off}) }.
 \end{equation}
 $C_d$ and $C_q$ can be determined from $Y-f$ (Sect.\,\ref{sect:ElParameters}), and $V_{bd} $ from $I-V$\,measurements (Sect.\,\ref{sect:Vbd}).
 If the approximation $V_{off} \approx V_{bd}$ is made, which is valid for $V_{OV} \gg V_{bd} - V_{off}$, $ECF \cdot DCR_p$ can be obtained using Eq.\,\ref{equ:DCR-Idark}.
 As $ECF$ is typically $\lesssim 1.2$, $ECF \cdot DCR_p$ is already a quite good approximation to $DCR$.
 An alternative, which is used in Refs.\,\cite{Garutti:2018, Klanner:2018} for the study of radiation damage, is to assume that $ECF$ and $V_{bd} - V_{off}$ do not change with irradiation, and determine $V_{bd}$ from the $I-V$ measurements.
 The validity and accuracy of these assumptions has not been checked so far.

\begin{figure}[!ht]
   \centering
   \begin{subfigure}[a]{0.6\textwidth}
    \includegraphics[width=\textwidth]{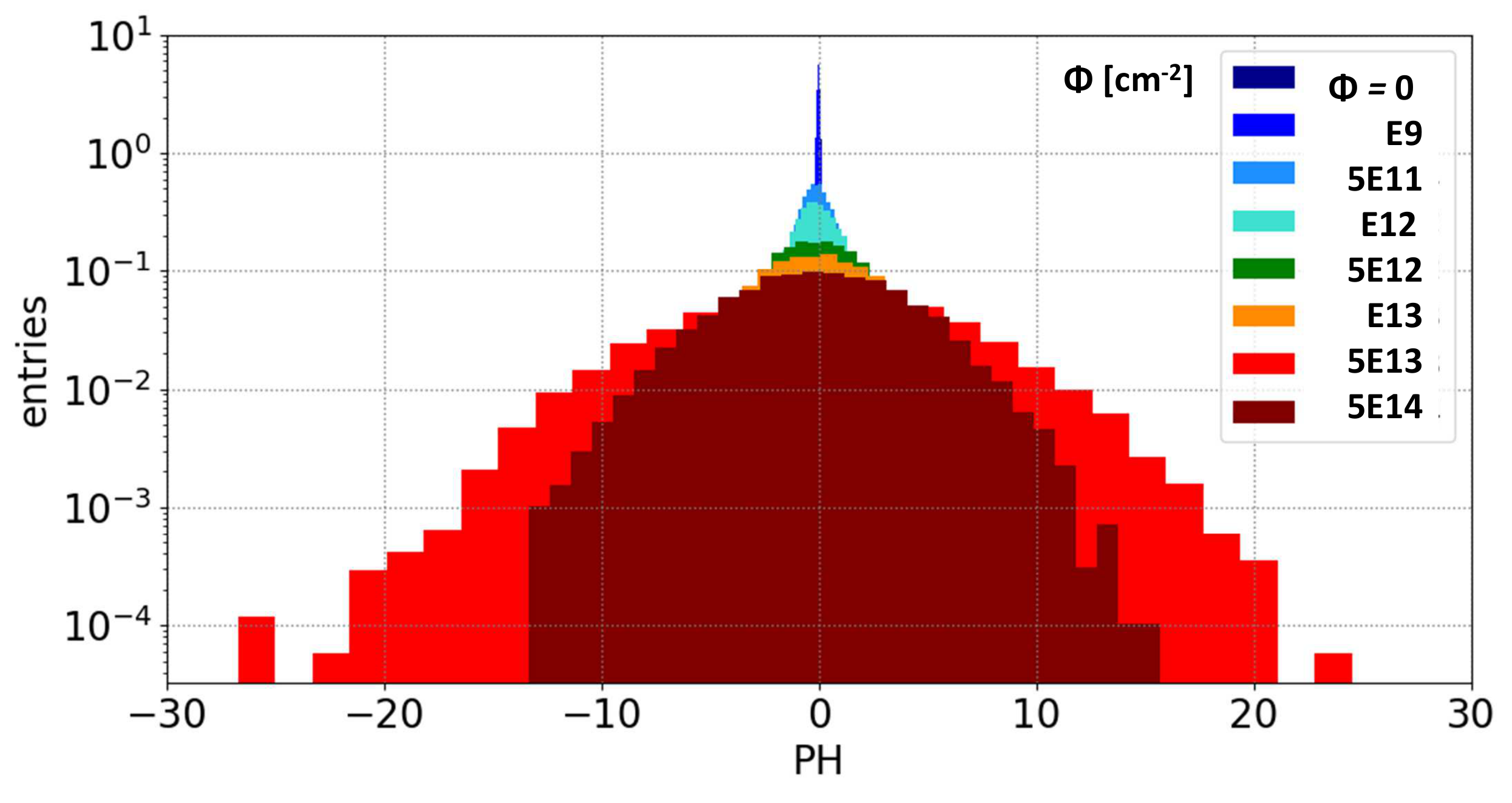}
    \caption{ }
    \label{fig:Vardamage}
   \end{subfigure}%
    ~
   \begin{subfigure}[a]{0.4\textwidth}
    \includegraphics[width=\textwidth]{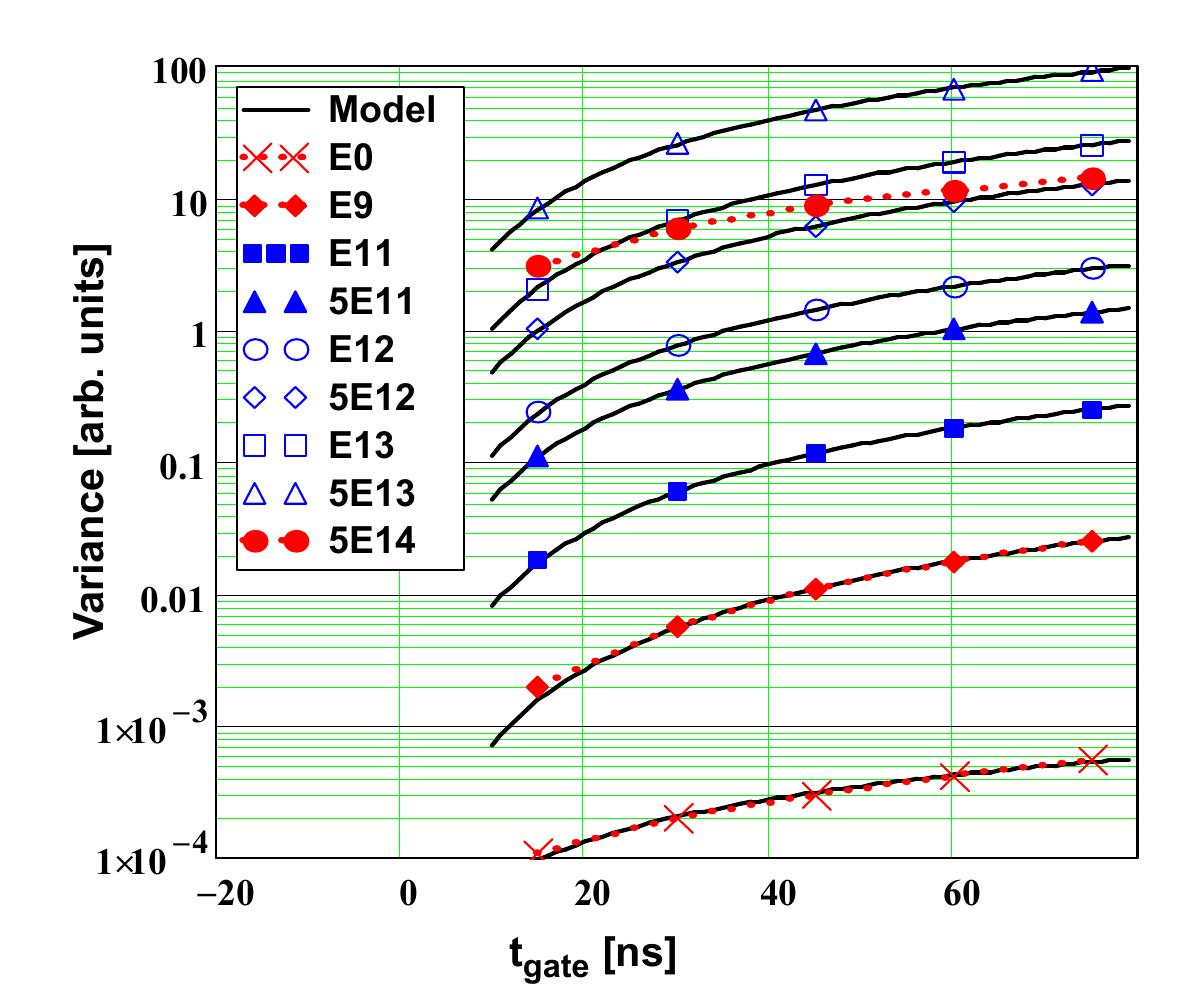}
    \caption{ }
    \label{fig:Vargate}
   \end{subfigure}%
   \caption{(Colour online) (a) Charge spectra measured at $- 30\,^\circ$C in the dark for KETEK MP15 SiPMs after irradiation to different neutron fluences, $\Phi $. Measurements and figure are from S.\,Cerioli, Hamburg University.
   (b) Variances, $\sigma _{dark} ^2$, of the charge spectra as a function of the gate width, $t_{gate}$. The symbols are the data, and the lines the fits by Eq.\,\ref{equ:SigDark} with $\tau _r$ and the term in the left parenthesis of Eq.\,\ref{equ:SigDark} as free parameters. Solid lines represent good, and dotted lines poor description of the data by the fit.}
  \label{fig:VarDark}
 \end{figure}

 Next, the determination of $DCR_p$ from the measurement of the rms-spread, $\sigma _{dark}$, of the charge (or amplitude) distribution measured without illumination, will be discussed.
 Fig.\,\ref{fig:Vardamage} shows charge spectra measured with a gate width $t_{gate} = 75$\,ns without illumination for the KETEK PM15 SiPM irradiated by neutrons to different fluences up to $5 \times 10^{14}$\,cm$^{-2}$.
 As discussed in detail in Ref.\,\cite{Garutti:2018}, the dominant effect of radiation damage is the increase of $DCR$ by many orders of magnitude.
 One sees that $\sigma _{dark} $ first increases with fluence, and above a fluence of $5 \times 10^{13}$\,cm$^{-2}$  decreases. For high $DCR$ values many pixels are already busy with Geiger discharges, and this high occupancy is responsible for the decrease of $\sigma _{dark} $.
 The formula used to extract $DCR_p$ from $\sigma _{dark}$ is:

 \begin{equation}\label{equ:SigDark}
   \sigma_{dark} ^2 = \big((q_0 \cdot G)^2 \cdot ENF \cdot ECF^2 \cdot DCR_p\big) \cdot \big(t_{gate} - \tau _r \cdot (1 - e^{- t{gate} / \tau _r}) \big).
 \end{equation}

 It is derived in the appendix, under the assumption that the SiPM current pulse for a Geiger discharge at time $t_0$ is described by $I(t) \propto e^{-(t-t_0)/\tau _r}$ for $t \geq t_0$.
 An extension to other pulse shapes is straight-forward inserting the functional form of $f(t)$ in Eq.\,\ref{equ:Q(tgate)}.
 To verify the predicted $t_{gate}$\,dependence, Fig.\,\ref{fig:Vargate} compares the measured $\sigma _{dark}^2 ( t_{gate})$ (symbols) to fits by Eq.\,\ref{equ:SigDark} (solid lines) with $\tau _r$ and the term in the parenthesis on the left side, as free parameters.
 The dependence of $ECF$ and $ENF$ on $t_{gate}$ has been neglected in the fits.
 Up to a fluence of $10^{13}$\,cm$^{-2}$ the data are well described, and allow to determine  $\tau _r$ with an accuracy of about 10\,\%.
 For fluences exceeding $10^{13}$\,cm$^{-2}$, the quality of the fit worsens because of the high pixel occupancy at high $DCR$.
 For low $DCR$, $\sigma _{dark}$ is dominated by electronics noise, which has to be subtracted quadratically from $\sigma _{dark}$.
 If the electronics noise dominates, the method becomes unreliable.
 A formula, which takes into account the reduction of $\sigma _{dark}$ due to pixels occupied by dark counts, still has to be derived.
 In order to determine $DCR_p$ from Eq.\,\ref{equ:SigDark}, assumptions for $G$, $ECF$ and $ECN$ have to be made.
 For the determination of $G^\ast$, when peaks corresponding to different Geiger discharges cannot be distinguished, Eq.\,\ref{equ:GN} can be used.
 For the determination of $ECF$ and $ENF$ no method is known to the author, if peaks corresponding to different number of Geiger discharges can not be distinguished.
 However, in most practical cases the problem of merging peaks is either the result of ambient light or of radiation damage.
 In these cases, $ECF$ and $ENF$ can be measured initially, and the assumption made that the values do not change for the conditions in which the SiPM is finally used.
 The validity of these assumptions for radiation damage still has to be demonstrated.

 In Ref.\,\cite{Garutti:2018}, $DCR_p$ is determined as a function of the neutron fluence using the measured $I_{dark}$ and Eq.\,\ref{equ:SigDark}, assuming for $ECF$ and $ENF$ the values of the non-irradiated SiPM.
 The results are compared to the $DCR_p$ results using  Eq.\,\ref{equ:DCR-Idark}, assuming $ECF$ and $V_{bd} - V_{off}$ from the non-irradiated SiPM and the values of $V_{bd}$ from the $I-V$ analyses for the different neutron fluences.
 In the range of the validity of the $\sigma _{dark}$ method an agreement to better than 30\,\% is observed, which is considered satisfactory for $DCR$ values exceeding several GHz.
 A detailed  comparison of the methods and their sensitivity to the assumptions used is still missing.

 To summarise this subsection on the $DCR$ and $DCR_p$ determination:
 If the $DCR$ is sufficiently low and the peaks for different number of Geiger discharges can be distinguished, the different methods, counting the dark-counts in the transients, analysing the time-difference $\Delta t$, and the $f_{0.5, \, dark}$ method, are straight-forward and  give reliable results.
 For $DCR$ values approaching or exceeding $1/\tau _r$, the pixel-recharging time constant (typically between 15 and 200\,ns), the situation becomes significantly more complicated.
 Based on ongoing studies, the preliminary conclusion is that for high $DCR$s using $I_{dark}$ and Eq.\,\ref{equ:DCR-Idark} is the most reliable method to determine $DCR_p$.
 This method however, requires the knowledge of  $G$ and $ECF$.
 For $G$ it is recommended to use $G = (C_q + C_d) \cdot (V_{bias} - V_{off})$, with $C_d + C_q$ from $Y-f$ (admittance-frequency) measurements for $V_{bias}$ 0.5 to 1\,V below $V_{bd}$.
 For $V_{off}$, $V_{bd}$ can be used, which however is a poor assumption for small $V_{OV}$ values if $V_{bd}$ differs significantly from $V_{off}$.
 If the difference $V_{bd}-V_{off}$ can be determined for a low DCR, then the assumption of a constant difference can be made, and $V_{bd}$ obtained from $I_{dark}- V_{bias}$ measurements.
 To better understand the effects of high $DCR$, the following study is recommended: For a SiPM, with properties precisely determined using the methods of individual Geiger discharges, different $DCR$ values can be simulated by DC-light of variable intensity illuminating uniformly the SiPM. In this way the different methods can be compared and the most suitable  determined.


 \subsubsection*{Cross-talk, after-pulses, ECF, ENF and optimal resolution}

 Fig.\,\ref{fig:2dCN} shows the 2-D distribution of the time between pulses, $\Delta t$, versus  pulse amplitude, which allows identifying the different physical effects responsible for the nuisance parameters.
 Analysing separately the different event classes allows to study their rate and properties.
 These studies as a function of $V_{bias}$ and temperature are essential for understanding the different effects and proposing technological modifications of the fabrication process to improve the SiPM performance.
 They also provide input  for the development of realistic SiPM models.
 An example is given in Fig.\,\ref{fig:Deltat} with the discussion on the extraction of after-pulses and delayed cross-talk.

 However, as long as saturation effects can be ignored, for most users the knowledge of $DCR$, $G^\ast$, $ECF$ and $ENF$ as a function of $V_{bias}$ of the SiPMs will be sufficient to characterise the SiPM and determine the  optimal operating conditions.
 These parameters depend not only on the SiPM properties, but also on the readout used.
 As an example: A shorter integration of the SiPM current, results in a reduction of $G^\ast$, $ECF$ and $ENF$.

 The most direct way to determine these parameters uses two charge spectra:  One recorded in the dark ($Q_{dark}$), and one with low-intensity light (Q), so that the fraction of events in the peak for zero Geiger discharges can be measured precisely.
 In the following it is assumed that the mean of the zero-Geiger discharge peak corresponds to zero charge.
 $DCR$ is obtained using Eq.\,\ref{equ:fhalf} from the $Q_{dark}$ spectrum and $G^\ast$ from the distance between the peaks of the $Q$ spectrum, as discussed in Sect.\,\ref{sect:Gain}.
 The mean number of primary Geiger discharges due to photons from the light source, $\langle N_{pG, \, photo}\rangle$, is obtained from Eq.\,\ref{equ:NpGphoto}.
 $ECF$ is obtained from the ratio of the mean of the measured charge distribution $\langle Q \rangle$ to the expectation for a Poisson distribution $q_0 \cdot G^\ast \cdot \langle N_{pG, \, photo}\rangle$, and $ENF$ from the ratio of the square of the rms-spread,  $\sigma _Q ^2$, to the Poisson expectation $(q_0 \cdot G^\ast)^2 \cdot { \langle N_{pG, \, photo} \rangle }:$
 \begin{equation}\label{ECFENF}
   ECF = \frac{\langle Q \rangle} {q_0 \cdot G^\ast \cdot \langle N_{pG, \, photo}\rangle} \hspace{5mm} \mathrm{and} \hspace{5mm}
   ENF = \frac{ \sigma_Q^2 } { (q_0 \cdot G^\ast )^2 \cdot \langle N_{pG, \, photo}\rangle  }.
 \end{equation}
 The rms-spread $\sigma _{dark}$ allows to estimate the contribution of dark counts to the $ENF$.

 To illustrate the use of $ENF$, the calculation of the operating voltage at which the photon resolution is optimal is presented.
 Inserting $\langle N_{pG} \rangle$ from Eq.\,\ref{equ:NpG} into Eq.\,\ref{equ:meanQ} and \ref{equ:sigQ} one obtains for the relative resolution
 \begin{equation}\label{equ:Resolution}
   \frac{\sigma _Q} {\langle Q \rangle} = \sqrt{\frac{ENF} {N_{\gamma } \cdot PDE } }.
 \end{equation}
 Both $ENF$ and $PDE$ increase with $V_{bias}$. Whereas $PDE$ eventually saturates, $ENF$ continues to increase, and the relative resolution has a minimum.
 An example of such a dependence is given in Ref.\,\cite{Chmill:2017}.
 As both $ENF$ and $G^\ast$ depend on the effective charge integration time, Eq.\,\ref{equ:Resolution} can also be used to optimise the readout electronics.

 \subsection {Non-linearity and saturation}
  \label{sect:non-linearity}

 For calorimetric measurements, e.\,g. in collider and astro-physics experiments or in PET, the dynamic range, i.\,e. the range of $N_\gamma $, where precise measurements are possible, is an essential parameter.
 High gain, high $PDE$, single photon-detection and high dynamic range are conflicting requirements, which can hardly be achieved simultaneously.
 In this respect vacuum photomultipliers are superior to SiPMs.
 A major complication is that all the nuisance parameters discussed so far enter in one way or another into the dynamic range.
 In addition, the pulse shape is expected to depend on the number of simultaneous Geiger discharges, and because of the pixel recharging time constant, the dynamic range depends on the time distribution of the photons.
 In spite of the high relevance of the dynamic range, systematic studies so far are quite scarce.

\begin{figure}[!ht]
   \centering
   \begin{subfigure}[a]{0.5\textwidth}
    \includegraphics[width=\textwidth]{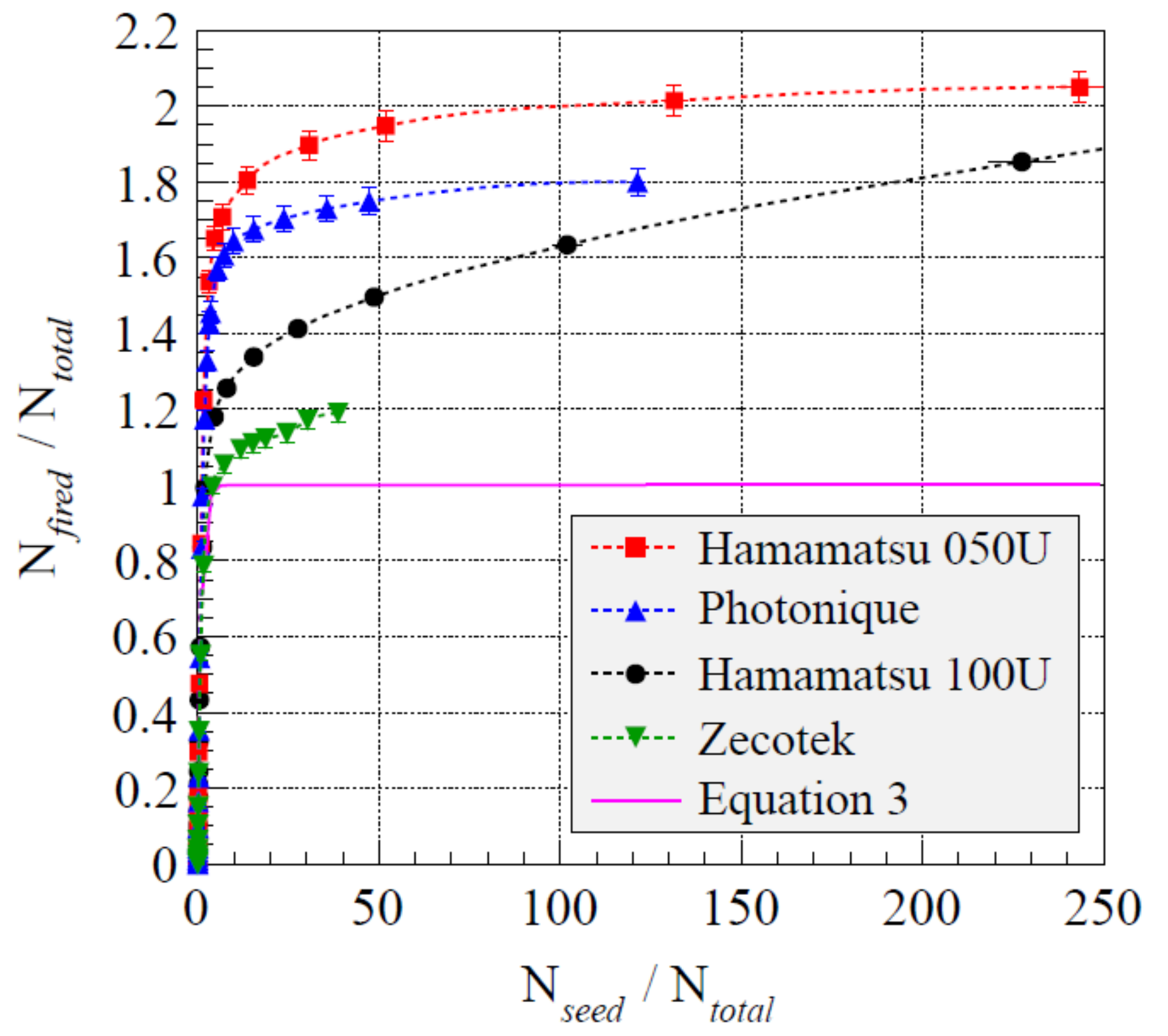}
    \caption{ }
    \label{fig:SatGruber-hi}
   \end{subfigure}%
    ~
   \begin{subfigure}[a]{0.5\textwidth}
    \includegraphics[width=\textwidth]{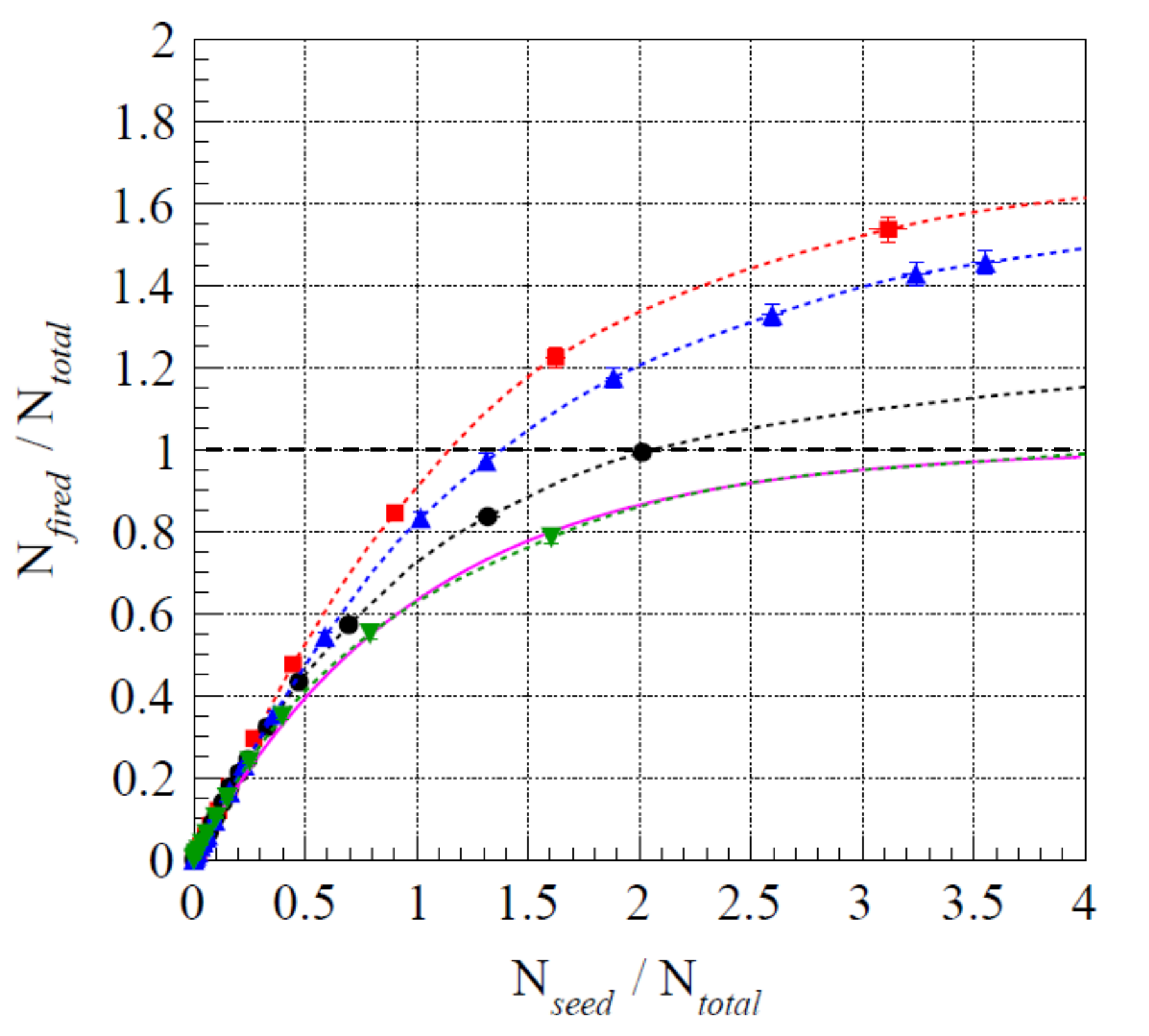}
    \caption{ }
    \label{fig:SatGruber-lo}
   \end{subfigure}%
   \caption{ Dynamic range results using amplitude measurements from Ref.\,\cite{Gruber:2014}. Shown is $\langle N_{G} / N_{pix} \rangle$ (in the figure $N_{fired} / N_{total}$) versus $N_{\gamma,\,rel} / N_{pix}$, the ratio of the expected number of Geiger discharges for a hypothetical linear SiPM without saturation to $N_{pix}$ (in the figure $ N_{seed} / N_{total} $) for
   (a) $\langle N_{G} \rangle / N_{pix} < 250$, and
   (b) expanded view with $\langle N_{G} \rangle / N_{pix} < 4$.
   In all cases a significant excess of the number of measured Geiger discharges above $ N_{pix}$ (lines for  $N_{G} / N_{pix} = 1 $) is observed. }
  \label{fig:SatGruber}
 \end{figure}

 In Ref.\,\cite{Gruber:2014} light from a laser with a wavelength of 404\,nm and a pulse-width of 32\,ps has been used to investigate the dynamic range of simultaneously arriving photons, up to a photon intensity at which $\approx 500 \times N_{pix}$ Geiger discharges would be triggered simultaneously, if the SiPM were linear.
 Four different SiPMs with 1\,mm$^2$ area and $N_{pix} = 100$, 400, 556 and 560 from Hamamatsu, Photonique and Zecotec were investigated.
 Measured was the amplitude of the SiPM pulse recorded with a digital scope as a function of the relative number of photons, $N_ {\gamma,\,rel} $, called $N_{seed}$ in the paper.
 $N_ {\gamma,\,rel} $ is proportional to the current in a PIN photo-diode with a linear response, normalised so that $N_ {\gamma,\,rel} = \langle N_G \rangle$ for low light intensities, where the SiPMs are known to be linear.
 The measurements were performed for $V_{bias} - V_{bd}$ values between 0.5\,V and 1.3\,V.
 The way $V_{bd}$ has been determined is not reported in the paper.
 Fig.\,\ref{fig:SatGruber} shows the results.
 It is observed that at high light intensities for all SiPMs the mean number of measured Geiger discharges, $\langle N_G \rangle$, significantly exceeds the number of pixels, $N_{pix}$, and $\langle N_{G} \rangle$ does not appear to reach a constant saturation value.
 Thus the expectation given in Eq.\,\ref{equ:NGsat} is not observed.
 Various explanations of this phenomenon are discussed in the paper, but the conclusion is: \emph{Up to now, no convincing explanation for this over saturation and enhanced dynamic range could be found.}

 Note, that it is expected that the pulse amplitude, which was used in the described measurements, depends on the number of Geiger discharges in a pixel.
 The reason is that $n$  micro-plasma tubes in a pixel correspond to $n$ resistors $R_d$ in parallel in the electrical model shown in Fig.\,\ref{fig:Emodel}, which results in a decrease of the time constant of the Geiger discharge and thus in a higher amplitude of the fast pulse.
 This effect has  been reported in Ref.\,\cite{Popova:2018}.
 Whereas the charge was found to be independent of $n$, the current amplitude increased with $n$.
 It also should be noted, that the fast component of the pulse has a rise time of typically several tens of ps and a full width below  1\,ns.
 Therefore the measured pulse amplitude is very much influenced by the bandwidth of the readout.

\begin{figure}[!ht]
   \centering
    \includegraphics[width=0.75\textwidth]{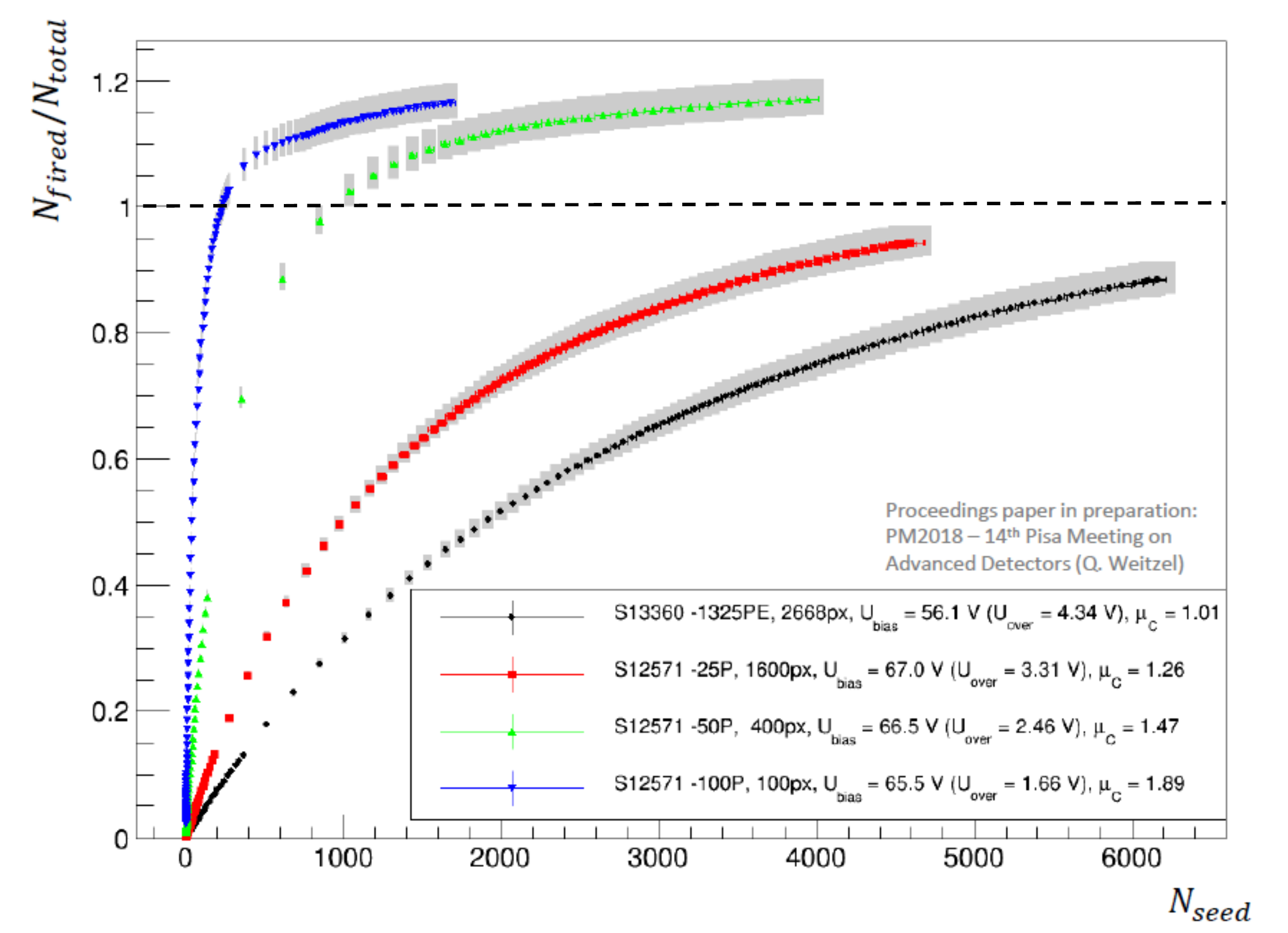}
     \caption{Dynamic-range measurements from Refs.\,\cite{Weitzel:2018,Krause:2018}. The same quantities as in Fig.\,\ref{fig:SatGruber} are shown, however the charge instead of the amplitude was measured for the four different Hamamatsu SiPMs.
     The insert gives bias voltage ($U_{bias}$), over-voltage $(U_{over}$), and $\mu_c$, the ratio of the saturation value of $\langle N_G \rangle / N_{pix}$  from the fit.}
   \label{fig:SaturationJGU}
  \end{figure}

 A  study with a picosecond laser, however using charge- instead of amplitude-measurements, for four SiPMs from Hamamatsu ($N_{pix} / pitch\,[\,\upmu$m] = 2668/25, 1600/25, 400/50 and 100/100) has been presented by G.\,Weitzel\,\cite{Weitzel:2018} and S.\,Krause\,\cite{Krause:2018}.
 The results are shown in Fig.\,\ref{fig:SaturationJGU}.
 $N_{total} (\equiv N_{pix})$ is the number of pixels, $N_{seed}$ the number of Geiger discharges expected in the absence of saturation, and $N_{fired} (\equiv \langle N_G \rangle )$ the measured mean number of Geiger discharges.
 $N_{seed}$ is obtained by scaling $N_\gamma $ so that $N_{seed} = N_{fired}$ for low $N_\gamma $, where the SiPM response is known to be linear.
 The data were fitted by $\mu _c \cdot \big( 1 - e^{-N_{seed}/(\mu _c \cdot N_{total})} \big) $, where $\mu _c = N_{G,\,sat} / N_{pix}$, and $N_{G,\,sat}$ the saturation value of $N_G$ for high light intensities.
 For the SiPMs with $50\,\upmu$m and $100\,\upmu$m pitch,  the values found for $\mu _c$ are significantly larger than 1 and similar to the findings of Ref.\,\cite{Gruber:2014}.
 The SiPMs with $25\,\upmu$m pitch shows  $\mu _c$ values compatible with or closer to 1.
 An explanation for the difference could be the merging of  micro-plasma channels from  Geiger discharges in the same pixel for small pixels.
 Clearly, more studies are needed to understand these  results.

\begin{figure}[!ht]
   \centering
   \begin{subfigure}[a]{0.5\textwidth}
    \includegraphics[width=\textwidth]{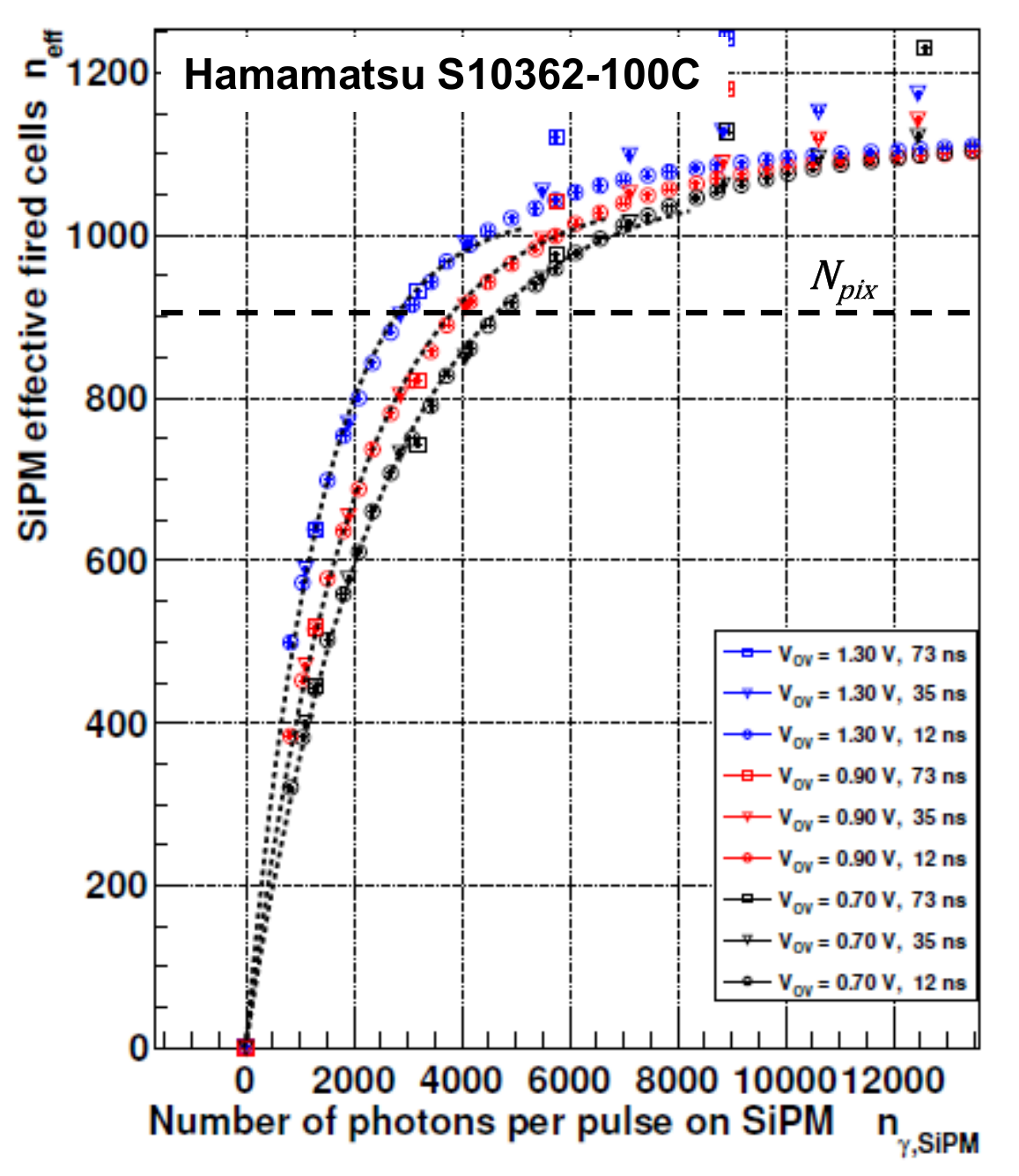}
    \caption{ }
    \label{fig:SatBretzHPK}
   \end{subfigure}%
    ~
   \begin{subfigure}[a]{0.5\textwidth}
    \includegraphics[width=\textwidth]{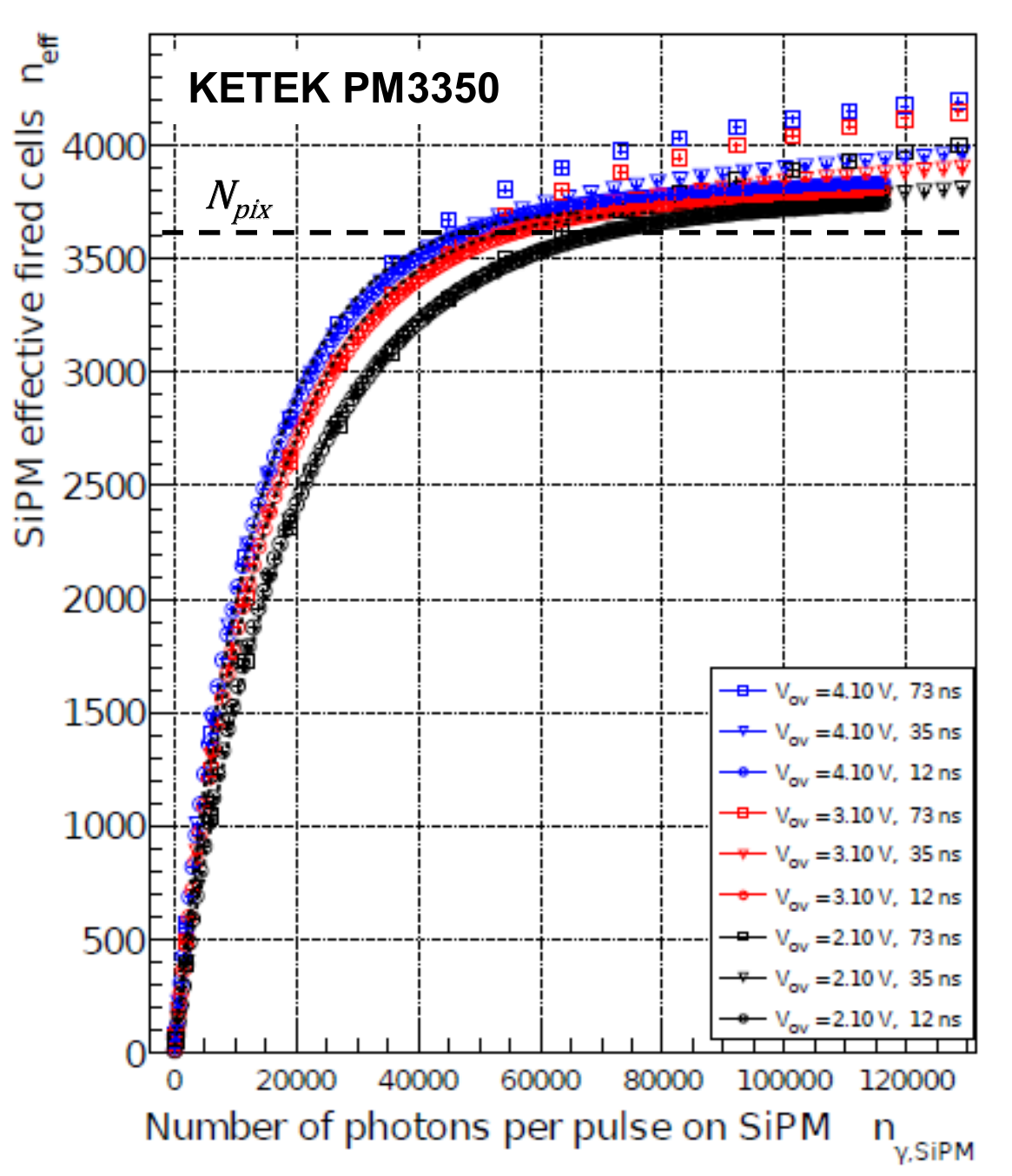}
    \caption{ }
    \label{fig:SatBretzKETEK}
   \end{subfigure}%
   \caption{ Effective number of Geiger discharges as a function of $N _\gamma $ for
   (a) the Hamamatsu SIPM measured with a 150\,ns gate,
   (b) the KETEK SIPM measured with a 400\,ns gate
   from Ref.\,\cite{Bretz:2016}.
   The insert gives the values of $V_{OV}$, and the duration of the light pulse. }
  \label{fig:SatBretz}
 \end{figure}

 In Ref.\,\cite{Bretz:2016} the dynamic range for light pulses of different durations is studied.
 As light source a LED is used, driven by a computer controlled pulse generator to generate light pulses of up to  $\tau _{light}= 100$\,ns duration and photon numbers hitting the SiPM, $N_ \gamma$, of up to $> 10^5$.
 The SiPMs investigated were fabricated by KETEK with $N_{pix} = 3600$ and $pitch = 50\,\upmu$m, and by Hamamatsu  with $N_{pix} = 900$ and $pitch = 100\,\upmu$m.
 Fig.\,\ref{fig:SatBretz} shows the dependence of $\langle N_G \rangle$ after correction for dark pulses (called $n_{eff}$) on $N_\gamma$.
 The value of $N_{pix}$ is shown as a dashed line.
 For the KETEK SiPM the charge was recorded with a 400\,ns gate and for the Hamamatsu SiPM with a 150\,ns gate.
 The authors conclude that in the linear range (which extends to $n_{eff} \approx 0.2 \cdot N_{pix}$), the response is independent of $\tau _{\,light}$.
 For higher $N _\gamma $ values the response increases with $\tau _{light}$, as does the saturation value.
 The ratio of the saturation value to $N_{pix}$ for the SiPM with $100\,\upmu$m pixels is larger than for the one with $50\,\upmu$m, which agrees with the observations of Refs.\,\cite{Weitzel:2018,Krause:2018}.

 To summarise: The dynamic range of SiPMs is determined by four factors: The number of pixels, $N_{pix}$, the pixel recharging time, $\tau _r$, the correlated noise, and the pixel occupancy due to dark counts.
 As long as the number of photons generating simultaneously primary Geiger discharges and the pixel occupancy by dark counts $DCR \cdot \tau _r / N_{pix} \lesssim 0.2 $, the response is close to linear and independent of the arrival time of the photons.
 If this is not the case and  the probability of two or more Geiger discharges in a pixel during the time interval  $\tau _r$ is significant, the response becomes non-linear and dependent on the arrival-time distribution of the photons and the readout electronics.
 In this case a quantitative understanding of the response and its parametrisation based on a physical model is complicated,.
 So far, in the author's opinion, the situation is not fully understood, however as discussed in Ref.\,\cite{Bretz:2016}, for a given situation, the response function can be measured, and phenomenological parameterisations found and used to correct for the non-linearity.
 The observation of an increase in the mean number of Geiger discharges, $\langle N_G \rangle$, as a function of $N_\gamma $ beyond $N_{pix}$ and the observation that the pulse shape changes with the number of Geiger discharges in a single pixel, may point a way towards extending the measurement capabilities of SiPMs into the domain of high $N_ \gamma$, where its response is  highly non-linear.
 So far a detailed general study of the non-linearity and the worsening of the resolution caused by high pixel occupancies from dark counts caused by radiation damage, is also lacking.
 This is of particular relevance for the upgrade of the experiments at the Large Hadron Collider, LHC at CERN, where SiPMs will be exposed to high  fluences of hadrons.

 \section{Conclusions and outlook}
 \label{sect:Conclusions}

 SiPMs have already found a broad range of applications, which is illustrated in the different contributions to this Special Issue of Nuclear Instruments and Methods devoted to SiPMs.
 Given the many new ideas for future applications, it is certain that the use of SiPMs will continue to expand.
 Well documented methods of characterisation, from which concise specifications can be derived, will become more and more important.
 The paper is an attempt to give an overview and clearly define the parameters required to describe the performance of SiPMs, discuss  different characterisation methods, point out some of their limitations and give a number of recommendations.

 The main aims of the efforts on SiPM characterisation are:
  \begin{enumerate}
    \item Provide a basis for specifications by the vendors, which allow users to choose  the SiPM best suited for the intended application.
    \item Enable quality control and sample selection.
    \item Provide a basis for the development of the calibration and analysis methods for a given application.
    \item Improve the basic understanding of SiPMs, which is the input for further improving their performance.
  \end{enumerate}

 The characterisation methods which can be successfully used depend on the application regime, which  are grouped in four classes:
 \begin{enumerate}
   \item \emph{Low light level, temperature range} $- 30 \,^\circ $\rm C \, \emph{to} $ + 30 \,^\circ $C, \emph{no radiation damage, no ambient light}:
     In these conditions pulses from 0, 1, 2, etc. Geiger discharges can be separated.
     From  spectra recorded in the dark and with pulsed light, recorded either by a QDC, by integrating  the current transient or by measuring the pulse amplitude after pulse-shaping, $DCR$ (dark count rate  ), $G^\ast$ (gain), $V_{off}$ (turn-off voltage), relative $PDE$ (photon-detection efficiency), $ECF$ (excess charge factor) and $ENF$ (excess noise factor) can be determined.
     Various methods are described in the paper.
     The measurements should be done as a function of $V_{bias}$ (bias voltage) and for a few temperatures.
     The wavelength of the pulsed light source should be close to the wavelength relevant for the application.
     If necessary, the absolute $PDE$ should be measured, which however is quite involved.
     In our view no detailed analysis of correlated pulses is required for most applications.
     The knowledge of $ECF$ is sufficient to obtain the absolute number of photons producing primary Geiger discharges, and the optimum operating point with respect to photon resolution can be obtained from $DCR$, $ECF$ and $ENF$.
     If the intention is to operate the SiPM at high over-voltages, the dark current, $I_{dark}$, as a function of $V_{bias}$ should be investigated.
     It is not recommended to use the SiPM at $V_{bias}$ values at which the slope of $\ln (I_{dark})$ shows a second increase.
   \item \emph{Like 1., however for cryogenic temperatures}:
    Given the large interest to use SiPMs at cryogenic temperatures in Dark Matter, neutrino-less double beta decay and neutrino oscillation experiments (Ref.\,\cite{Retiere:2018}) significant efforts are devoted to developing and characterising SiPMs, which work at temperatures down to 100\,K and even lower.
    At these low temperatures many properties of doped silicon change significantly compared to room temperature.
    This has a major impact on their performance, as discussed in Ref.\,\cite{Collazuol:2011}.
    The methods used so far are similar to those at higher temperatures.
    An example is Ref.\,\cite{Acerbi:2017, Aalseth:2017}.
   \item \emph{Similar to 1., however for high light intensities}:
    In addition to the calibration suggested in 1., spectra should be recorded at high photon numbers, $N_\gamma $, and both mean values, $\langle Q \rangle $, and rms-spread, $\sigma _Q$, determined.
    The relative number of photons is obtained from a linear photon-detector, e.\,g. a photo-diode looking at the light source.
    The normalisation is obtained at low light intensities, where the SiPM is expected to be linear, from $N_\gamma ^{norm} = \langle N_{pG} \rangle$ using $\langle N_{pG} \rangle$ from Eq.\,\ref{equ:NpG}.
    In this way the calibration curve, $Q(N_\gamma ^{norm})$, is obtained, which can be used to correct the measured $Q$ for the non-linearity.
    As the non-linearity depends on the arrival time of the photons, it is important that the pulse shape of the calibration pulse is similar to the pulse shape of the intended application.
    To a lesser extent, the non-linearity also depends on the wavelength.
    Therefore the wavelength used for the calibration should be similar to the one of the intended application.

   \item \emph{High dark count rate   due to radiation damage or ambient light}:
    If possible, the characterisation described in 1. should be performed in a situation, where the $DCR$ is low, e.\,g. before irradiation.
    In addition, the dark current, $I_{dark}$, and the current with DC-light, $I_{light}$, should be measured as a function of $V_{bias}$.
    The comparison of $I_{photo}=I_{light}-I_{dark}$ for the low and high $DCR$ situation already gives a good idea on the possible reduction of $PDE$ due to high pixel occupancy by dark counts.
    Methods of how to determine $DCR$ from $I_{dark}$, and how from $\sigma _{dark}$ (spread of the recorded spectrum without illumination) are presented in the paper.
    The method using $I_{dark}$ appears to be the more reliable one.
    In addition, a method of how to determine the relative $PDE$ and $G^\ast$ from the mean, $\langle Q \rangle$, and rms-spread, $\sigma _Q$ of the spectrum with illumination in the absence of saturation effects, is presented.
    An extension of this method including saturation effects, which is relevant at high $DCR$ values or high light intensity, still has to be developed.
 \end{enumerate}

 In spite of large and highly successful efforts to characterise SiPMs, a lot of work remains to be done:
 Examples are the characterisation at cryogenic temperatures, the determination of the non-linearities at high photon intensities and the improvement of the measurement and analysis methods of highly radiation-damaged SiPMs.

 \section{Appendix}
 \label{sect:Appendix}

   In this appendix Eq.\,\ref{equ:SigDark}, the relation between the variance, $\sigma _{dark} ^2$, and the primary dark count rate, $DCR_p$, for current pulses $I(t) = (q_0 \cdot G / \tau)\cdot e^{-t/\tau}$  occurring randomly at the rate $DCR_p$ and integrated in the time interval $t_{gate}$, including the effects of correlated noise is derived.
  First, a single (1) SiPM pulse occurring at $t = 0$ is considered, with the current transient
  \begin{equation}\label{equ:I(t)}
    I_1(t) = \left\{
           \begin{array}{rll}
             0 & & \hbox{\rm{for}\,\, $t < 0$,} \\
    \vspace{1mm}
             f(t) & = \frac{1} {\tau } \cdot e^{-t/\tau} & \hbox{\rm{for}\,\, $t \geq 0$.}
           \end{array}
         \right.
  \end{equation}
  The integral $Q_1$ of the current for a gate starting at $t = t_1$ with width  $t_{gate}$ is
  \begin{equation}\label{equ:Q(tgate)}
    Q_1(t_1) = \left\{
           \begin{array}{rll}
             0 & & \hbox{\rm{for}\,\, $t_1 < -t_{gate}$,} \\
    \vspace{1mm}
             \int _0 ^{t_{gate}+t_1} f(t) \, \rm{d}$t$& = 1 - e^{-(t_{gate}+t_1)/\tau} & \hbox{\rm{for}\, $ -t_{gate} \leq t_1 < 0$,} \\
    \vspace{1mm}
             \int _{t_1} ^{t_{gate}+t_1} f(t) \, \rm{d}t & = e^{-{t_1/\tau}} \dot (1- e^{- t_{gate}/\tau}) & \hbox{\rm{for}\, $ t_1 \geq 0$.}
           \end{array}
         \right.
  \end{equation}
 Next,  the mean charge $\langle Q_1(\Delta t_1) \rangle $ and the variance $\sigma_1^2(\Delta t_1) = \Big\langle \Big(Q(t_1) - \langle Q_1( \Delta t_1) \rangle \Big)^2 \Big\rangle$ for the $t_1$ interval from $-t_{gate}$ to $t_0$, denoted $\Delta t_1 = t_{gate} + t_0$, are calculated.
 The value of $t_0$ is not relevant, as only the limes $\Delta t_1 \rightarrow \infty$ is relevant.
 One finds
 \begin{equation}\label{equ:Mean}
   \langle Q_1(\Delta t_1)\rangle = \lim_{\Delta t_1 \to \infty}\frac {\int _{\Delta t_1}  Q(t_1) \, \mathrm{d}t_1} {\Delta t_1} =
   \lim_{\Delta t_1 \to \infty} \frac{t_{gate}+\tau \,e^{-t_0/\tau} \, (e^{-t_{gate}/\tau}-1)} {\Delta t_1} = \frac{t_{gate}} {\Delta t_1}
 \end{equation}
 For the variance a similar calculation  gives
  \begin{equation}\label{equ:Var}
   \sigma _1 ^2 ( \Delta t_1) = \lim_{\Delta t_1 \to \infty}\frac {\int _{\Delta t_1} \big( Q(t_1) - \langle Q _1 (\Delta t_1) \rangle \big)^2\, \mathrm{d}t_1} {\Delta t_1} = \frac{t_{gate}-\tau \,(1-e^{-t_{gate}/\tau})} {\Delta t_1}.
 \end{equation}
 For the dark count rate  $DCR_p$, there will be on average $N_{DC} = DCR_p \cdot \Delta t_1$ dark counts in the time interval $\Delta t_1$, and the pulse height distribution will be the convolution of $N_{DC}$ single pulses with the mean
  \begin{equation}\label{equ:MeanDCR}
   N_{DC} \cdot \langle Q_1 \rangle = DCR_p \cdot t_{gate},
  \end{equation}
    and the variance
  \begin{equation}\label{equ:VarDCR}
   N_{DC} \cdot \sigma _1 ^2 = DCR_p \cdot \big(t_{gate}-\tau \cdot (1-e^{-t_{gate}/\tau})\big).
  \end{equation}
 For finite $\Delta t_1$ values one has to  take into account that $N_{DC}$ is distributed according to a Poisson distribution, however in the limit $\Delta t_1 \rightarrow \infty$ the Poisson distribution approaches a $\delta $-function at $N_{DC}$ and its contribution to the variance vanishes.


 Eq.\,\ref{equ:VarDCR} describes the variance for $q_0 \cdot G = 1$ in the absence of correlated noise from cross-talk and after-pulses.
 The effect of $ECF$ and $ENF$ is taken into account by replacing $ \langle N_{pe} \rangle$ in Eq.\,\ref{equ:sigQ} by $N_{DC} \cdot \sigma _1 ^2$ from Eq.\,\ref{equ:VarDCR}, from which Eq.\,\ref{equ:SigDark}
  \begin{equation}\label{equ:SigDCR}
   \sigma _{dark} ^2 = \big ((q_0 \cdot G)^2 \cdot ENF \cdot ECF ^2 \cdot DCR_p \big) \cdot \big(t_{gate}-\tau \cdot (1-e^{-t_{gate}/\tau})\big)
  \end{equation}
  is obtained.
  For $t_{gate} \gg \tau$ the term in the right parenthesis is  $\approx t_{gate} - \tau $, and for $t_{gate} \ll \tau$ it is $\approx t_{gate} ^2 /2 \tau$.

 \section*{Acknowledgement}
 \label{sect:Acknowledgement}

 I am  grateful to the reviewers, who made many suggestions and thereby helped to improve the quality of the paper.
 I also want to thank the members of the Hamburg Detector Laboratory working on SiPMs for the measurements they performed, some of which are discussed in this paper, and for many comments and fruitful discussions.

\input{bibliography}

 \label{sect:Bibliography}




\end{document}

%% file: bibliography.tex
\section{List of References}